\documentstyle[12pt,epsfig]{article}
\begin{document}

\thispagestyle{empty}
\renewcommand{\thefootnote}{\fnsymbol{footnote}}

\setcounter{page}{0}
\begin{flushright} 
DESY 99--016\\
February 1999
\end{flushright}

\begin{center}
\vspace*{0.5cm}
{\Large\bf
REGGE DESCRIPTION OF SPIN-SPIN ASYMMETRY IN PHOTON DIFFRACTIVE
DISSOCIATION
}\\
\vspace{1.4cm}
{\sc  S.~I.~Manayenkov\footnote{This work was supported 
by Volkswagen Stiftung}}\\
\vspace*{0.3cm}
{\it Petersburg Nuclear Physics Institute,\\
Russian Academy of Science,\\
Gatchina, St.Petersburg district, 188350, Russia\\}
\vspace*{0.3cm}
{\bf Abstract}\\
\end{center}
\parbox[t]{\textwidth}
{We explore the possibility whether the gluon helicity distribution
$\Delta G(x)$ can be extracted from a
comparison of experimental
data on the longitudinal spin-spin asymmetry $A_{LL}$ in  $\gamma p$
diffractive deep inelastic scattering with calculations performed
within the framework of perturbative QCD (pQCD). The data could
be
obtained at the
future HERA collider in scattering of
polarized electrons/positrons off polarized protons.
In this paper we look for such kinematical regions where contributions
to $A_{LL}$ from soft processes (reggeon exchanges) are suppressed
to guarantee an applicability of pQCD. It is shown that
for the square of the center-of-mass
energy $s_{\gamma p} \geq 10^3$ GeV$^2$,
the hadronic diffractive mass $M_X \leq 10$ GeV/c$^2$, the momentum
transferred
to the proton $\Delta _T\leq 0.5$ GeV/c, and $Q^2\geq4$ (GeV/c)$^2$ the
longitudinal
spin-spin asymmetry due to reggeon exchanges is less than $10^{-4}$.
This value is presumably lower than
the asymmetry which can be measured with modern experimental technique.
This means that the pQCD prediction can be reliably compared
with data in this kinematical region.}
\vspace*{0.3cm}
\newpage
\renewcommand{\thefootnote}{\arabic{footnote}}
\setcounter{footnote}{0}

\newpage
\renewcommand{\thefootnote}{\arabic{footnote}}
\setcounter{footnote}{0}

\section{Introduction}

Study of double spin asymmetries in deep inelastic scattering (DIS) of
leptons on nucleons is of great importance as it provides valuable
information about spin-dependent parton distributions in the nucleon. An
extraction of the gluon helicity distribution from DIS is the most
difficult procedure and gives large uncertainties as the gluon is an
electrically neutral particle. Recently it has been argued that the
diffractive   $\rho$-meson, open charm electroproduction
 \cite{GNZ}, \cite{LMRT}, \cite{BRG}, \cite{R}
 and the production
of di-jets \cite{NNNZ}, \cite{RRM}, \cite{RR}, \cite{BGSV} in hard
collisions of polarized electrons with
protons can be
used for the investigation of the unpolarized gluon density and the gluon
helicity distribution in the
proton. Such predictions have been made in the framework of 
perturbative QCD when typical transverse momenta of produced particles
and total created masses are large enough for 
perturbative QCD to be applicable.
To decrease statistical errors we have to consider
events with hadron transverse momenta and total created masses of the
order of 1 GeV where correction of soft processes can be appreciable. So we
need an estimate of the corrections to perturbative QCD predictions. 

The physical picture of photon diffractive dissociation is as follows.
The virtual photon produced by a scattered lepton dissociates into a
quark-antiquark pair which gives final hadrons after rescattering on
the proton. In perturbative QCD, scattering of the $q \bar{q}$-pair off
the proton is described by the gluon ladder graphs which correspond to
the
hard part of pomeron exchange. To estimate the soft part
contribution to the amplitude of $q \bar{q}$-pair scattering off the
proton we shall make use of the Regge phenomenological approach. In the
sixties and seventies the Regge complex angular momentum theory has been 
very successfully applied to the description of elastic scattering and
charge exchange reactions for different hadrons (see, for
example, reviews
\cite{KS}, \cite{IW}). We shall
apply the parameters of the Regge trajectories and the hadron-reggeon
vertices presented in \cite{IW}, \cite{BST}, \cite{BLSTM}
which have been
found from the phenomenological analysis of 
experimental data on hadron-hadron collisions at high energies.   
To extract quark-reggeon vertices we make use of the nucleon wave function
in the naive quark model \cite{JJJ}. The obtained quark-reggeon vertex
parameters depend of course on the applied model of the nucleon but we
hope that for our rough estimates of the order of magnitude of the soft
part contribution to the longitudinal spin-spin asymmetry this approach
gives
reasonable results. It is true mainly due to the fact that the main aim 
of the present paper consists in finding such kinematical conditions
where the soft process contribution to
the asymmetry is much less than the prediction of perturbative QCD
and hence we do not need the real theoretical calculation of the Regge
pole contribution but some estimate only.

In the Regge phenomenology all invariant amplitudes of quark proton
scattering due to exchange with a reggeon $R$ contain the  factor
$(s/s_0)^{\alpha _R(0)-1}$ where $s$ is the square of the total
center-of-mass energy of the colliding particles, $s_0 \sim 1$ GeV$^2$
denotes the parameter in the Regge theory and $\alpha _R(t)$ is a Regge
trajectory depending
on $t$ (the square of the reggeon momentum). For the
vacuum (pomeron) trajectory $\alpha _R(0) \approx 1$, for $f,\;\rho,\;
\omega,\;A_2$ reggeons $\alpha _R(0) \approx 0.5$ and for the $\pi$ and
$A_1(1260)$ trajectories $\alpha _R(0)\leq 0$. We see that there is the
hierarhy
of reggeon contributions valid at large $s$. Since both for quark and
antiquark scattering on the proton the square of the total center-of-mass
energy is proportional to $s_{\gamma p}$ (the Mandelstam variable for the
$\gamma p$-collision) it is convenient to introduce the small parameter
$\epsilon=\sqrt{s_0/s_{\gamma p}}$ for the classification of reggeon
exchange
contributions. We decompose the cross section and the longitudinal
spin-spin asymmetry, $A_{LL}$ into a power series in $\epsilon$ and study
properties of terms $\sim \epsilon ^0$, $\epsilon ^1$ and $\epsilon ^2$.  

We apply our formul\ae$\;$  
to $\gamma p$-scattering at energies achieved at
the HERA collider. We have found that though the pure pomeron
contribution to $A_{LL}$ is a quantity $\sim \epsilon ^0$ it is
numerically much smaller than the contributions $\sim \epsilon ^1$ and
$\epsilon ^2$ at $s_{\gamma p}=10^2$ to $10^5$ GeV$^2$ and for the
zero momentum transfer. More over the contributions
$\sim \epsilon ^1$ to $A_{LL}$ representing the interference terms between
pomeron and  $f,\;\rho,\;\omega,\;A_2$ exchanges are numerically much
less than the terms $\sim \epsilon ^2$ at the HERA energies. Here we
suppose the pomeron contribution to be a sum of the pole term and the cuts
in the complex plane of the angular momentum due to  exchanges with the
$n$ pomerons ($n \geq 2$). The $f,\;\rho,\;\omega,\;A_2$ contributions are
also assumed to be  sums of $f,\;\rho,\;\omega,\;A_2$ exchanges and some
number of pomeron exchanges. The explanation of this numerical
hierarhy being inverse to the parameteric hierarhy $\epsilon ^0 \gg
\epsilon ^1 \gg \epsilon ^2$ is as follows. The fit \cite{BLSTM},
\cite{BLST} of
experimental data has shown that the spin-dependent parts of
the quark-quark-pomeron  and nucleon-nucleon-pomeron   
vertices are small compared with the scalar parts. Besides the pomeron
contribution to the numerator in the formula for $A_{LL}$ at $t=0$
contains the sum of the squares of the invariant amplitudes 
which vanish in the one and two pomeron exchange approximations. The
nonzero invariant amplitudes contain the third or 
higher powers of the small spin-dependent pomeron vertices and this leads 
to the very small ($\leq 10^{-12}$) contribution to $A_{LL}$. For nonzero
momentum transfers (at $\Delta _T \sim 1$ GeV/c) $|A_{LL}|$ is greater by
few orders of magnitude than at $t=0$. 
The contributions to $A_{LL}$ $\sim \epsilon ^1$ and $\sim \epsilon ^2$ 
became numerically comparable with each other at $|t| \sim 1$ (GeV/c)$^2$
but the contribution $\sim \epsilon ^0$ is much smaller than the first and 
second order contributions to $A_{LL}$.

Formally the amplitudes of $\pi$ and $A_1(1260)$ exchanges are quantities 
$\sim \epsilon ^2$ and are to be smaller than the amplitudes of
$f,\;\rho,\;\omega,\;A_2$ exchanges which are $\sim \epsilon ^1$. This is
true for $s_{\gamma p} \rightarrow \infty$ but it is wrong numerically at
the HERA energies. Indeed, we have for the square of the quark-proton
center-of-mass energy the relation $s=z s_{\gamma p} $ where $z$ is the
fraction of the virtual photon momentum carried by the quark (in the
light front system of frame). 
It turns out that the dominant contribution of quark-proton scattering 
to $A_{LL}$ originates from the region in which $z$ is close to
its minimal value $z_{min} \approx (m_q^2+k_T^2)/M_X^2$. 
We denote by $M_X$ the
mass of hadrons produced in hard $\gamma p$-scattering,  $k_T$ and $m_q$
are the transverse momentum and the mass of the quark, respectively.
For experiments at HERA $z_{min}$ can be $\sim 10^{-2}$ - $10^{-3}$, hence
$s$ can be
$\sim 1$ GeV$^2$ even for $s_{\gamma p} \sim 10^2$ - $10^3$  GeV$^2$. It
is clear that the contributions of $\pi$, $A_1(1260)$ reggeons to $A_{LL}$
at $s \sim 1$ GeV$^2$ can be of the same order of magnitude as the 
$f,\;\rho,\;\omega,\;A_2$ contributions. 
Our numerical calculations show that we can neglect the $A_1(1260)$
contribution to $A_{LL}$ but the $\pi$-reggeon gives the appreciable
contribution and sometimes the dominant contribution to $A_{LL}$ even at 
$s_{\gamma p} \geq 10^3$ GeV$^2$ 
due to the large value of the $\pi N N$ coupling constant.

Considering, for example, experimental
events with relatively low $M_X$ (say, $M_X \leq 3$ GeV/c$^2$) or
large $k_T$ ($k_T \geq 1$ GeV/c) we increase a value of $z_{min}$ and
hence we  increase the minimal center-of-mass  
energy of quark-nucleon scattering (and antiquark-nucleon scattering
too). As a result we  reduce the contributions of secondary reggeons both
with the natural parity  ($f,\;\rho,\;\omega,\;A_2$) 
and with the unnatural parity ($\pi$, $A_1$).
Their contribution to $A_{LL}$ can be suppressed down to
$10^{-4}$. This value is presumably the low limit for the
longitudinal spin-spin asymmetry which can be measured by modern
experimental techniques. The perturbative QCD contribution to
$A_{LL}$ can be calculated most unambiguously from the theoretical
point of view. 
If it is larger than $10^{-4}$, then the perturbative QCD prediction
can be reliably compared with an experimental value of the
longitudinal spin-spin asymmetry which can be obtained at
the future HERA collider in scattering of the polarized
electrons/positrons off the polarized protons. 

The paper is organized as follows. In Sections~2 and 3 we consider
spectator graphs only and discuss contributions to the longitudinal   
spin-spin asymmetry of  pure pomeron exchanges  and  
contributions of secondary Regge trajectories, respectively. Properties of
nonspectator graphs are considered in Section~4. Results of numerical
calculations of the asymmetry and discussion of cuts suppressing its value
are presented in Section~5. Main results are summarized in the Conclusion.
The most complicated formul\ae$\;$ applied for the numerical calculations
are given in the Appendix.

\section{Spectator diagrams. Pomeron contribution }

We consider in the present paper unenhanced Regge diagrams only and divide
the pomeron exchange graphs into two sets. The first type diagrams,
called spectator graphs, are
shown in Fig.~1a and Fig.~1b. They describe scattering of a quark (or
antiquark)
off the nucleon, the other particle of the $q \bar{q}$-pair being a
spectator.
The graphs of the second type are presented in Figs.~2a and 2b. They
describe scattering both of the quark and antiquark on the nucleon and
are named nonspectator graphs. 
\vspace{0.5cm}
\begin{center}     
\begin{tabular}{cc}
\mbox{\epsfig{file=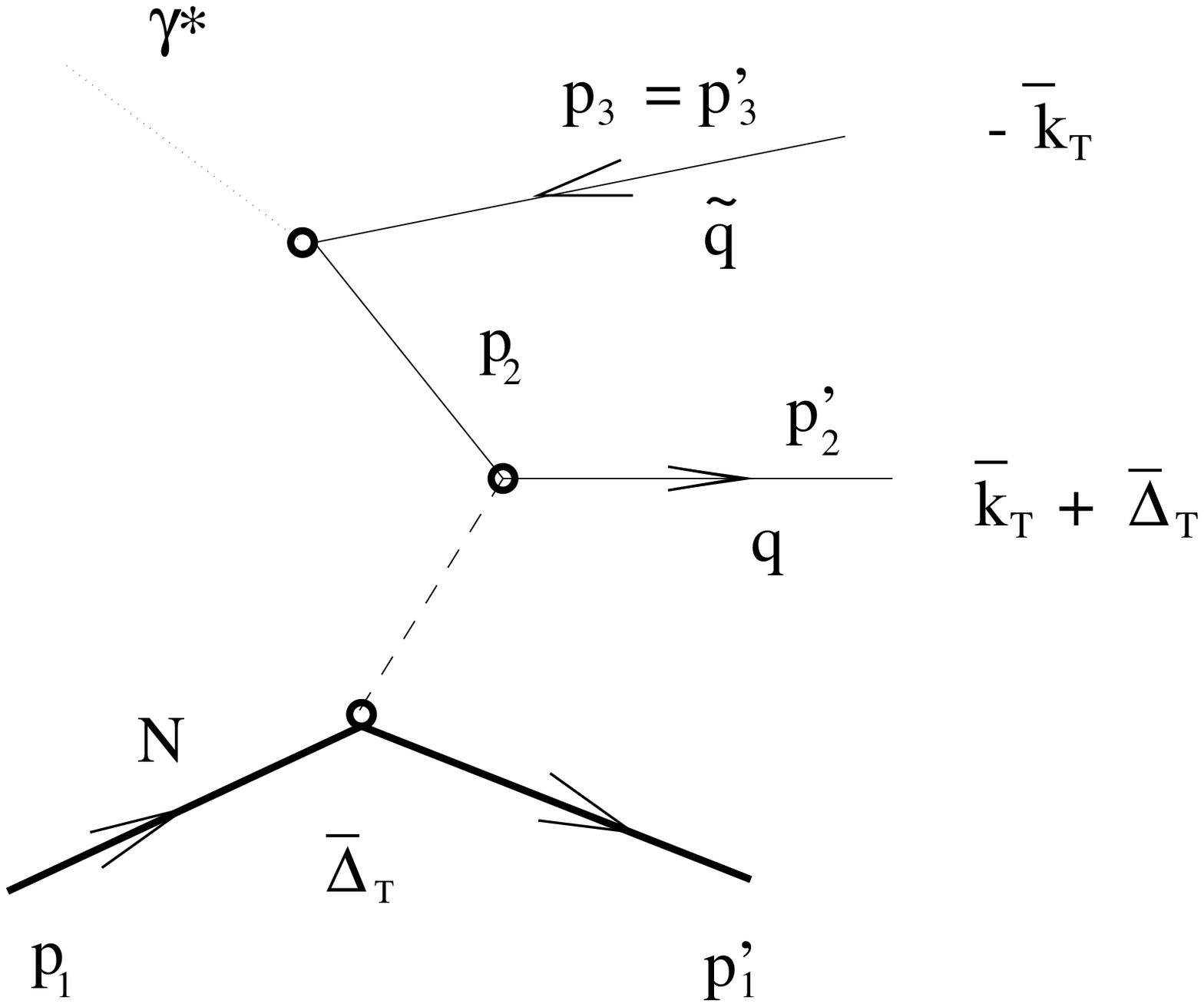,height=6cm,width=7.5cm}}
\vspace{2mm}
\noindent
\small
&
\mbox{\epsfig{file=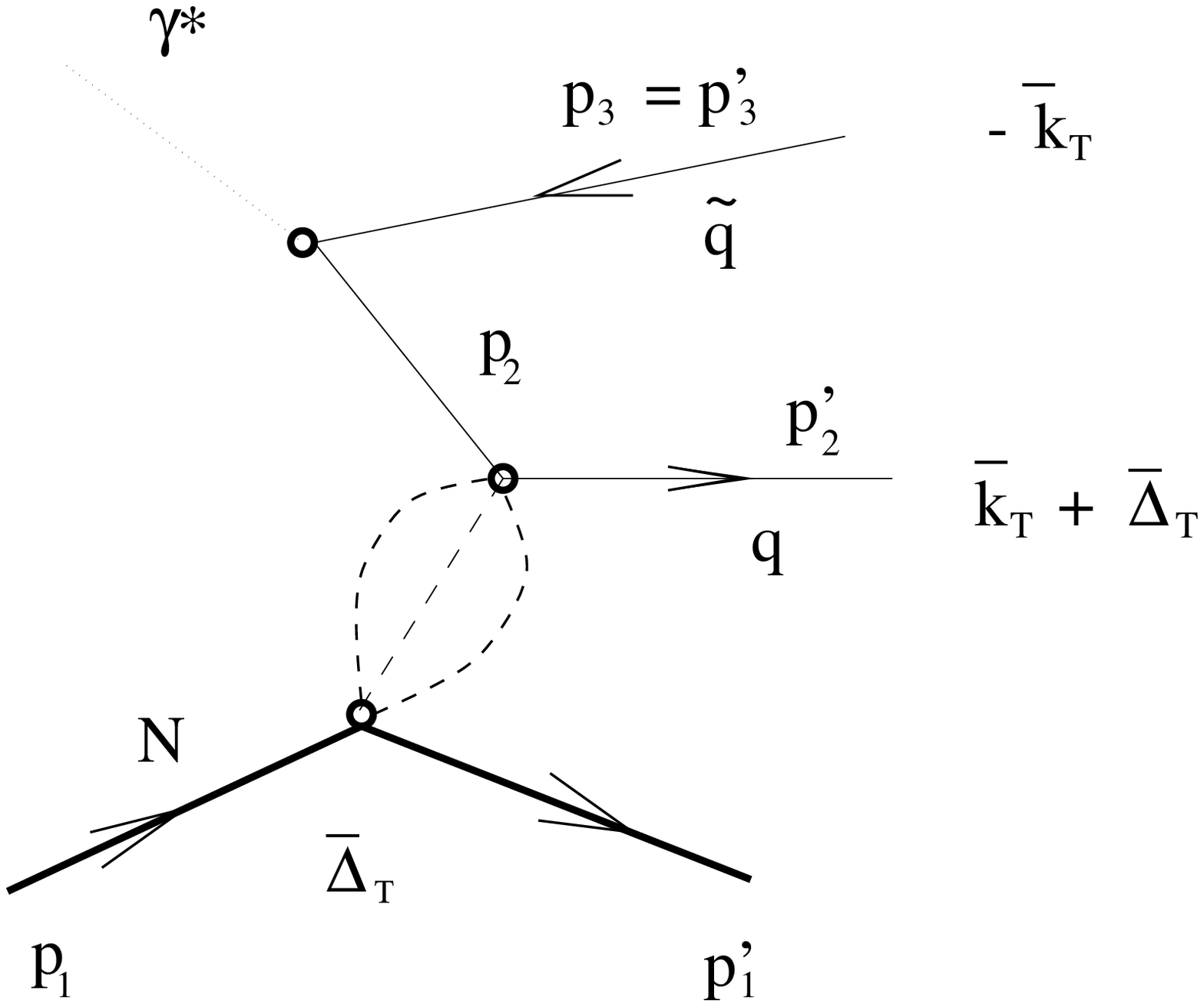,height=6cm,width=7.5cm}}
\vspace{2mm}  
\noindent
\small
\\
\begin{minipage}{7cm}
{\sf Fig.~1a:}
One reggeon exchange spectator graph. Dotted, thick straight and dashed
lines
correspond to photon, nucleon and reggeon, respectively. Straight lines
describe either quarks or antiquarks.
\end{minipage}
 \normalsize
&
\begin{minipage}{7.5cm}
{\sf Fig.~1b:}
Spectator diagram with three reggeon exchanges.
Lines have  the same meaning as in Fig.~1a.
 \end{minipage}
\normalsize
\end{tabular}
\end{center}  
\newpage
\vspace{0.5cm}
\begin{center}
\begin{tabular}{cc}
\mbox{\epsfig{file=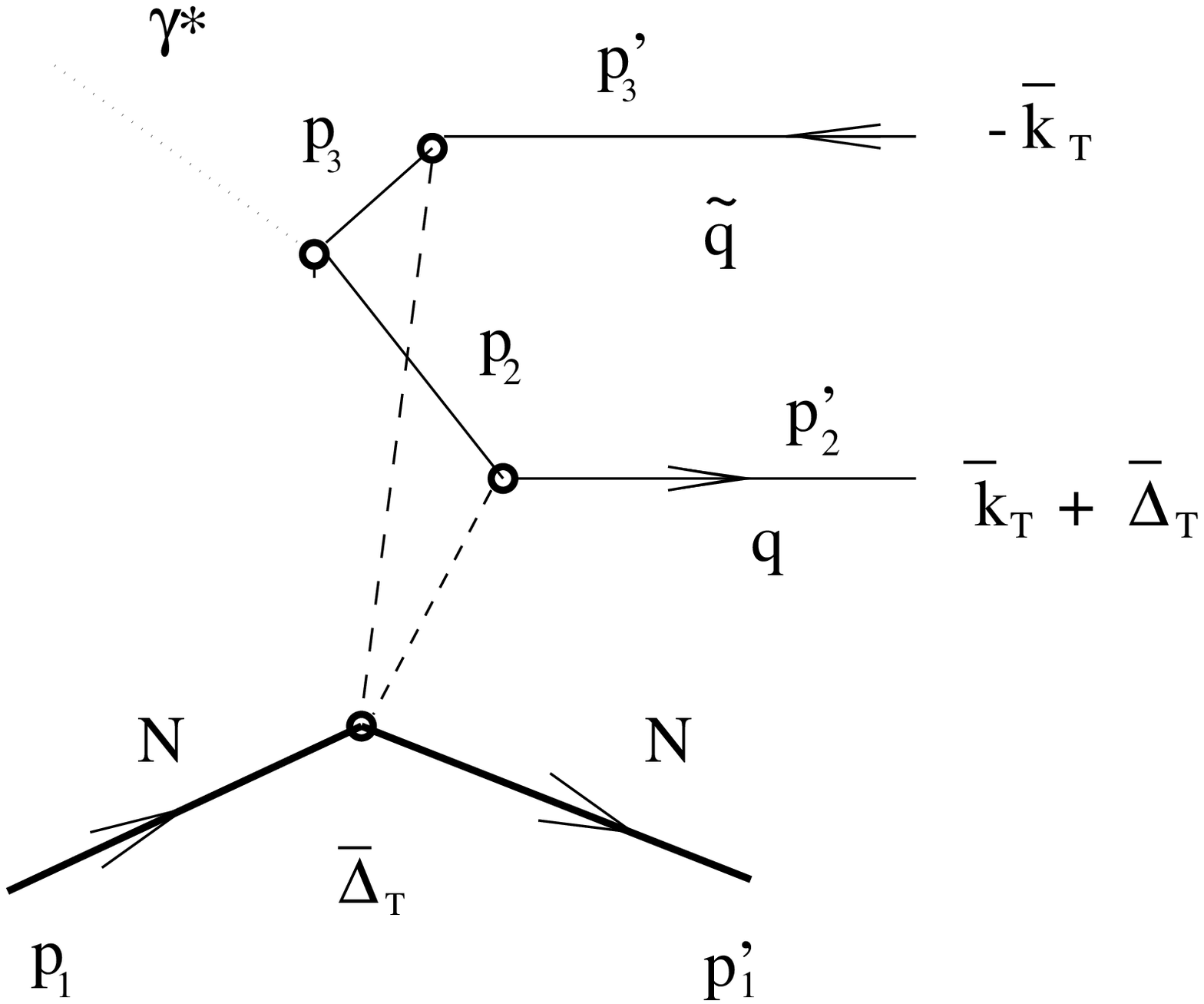,height=5cm,width=7.5cm}}
\vspace{2mm}
\noindent
\small
&
\mbox{\epsfig{file=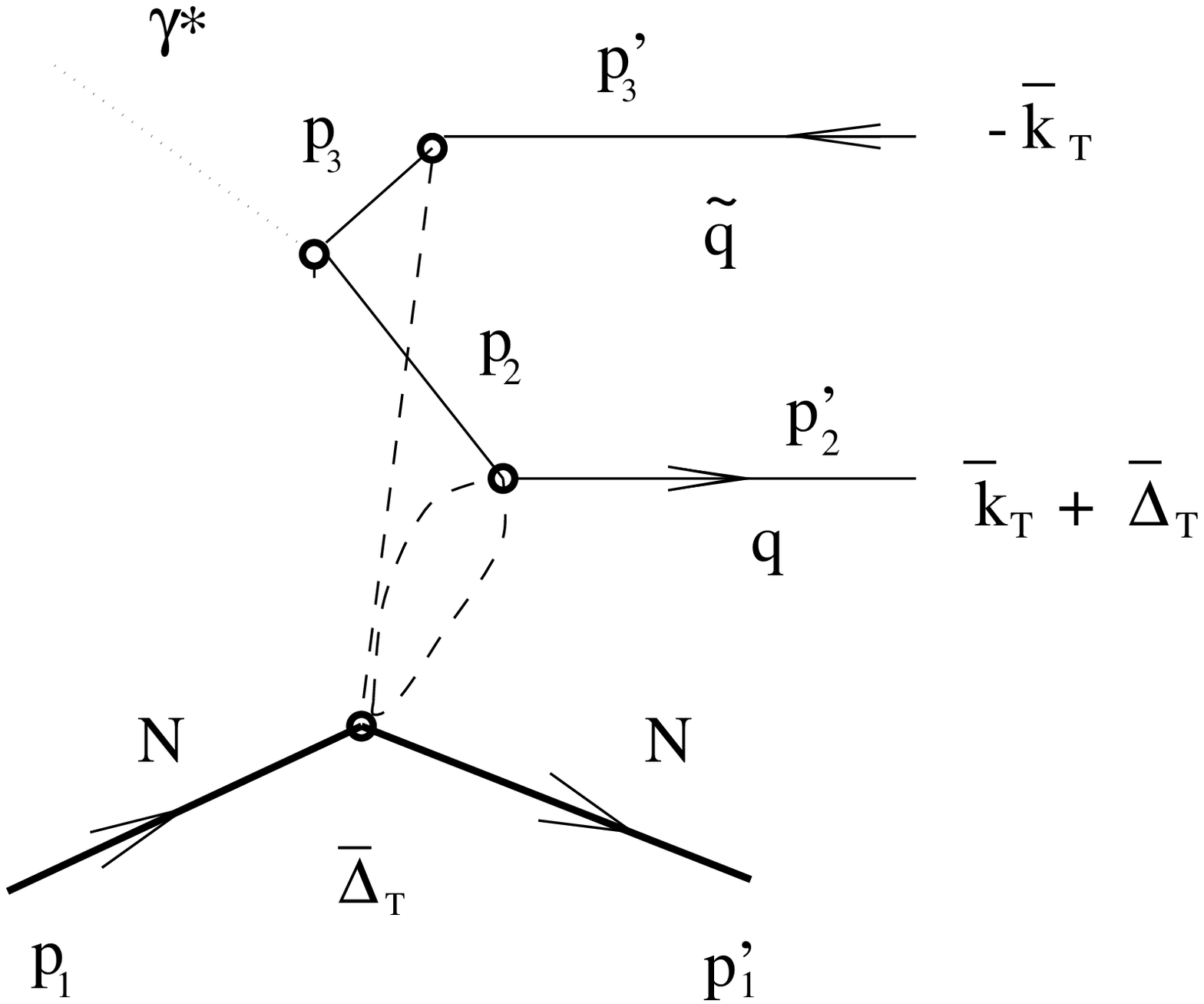,height=5cm,width=7.5cm}}
\vspace{2mm}
\noindent
\small
\\
\begin{minipage}{7cm}
{\sf Fig.~2a:}
Non-spectator graph with two reggeon exchanges.
Lines have  the same meaning as in Fig.~1a.
\end{minipage}
 \normalsize
&
\begin{minipage}{7.5cm}
{\sf Fig.~2b:}
Non-spectator graph with three reggeon exchanges.
Lines have  the same meaning as in Fig.~1a.
 \end{minipage}
\normalsize
\end{tabular}
\end{center}
The graphs presented in Figs.~1a and 1b 
describe the one and three pomeron exchanges,
respectively. We restrict ourselves to graphs containing not
more than three reggeon exchanges. In Fig.~1a $p_1,\;p_2,\;p_3$ are the 
incident momenta of the nucleon, quark, antiquark,
respectively. Here
and after we
denote the nucleon as particle number 1, the quark and antiquark
will be quoted as particles with numbers 2 and 3, respectively. The
momenta
of outgoing
particles are denoted by $p'_1,\;p'_2,\;p'_3$, transferred four-momentum
$\Delta$ is equal to $p'_1-p_1$. We can present the amplitude of
quark-nucleon scattering as a sum of the amplitudes of one, two, ..., n
pomeron exchanges
\begin{equation}
\hat{A}_P=\hat{A}^{(1)}_P+\hat{A}^{(2)}_P+\hat{A}^{(3)}_P+...\;.
\end{equation}
The first term in (1) corresponds to the Regge pole contribution
\begin{equation}
\hat{A}^{(1)}_P=p(\Delta )P(\Delta )G_P(\Delta ,s)\;,
\end{equation}
where $s$ is the square of the center-of-mass energy of the colliding  
particles, $G_P(\Delta ,s)$ denotes the pomeron propagator, and
$p(\Delta )$, and $P(\Delta )$ are quark-quark-pomeron ($qqP$) and
nucleon-nucleon-pomeron ($NNP$) vertices, respectively. The formula for
the $NNP$-vertex $P(\Delta )$ in the helicity representation reads
\begin{equation}
P^{\lambda'_1}_{\lambda_1}(\Delta )=(I_1)^{\lambda'_1}_{\lambda_1}
P_s(\Delta ^2)+iP_y(\Delta ^2)
(\vec{\sigma} _1 \cdot\vec{l} \times \vec{\Delta} _T)
^{\lambda'_1}_{\lambda_1}\;,
\end{equation}   
where $\vec{l}$ denotes a unit vector along the quark three-momentum,
$\lambda_1$ and $\lambda'_1$ are helicities of the initial and
scattered nucleon, respectively (at high energy we neglect the masses of
all
particles, hence the nucleon helicity becomes a good quantum number.).
The matrices $I_1$ and $\vec{\sigma} _1$ are the unit matrix  and the
Pauli
matrices acting on the helicity variables of the nucleon. The
functions $P_s(\Delta ^2)$ and $P_y(\Delta ^2)$ can be taken in the
gaussian form \cite{BLST}
\begin{equation}
P_s(\Delta ^2)=P_s \exp\{-r_s^2 \Delta _T^2/2\}\;,\;\;\;
P_y(\Delta ^2)=P_y \exp\{-r_y^2 \Delta _T^2/2\}\;,
\end{equation}
where $\Delta _T=|\vec{\Delta}_T|$ and the three-vector $\vec{\Delta_T}$
denotes a transverse part of $\vec{\Delta}$ (orthogonal
to the four-vectors $p_1$ and $p_2$) and 
$\Delta ^2 \approx
-\vec{\Delta}_T^2$. Parameterization (4)
with $r_s=r_y\equiv r_P$ permitted to achieve in \cite{BST}, \cite{BLSTM}
a reasonable
description of the differential cross sections and polarizations in    
hadron-hadron collisions at beam energies 10 to 100 GeV. The
values
of the parameters $P_s,\;P_y,\;r_P$ and parameters for $\rho,\; f,\;A_2,\;
\omega$ reggeons are taken from \cite{BST}, \cite{BLSTM} and presented
in Table 1.

\hspace{0.4cm}
\begin{center}
\begin{tabular}{|l|l|l|l|l|l|}
\hline \multicolumn{6}{|c|}{\bf Table 1.} \\ \hline
Reggeon, a& $P$& $\rho$& $f$&$A_2$&$\omega$ \\ \hline
$\alpha _a(0)$&1.075&0.49&0.45&0.35&0.425 \\ \hline
$\alpha '_a(0)$, GeV$^{-2}$&0.26&0.7&1.0&0.7&1.0 \\ \hline
$\sigma _a$&+1&-1&+1&+1&-1\\ \hline
$T_a$&0&1&0&1&0 \\ \hline
$A_s$, GeV$^{-1}$&2.28&0.41&2.5&0.41&1.94 \\ \hline
$A_y$, GeV$^{-2}$&0.22&-1.52&0.32&-1.21&1.03 \\ \hline
$a_s$, GeV$^{-1}$&0.76&0.41&0.83&0.41&0.65 \\ \hline
$a_y$, GeV$^{-2}$&0.22&-0.91& 0.32&-0.73&1.03 \\ \hline
$r^2_a$, GeV$^{-2}$&2.3&3.46& 5.2&2.0&9.1 \\ \hline
\end{tabular}
\end{center}

We consider quarks and antiquarks as point-like particles so the
quark-quark-pomeron ($qqP$) vertex $p(\Delta)$ looks like
\begin{equation}
p(\Delta)=p_s+ip_y (\vec{\sigma} _2 \cdot \vec{l}\times\vec{\Delta}_T)\;,
\end{equation}  
where $p_s$ and $p_y$ are some constants. To extract values of $p_s$ and
$p_y$ from the $NNP$-vertex we consider the nucleon as a three quark
system
described by the nonrelativistic quark model. Applying the well known spin
part of the proton wave functions \cite{JJJ}
\begin{eqnarray}
\nonumber
|p_{\pm \frac{1}{2}}>=
\frac{1}{\sqrt{18}} 
\biggl \{ \pm 2 |u_{\pm\frac{1}{2}} u_{\pm\frac{1}{2}}
d_{\mp\frac{1}{2}}>
\pm 2|u_{\pm\frac{1}{2}} d_{\mp\frac{1}{2}} u_{\pm\frac{1}{2}}> \\
\nonumber
\pm 2 |d_{\mp\frac{1}{2}} u_{\pm\frac{1}{2}} u_{\pm\frac{1}{2}}> 
\mp|u_{\pm\frac{1}{2}} u_{\mp\frac{1}{2}} d_{\pm\frac{1}{2}}> 
\mp|u_{\pm\frac{1}{2}} d_{\pm\frac{1}{2}} u_{\mp\frac{1}{2}}> \\
\mp|u_{\mp\frac{1}{2}} u_{\pm\frac{1}{2}} d_{\pm\frac{1}{2}}> 
\mp|d_{\pm\frac{1}{2}} u_{\pm\frac{1}{2}} u_{\mp\frac{1}{2}}> 
\mp|u_{\mp\frac{1}{2}} d_{\pm\frac{1}{2}} u_{\pm\frac{1}{2}}> 
\mp|d_{\pm\frac{1}{2}} u_{\mp\frac{1}{2}} u_{\pm\frac{1}{2}}> \biggr \}
\end{eqnarray}
one can get the relations of interest
\begin{equation}
p_s=\frac{1}{3} P_s\;,\;\;\;p_y=P_y\;.
\end{equation}  
In (6) $|p _{\frac{1}{2}}>$ ($|p _{-\frac{1}{2}}>$) describes the proton
with
its
spin parallel
(antiparallel) to a quantization axis. The analogous meaning have
quantities $|u_{\pm\frac{1}{2}}>$ and $|d_{\pm\frac{1}{2}}>$ for $u$- and
$d$-quarks.

The propagator $G_a(\Delta,s)$ ($a=P,\;\rho,\;f,\;A_2,\;\omega$)
describes
exchange of a reggeon $a$ with spin $\alpha _a(t)$ whose value
depends on the square of the reggeon momentum  ($t=\Delta ^2
\approx- 
\vec{\Delta} _T^2$). The formul\ae$\;$ for $G_a(\Delta, s)$ reads
\cite{BLST}

\begin{equation}
G_a(\Delta, s)=\eta _a(t) (s/s_0)^{\alpha _a(t)-1}\;,
\end{equation}
where $s$ is the square of the center-of-mass energy. The parameter $s_0$
in the Regge theory for proton-proton scattering is put usually equal to
$s_0=2m_p E_0$ with $E_0=1$ GeV where $m_p$ is the proton mass. 
A word of caution is in order. In a more accurate consideration the
quantity $\ln(s/s_0)$ should be replaced with the difference of the
incident particle rapidity with respect to the target and some standard
rapidity corresponding to $s_0$. Neglecting the motion and interaction of
quarks in the proton we have in the naive quark model
for quark-proton scattering at high energies the relation $s=s_{pp}/3$.
Here $s_{pp}$ denotes the square of the total energy in the center of
mass of two protons. To have the same rapidity difference for the
quark-proton system as for the proton-proton system we are to put $s_0=2
m_p E_0/3$ in (8) for quark-proton scattering.
The signature factor $\eta _a(t)$ is given
by the relation \cite{KS}, \cite{BLST}
\begin{equation}
\eta _a(t)=-\frac{1+\sigma _a \exp\{-i \pi \alpha _a(t) \}}{\sin \pi
\alpha _a(t)}\;,
\end{equation}
where $\sigma _a$ is a signature of the Regge trajectory. For the
trajectory on which the real particle (resonance) with spin equal to
$J$ lies, $\sigma _a=(-1)^J$, for the pomeron trajectory having the
vacuum quantum numbers (vacuum trajectory) $\sigma _P=1$. 
It is well known that at small $t$ Regge trajectories $\alpha _a(t)$ are
straight lines. Then putting the expression 
$\alpha _a(t)=\alpha _a(0)+\alpha '_a(0) t$ into (9) we get the 
approximate formula
\begin{equation}
\eta _a(t)=\eta _a(0) \exp \{t\frac{\pi}{2}\sigma _a \alpha '_a(0) \eta
_a^{*}(0) \}\;,
\end{equation}
where $\eta  _a^{*}$ denotes the complex conjugate quantity.
Since for the linear trajectory we have
\begin{equation}
(s/s_0)^{\alpha _a(t)}=(s/s_0)^{\alpha _a(0)}\exp\{ t\alpha
'_a(0)\ln(s/s_0)\}\;,
\end{equation}
we get instead of (8), taking into account (10), (11), the relation
\begin{equation}
G_a(\Delta, s) \approx \eta _a(0) (s/s_0)^{\alpha _a(0)-1}
\exp\{-\alpha '_a(0) 
[\ln(s/s_0)+\frac{\pi}{2}\sigma _a \eta _a(0)^{*}]\Delta _T^2\}
\end{equation}
which is convenient for numerical calculations.

The contributions of $n$ pomeron exchange corresponding to the graph
presented in
Fig.~1b in accordance with the Gribov diagram technique  are
\cite{VNG}, \cite{KAT}
\begin{eqnarray}
\nonumber
A^{(n)}_P=\frac{ i^{n-1}}{\pi^{n-1} n!} \int N_1^{n} N_2^{n}
G_P(\vec{\Delta}_1,s) G_P(\vec{\Delta}_2,s)...G_P(\vec{\Delta}_n,s)\\
\delta(\sum _{i=1}^n \vec{\Delta} _i-\vec{\Delta}) d^2\vec{\Delta}_1
d^2 \vec{\Delta}_2...d^2 \vec{\Delta}_n\;,
\end{eqnarray}
where $N_j^{n}\equiv
N_j^{n}(\vec{\Delta}_1,\vec{\Delta}_2,...,\vec{\Delta}_n)$ denotes the    
vertex for emission of $n$ pomerons by the jth particle ($j=1,\;2,\;3$)
and $\vec{\Delta}_n$ is a two dimensional vector orthogonal to the three
momenta of the colliding particles in the center-of-mass.
In the eikonal approximation the formula for the vertex $N_1^{n}$ reads
\begin{equation}
N_1^{n}(\vec{\Delta}_1,\vec{\Delta}_2,...,\vec{\Delta}_n)=
C_{sh}^{(n)}P(\vec{\Delta}_1) P(\vec{\Delta}_2)...P(\vec{\Delta}_n)\;.
\end{equation}
The graph corresponding to relation (14) with $C_{sh}^{(n)}=1$ is shown in
Fig.~3a. To take into account the possibility to produce a shower of
particles after each rescattering of the nucleon (such a process is shown
in Fig.~3b) one can put \cite{BLST} $C_{sh}^{(1)}=1$ and for $n\geq 2$
$C_{sh}^{(n)}=C_{sh}^{(2)} (C_0)^{n-2}$ with
$C_{sh}^{(2)}=\sqrt{1+\sigma^{in}/\sigma^{el}}$ and $C_0$ being a free
parameter. These parameters in our calculations have been put equal to 
$C_{sh}^{(2)}=\sqrt{1.3}$, $C_0=\sqrt{1.57}$ in accordance
with \cite{BST}, \cite{BLSTM} where they have been found from fitting
experimental
data.
We denote by $\sigma^{el}$ and $\sigma^{in}$ the total cross sections of
elastic and inelastic nucleon-nucleon scattering. For the
quark-quark-reggeon vertex we put $C_{sh}^{(n)}=1$ ignoring the possibility
of shower production in soft scattering of the point-like quarks
and antiquarks.
\vspace{0.5cm}
\begin{center}
\begin{tabular}{cc}
\mbox{\epsfig{file=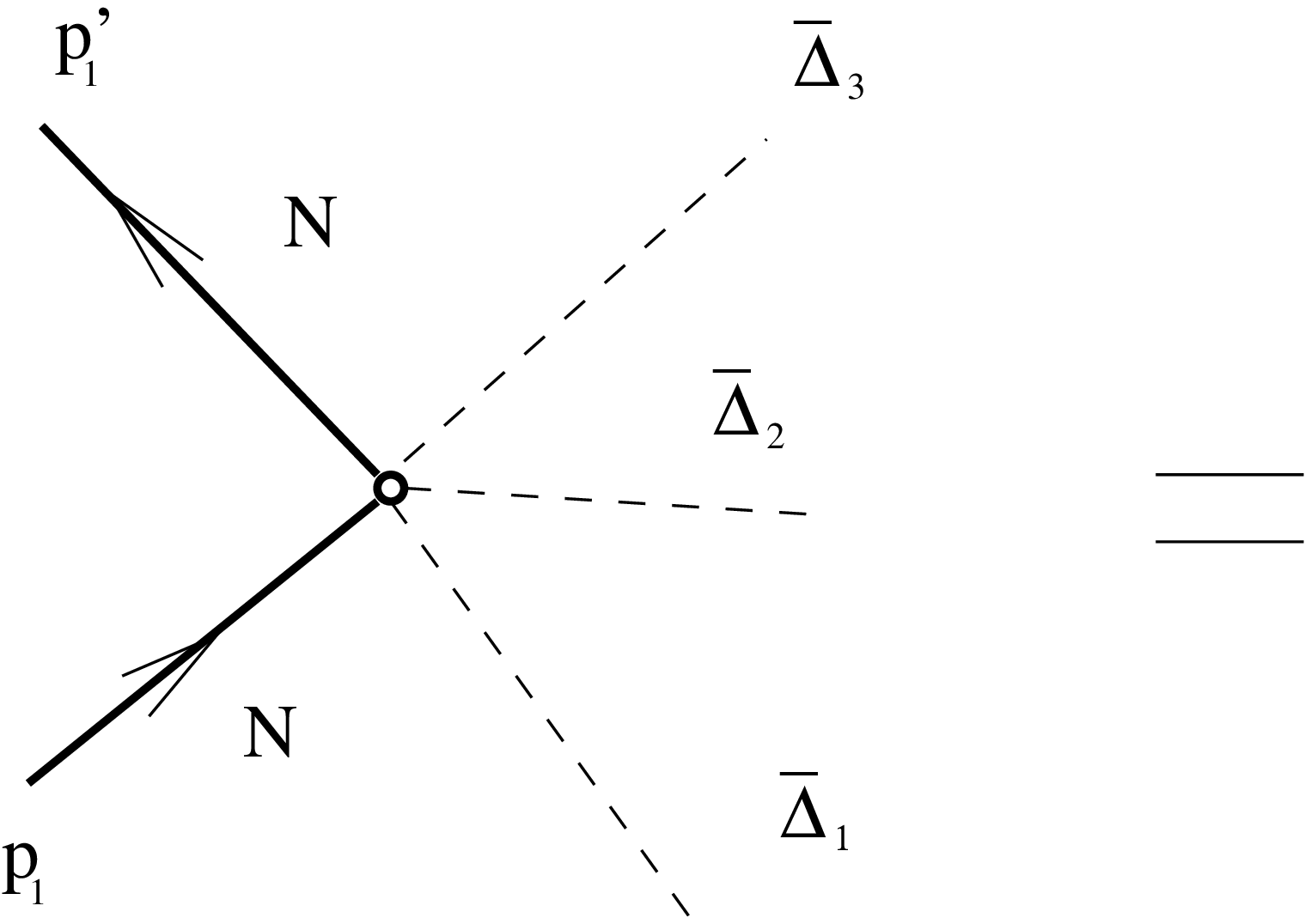,height=5.5cm,width=7.5cm}}
\vspace{2mm}
\noindent
\small
&
\mbox{\epsfig{file=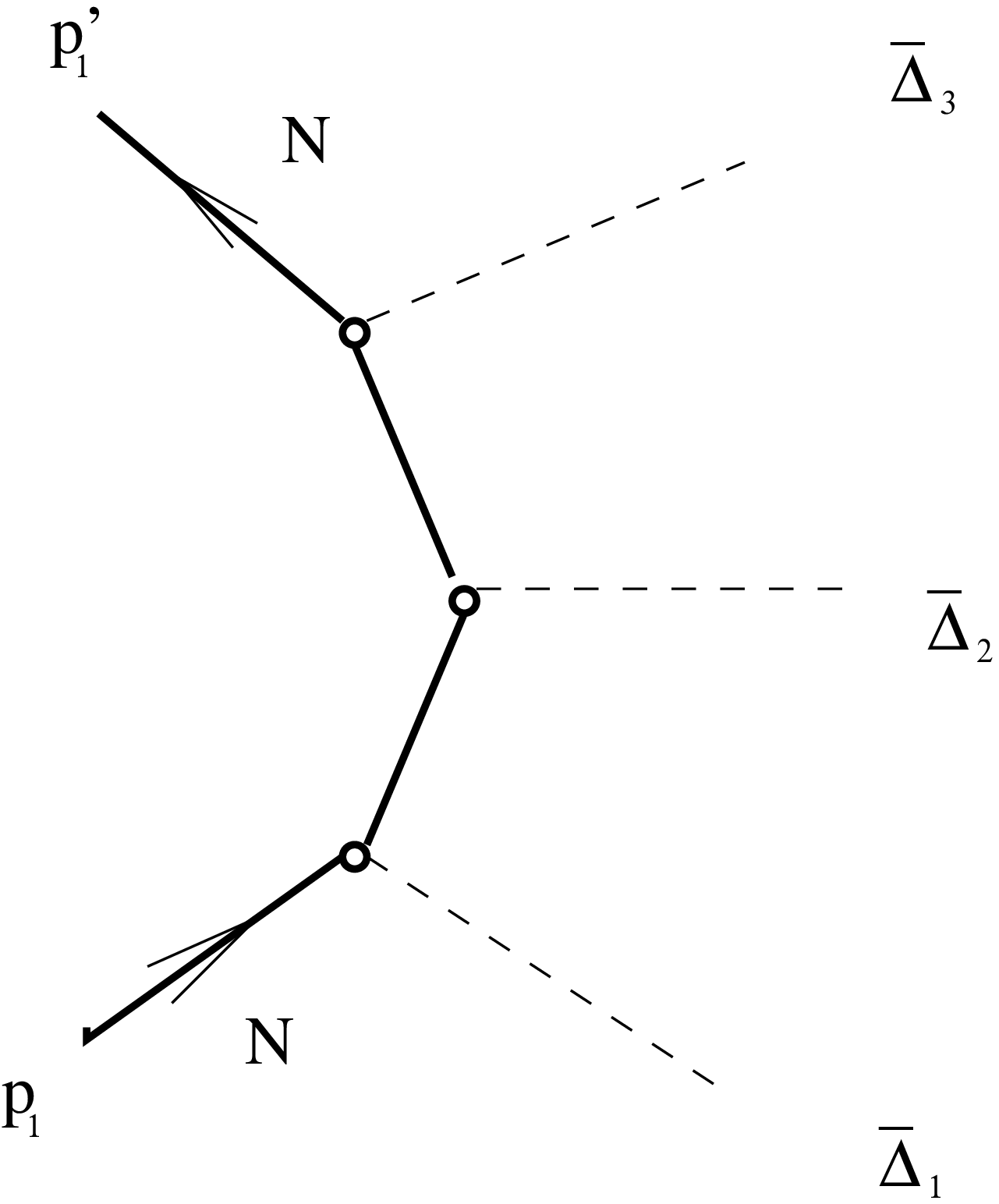,height=5.5cm,width=7.5cm}}
\vspace{2mm}   
\noindent  
\small
\end{tabular}
\begin{minipage}{14.5cm}
\vspace{7mm}
{\sf Fig.~3a:}
Nucleon-reggeon vertex for emission of three reggeons in eikonal
approximation.
Lines have  the same meaning as in Fig.~1a.
\end{minipage}
\end{center}
\vspace{0.3cm}
\normalsize

\vspace{0.5cm}
\begin{center}
\begin{tabular}{cc}
\mbox{\epsfig{file=fig3a.eps,height=5.5cm,width=7.5cm}}
\vspace{2mm}
\noindent
\small
&
\mbox{\epsfig{file=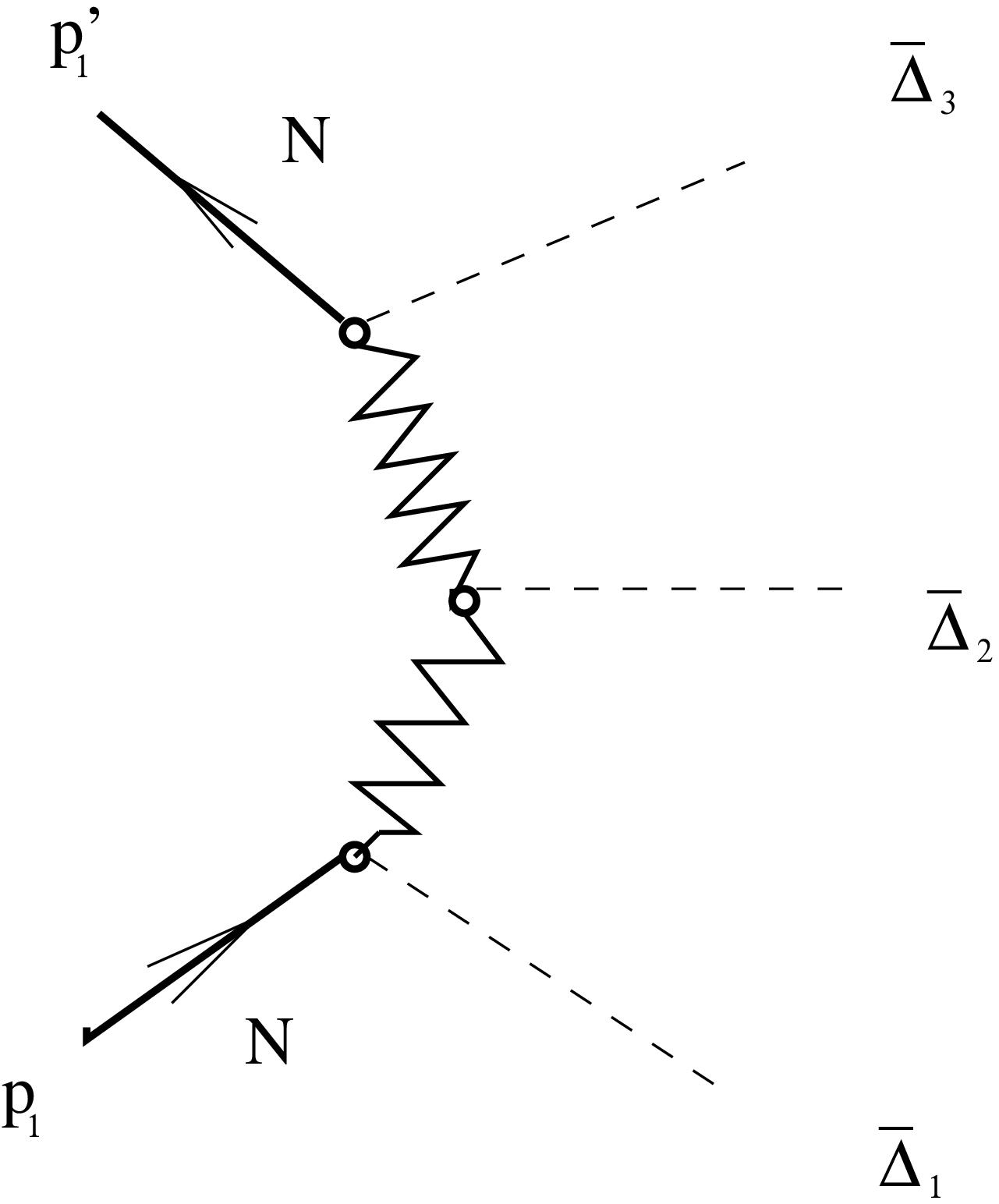,height=5.5cm,width=7.5cm}}
\vspace{2mm}
\noindent
\small
\end{tabular}  
\begin{minipage}{14.5cm}
\vspace{7mm}  
{\sf Fig.~3b:}
Nucleon-reggeon vertex for emission of three reggeons.
Zigzags correspond to showers of intermediate particles
between emissions of reggeons.
Other lines have  the same meaning as in Fig.~1a.
\end{minipage}
\end{center}
\vspace{0.3cm}
\normalsize
Formula (14) has to be corrected when $P(\vec{\Delta})$ depends on spin
variables and hence $P(\vec{\Delta} _i)$ and $P(\vec{\Delta} _j)$ do not
commute.
The vertex $N_1^{(n)}$ should be symmetrized under all the permutations of
momenta of reggeons $\vec{\Delta}_1,\;\vec{\Delta}_2,...,\;\vec{\Delta}_n$
\cite{DUN}. We would like to stress that we discuss the case of 
bosonic Regge trajectories. For pure pomeron exchanges this is obvious
as pomerons are identical bosons. Hence we can write for the $NNP$-vertex
\begin{eqnarray}
\nonumber
N_1^{n}(\vec{\Delta}_1,\vec{\Delta}_2,...,\vec{\Delta}_n)=
C_{sh}^{(n)}\bigl\{P(\vec{\Delta}_1) P(\vec{\Delta}_2)...
P(\vec{\Delta}_n)\bigr\}\\
\equiv \frac{C_{sh}^{(n)}}{n!} \sum P(\vec{\Delta}_1) P(\vec{\Delta}_2)...
P(\vec{\Delta}_n)\;,
\end{eqnarray}
where the brackets $\{\}$ in (15) mean a sum over all permutations of the
momenta divided by $n!$ 
The symmetry property should be valid for the $qqP$-vertex, hence we write
\begin{equation}
N_j^{n}(\vec{\Delta}_1,\vec{\Delta}_2,...,\vec{\Delta}_n)=
\bigl\{p(\vec{\Delta}_1) p(\vec{\Delta}_2)...
p(\vec{\Delta}_n)\bigr\}\;
\end{equation}
with $j=2,\;3$ (recall that $j=2$ and $j=3$ correspond to the quark and
antiquark, respectively).  
The general expression for the amplitude of elastic quark-nucleon
scattering reads
\begin{eqnarray}
\nonumber
\hat{A}(\vec{\Delta}_T)=A_1+A_2(\vec{\sigma }_2 \cdot \vec{n})+
A_6(\vec{\sigma }_1 \cdot \vec{n})\\
+A_3(\vec{\sigma }_2 \cdot \vec{n})(\vec{\sigma }_1 \cdot \vec{n})+
A_4(\vec{\sigma }_2 \cdot \vec{m})(\vec{\sigma }_1 \cdot \vec{m})+
A_5(\vec{\sigma }_2 \cdot \vec{l})(\vec{\sigma }_1 \cdot \vec{l})\;,
\end{eqnarray}  
where $A_j$ denote the invariant amplitudes; $\vec{l},\;\vec{m}$ are the
unite vectors along $\vec{p_2}$ and
$\vec{\Delta} _T$, respectively, and $\vec{n}=\vec{l}\times\vec{m}$.
The amplitude of elastic antiquark-nucleon scattering will be denoted by 
$\hat{B}$. It is related to the invariant amplitudes $B_j$ by formula
(17) in which we are to make the substitutions $A_j \rightarrow B_j$, 
$\vec{\sigma} _2 \rightarrow \vec{\sigma} _3,\; \vec{p_2}\rightarrow 
\vec{p_3}$. For pomeron exchanges 
$B_j=A_j$, relations between $B_j$ and $A_j$ for general case will be
discussed later. 
In accordance with (1) we present $A_j$ as a sum of amplitudes $A_j^{(n)}$
describing the n pomeron exchange contributions.
Putting in (13) formul\ae$\;$ (12), (15), (16), (3), (4), (5) 
we get the expressions for 
amplitudes
\begin{equation}
A_j^{(n)}=C_{sh}^{(n)}\bigl [\eta _P(0)(s/s_0)^{\alpha _P(0)-1} \bigr ]^n
\exp\{-\lambda_P\vec{\Delta}_T^2 /n\} a_j^{(n)}\;,
\end{equation}
where the parameter $\lambda _a$ for any reggeon $a$ is given by
\begin{equation}
\lambda_a=\frac{r^2_a}{2}+\alpha' _a(0)\bigl[\ln(s/s_0)+\frac{\pi}{2}
\sigma_a 
\eta^{*}_a(0) \bigr]\;.
\end{equation}
The expressions for the nonzero
amplitudes $a_j^{(n)}$ look like
\begin{eqnarray} 
\nonumber               
a_1^{(1)}=p_s P_s\;,\\
\nonumber
a_2^{(1)}=i\Delta _T p_y p_s\;,\\
\nonumber
a_6^{(1)}=i\Delta _T P_y P_s\;,\\
a_3^{(1)}=-\Delta _T^2 p_y P_y\;.
\end{eqnarray}
The formul\ae$\;$ for the two pomeron exchanges read
\begin{eqnarray}
\nonumber
a_1^{(2)}=\frac{i}{4\lambda_P}\bigl \{(p_s P_s)^2-
\frac{1}{2 \lambda_P}[(p_y P_s)^2+(p_sP_y)^2]
(\lambda_P \Delta _T^2/2-1)\\
\nonumber
+\frac{(p_yP_y)^2}{4\lambda ^2_P}(\lambda
^2_P
\Delta _T^4/4-\lambda_P \Delta _T^2+2) \bigr \}\;,\\
\nonumber
a_2^{(2)}=-\frac{\Delta _T}{8\lambda ^2_P} 
p_y p_s[2\lambda_P P_s^2-P_y^2(\lambda_P \Delta _T^2/2-1)]\;,\\
\nonumber
a_6^{(2)}=-\frac{\Delta _T}{8\lambda ^2_P}
P_y P_s[2\lambda_P p_s^2-p_y^2(\lambda_P \Delta _T^2/2-1)]\;,\\
a_3^{(2)}=-i\frac{\Delta _T^2}{8\lambda_P }p_y P_y p_s P_s\;.
\end{eqnarray}
For the three vacuum pole exchange we have
\begin{eqnarray}
\nonumber
a_1^{(3)}=-\frac{1}{18\lambda ^2_P}\bigl \{(p_s P_s)^3+
\frac{1}{ \lambda_P}(1-\lambda_P \Delta _T^2/3)[p_sp^2_y
P^3_s+p^3_sP_sP^2_y]\\
\nonumber
+\frac{p_sp^2_yP_sP^2_y}{2\lambda ^2_P}(3-\frac{4}{3}\lambda _P   
\Delta _T^2+\frac{2}{9}\lambda_P^2 \Delta _T^4) \bigr \}\;,\\
\nonumber
a_2^{(3)}=-i\frac{\Delta _T}{54\lambda ^2_P}\bigl \{3 p_s^2p_yP_s^3+
\frac{1}{3\lambda _P}(2-\lambda_P\Delta _T^2/3)
(p^3_yP^3_s+9p_s^2p_yP_sP_y^2)\\
\nonumber
-\frac{3}{\lambda _P}p_s^2p_y P_sP_y^2
+\frac{p^3_yP_sP_y^2}{3\lambda _P^2}
(\frac{9}{4}-\frac{2}{3}\lambda_P \Delta _T^2+\frac{2}{27}
\lambda_P^2\Delta _T^4) \bigr \}\;,\\
\nonumber
a_6^{(3)}=-i\frac{\Delta _T}{54\lambda ^2_P}\bigl \{3 p_s^3P_s^2P_y+
\frac{1}{3\lambda _P}(2-\lambda_P \Delta _T^2/3)
(p^3_sP^3_y+9p_sp_y^2P_s^2P_y)\\
\nonumber
-\frac{3}{\lambda _P}p_sp_y^2 P_s^2P_y
+\frac{p_sp_y^2P_y^3}{3\lambda _P^2}
(\frac{9}{4}-\frac{2}{3}\lambda_P \Delta _T^2+\frac{2}{27}
\lambda_P^2 \Delta _T^4) \bigr \}\;,\\
\nonumber
a_3^{(3)}=\frac{\Delta _T^2}{18\lambda ^2_P}\bigl \{p_s^2p_yP_s^2P_y+
\frac{1}{9\lambda _P}(2-\lambda_P \Delta _T^2/3)(p_y^3P_s^2P_y+
p_s^2p_yP_y^3)\bigr \}\\
\nonumber
+\frac{p_y^3P_y^3}{972\lambda ^5_P}\bigl \{4+
\frac{13}{3}\lambda_P \Delta _T^2-\frac{8}{9}\lambda_P^2 \Delta _T^4+
\frac{2}{27}\lambda_P^3 \Delta _T^6 \bigr \}\;,\\
a_4^{(3)}=\frac{p_y^3P_y^3}{486\lambda
^5_P}(2+\frac{1}{6}\lambda_P \Delta _T^2)\;.
\end{eqnarray}

We can easily see from formul\ae$\;$ (17), (18), (20), (21) that at
$\Delta
_T=0$
all the spin-dependent amplitudes of the one and two pomeron
exchange contributions vanish, hence there are no  polarization
phenomena for this case. Taking into account (7) we get from (17), (18) 
and (22) for $\Delta _T=0$  
\begin{eqnarray}                
\nonumber
\hat{A}_P^{(3)}=-C_{sh}^{(3)} \eta ^3_P(0)(s/s_0)^{3\alpha _P(0)-3} \\
\bigl
\{
\frac{P_s^6}{486\lambda_P^2}+
\frac{5}{243\lambda _P^3}P_y^2P_s^4+\frac{1}{36\lambda ^4_P}P_y^4P_s^2
-\frac{P_y^6}{243 \lambda_P^5} 
(\vec{\sigma }_{1T} \cdot \vec{\sigma }_{2T})
\bigr \}\;,
\end{eqnarray}
where we make use of the short notation
\begin{eqnarray}
\nonumber
(\vec{\sigma }_{1T}\cdot\vec{\sigma }_{2T}) \equiv 
(\vec{\sigma}_1\cdot\vec{m})(\vec{\sigma}_2\cdot\vec{m})+
(\vec{\sigma}_1\cdot\vec{n})(\vec{\sigma}_2\cdot\vec{n})=
(\sigma _{1x}\cdot\sigma _{2x})+
(\sigma _{1y}\cdot\sigma _{2y})\;.
\end{eqnarray}
We shall show that the last term in (23) gives the longitudinal
spin-spin asymmetry in the $\gamma p$ collisions. Indeed, the wave
function of the $q\bar{q}$-pair produced by the virtual photon with
helicity $m=\pm 1$ looks like \cite{R}
\begin{equation}
\Psi ^{(m)}_{\gamma}(\vec{k}_T,z)=\frac{
(\vec{k}_T \cdot \vec{e}^{(m)})[(2\lambda
_2)(1-2z)-m]}{Q^2z(1-z)+\vec{k}_T^2 +\mu _q^2}\delta^{\lambda
_2,-\lambda _3}\;,
\end{equation}
where $\lambda _2$ and $\lambda _3$ denote the helicities of the quark and
antiquark, respectively. Vectors $ \vec{e}^{(m)}$ of the photon
polarization are  $\vec{e}^{(\pm 1)}=( \vec{e}_x \pm i\vec{e}_y)/
\sqrt{2}$, $\vec{e}_x$, $\vec{e}_y$ being unite vectors orthogonal to
the $z$-axis directed along the photon three-momentum, $q^2=-Q^2$ is the
square
of the heavy photon four-momentum, $z$ denotes the photon three momentum
fraction
carried by the quark (more precisely $z=(p_2^0+p_2^3)/(q^0+q^3)$ where
$p_2^0$ ($q^0$) denotes the quark (photon) energy and $p_2^3$ ($q^3$) is
the $z$-component of the quark (photon) three-momentum),  $\vec{k}_T$
is the transverse part of the
quark momentum $\vec{p}_2$ before scattering and $\mu _q$ denotes the mass
of the constituent quark. We can neglect $\mu _q$ except in the case when
$k_T^2$ and $Q^2z(1-z)$ are small ($\ll 1$ (GeV/c)$^2$). The density
matrix of
the $q\bar{q}$-pair corresponding to the wave function (24) looks like
\begin{eqnarray}
\rho_{23}^{(m)}=\frac{V}{4}\bigl \{I+
(\vec{\xi}_2 \cdot \vec{\sigma}_2)+(\vec{\xi}_3 \cdot \vec{\sigma }_3)+
\eta_{lj}\sigma _{2l}\sigma _{3j} \bigr \} \;,\\
\nonumber
V=2\vec{k}_T^2[z^2+(1-z)^2]/[Q^2z(1-z)+\vec{k}_T^2 +\mu _q^2]^2\;.
\end{eqnarray}
The nonzero components of $\vec{\xi}_2$, $\vec{\xi}_3$, $\eta_{lj}$ are
\begin{eqnarray}
\nonumber
\xi_{2z}=-\xi_{3z}=\frac{m(2z-1)}{z^2+(1-z)^2}\;,\\
\eta_{xx}=\eta_{yy}=\frac{2z(1-z)}{z^2+(1-z)^2}\;,\;\;\;\;
\eta_{zz}=-1\;.
\end{eqnarray}
If we put in (25) $V=1$ we get the spin density matrix of the
$q\bar{q}$-pair normalized to unity. The general form at $\Delta _T =0$ of
the spectator amplitude of $q\bar{q}$-pair scattering on the nucleon shown
in Figs.~1a, 1b is 
\begin{eqnarray}
\nonumber
M^q_{sp}(0)=e e_q \bigl \{A_1+B_1+A_4 
(\vec{\sigma }_{1T} \cdot \vec{\sigma }_{2T})+
B_4 (\vec{\sigma }_{1T}  \cdot \vec{\sigma }_{3T}) \\ 
+A_5(\vec{l}  \cdot \vec{\sigma }_{1})(\vec{l}  \cdot \vec{\sigma }_{2})+
B_5(\vec{l}  \cdot \vec{\sigma }_{1})
(\vec{l}  \cdot \vec{\sigma}_{3})\bigr \}\;,
\end{eqnarray}
where $e$ denotes the electric charge of the positron, $e e_q$ is
the  electric charge of a quark and $A_j$,
$B_j$ are the amplitudes of quark-nucleon and
antiquark-nucleon
scattering, respectively. For the pomeron exchange amplitudes given
by relations (17), (18), (20), (21), (22) $A_5=B_5=0$ but they are nonzero
if we include in the consideration secondary Regge trajectories or take
into account four, five, etc. pomeron exchanges.

Our amplitudes are normalized so that the differential cross section of
elastic  scattering of 
unpolarized quarks on unpolarized nucleons is given by
\begin{eqnarray}
\nonumber
\frac{d \sigma}{d t}=4 \pi \sum _{j=1,6}|A_j|^2\;,
\end{eqnarray}
hence the cross section for scattering of $q\bar{q}$-pair
on
the nucleon is
\begin{equation}
\frac{d \sigma }{d t d M_X^2}=4 \pi n_c \sum _{q=u,d,...}\int
tr
\bigl
\{(M^q_{sp})^+(0)
M^q_{sp}(0) \rho ^{(m)}_{23} \rho _1 \bigr \} \delta \Bigl
(M^2_X-\frac{\vec{k}_T^2+\mu_q^2}{z(1-z)} \Bigr )dz d^2\vec{k}_T \;,
\end{equation}
where $M^+_{sp}(0)$ denotes a hermitian conjugate quantity, 
$n_c=3$ is the number of 
the quark colours, $M_X$ denotes the
mass of the $q\bar{q}$-pair in the final state and $\rho _1$ is the
spin density matrix of
the proton with the longitudinal polarization 
$\vec{\zeta}_P=(0,0,\zeta_P)$
\begin{equation}
\rho _1=\frac{1}{2}(I+\vec{\zeta}_P \cdot \vec{\sigma }_1)\;.
\end{equation}
Putting (25), (26), (27), (29) into (28) and integrating over
$d^2\vec{k}_T$ which gives due to the
$\delta$-function $\vec{k}_T^2=M^2_0z(1-z)$ we get
\begin{eqnarray}
\nonumber
\frac{d \sigma}{d td M_X^2}=8 \pi ^2 \frac{e^2 e_q^2 n_c
M_X^2}
{(Q^2+M_X^2)^2} \sum_{q=u,d,...}\int _0^1
\Bigl \{[|A_1+B_1|^2+2|A_4|^2+2|B_4|^2\\
\nonumber
+|A_5|^2+|B_5|^2-2\Re(A_5B_5^*)][z^2+(1-z)^2]+
8\Re(A_4B_4^*)z(1-z)\\
+2m \zeta_ P(2z-1)
[|B_4|^2-|A_4|^2+\Re ((A_1+B_1)(A^*_5-B^*_5))] \Bigr \} 
M_0^2 \theta(M_0^2)dz\;
\end{eqnarray}
where $M_0^2=M_X^2-\mu_q^2/[z(1-z)]$ and $\theta(x)$ denotes the Heavyside
function ($\theta(x)=1$ for $x \geq 0$ otherwise $\theta(x)=0$).
The formula for the longitudinal spin-spin asymmetry $A_{LL}$ follows
from its definition
\begin{eqnarray}
A_{LL}=\bigl \{\sigma  (+,+)-\sigma  (+,-) \bigr \}/
\bigl \{\sigma  (+,+)+\sigma  (+,-) \bigr \}\;,
\end{eqnarray}
where we have applied the short notation
\begin{eqnarray}
\nonumber
\sigma  (+,\pm)=\frac{d \sigma  (m=1,\zeta_P=\pm1)}{dt dM^2_X}\;,
\end{eqnarray}
and relation (30)
\begin{eqnarray}
\nonumber
A_{LL}= 2 \sum _{q=u,d,...}\int _0^1
(2z-1) 
[|B_4|^2-|A_4|^2+\Re ((A_1+B_1)(A^*_5-B^*_5))] 
M_0^2 \theta(M_0^2)dz  \\
\nonumber
/ \sum _{q=u,d,...}\int _0^1 \Bigl \{
[|A_1+B_1|^2+2|A_4|^2+2|B_4|^2+|A_5|^2+|B_5|^2\\
-2\Re(A_5B_5^*)][z^2+(1-z)^2]+
8\Re(A_4B_4^*)z(1-z) \Bigr \} 
M_0^2 \theta(M_0^2)dz  \;.
\end{eqnarray}
 
We see from (32) that the asymmetry does not depend on $Q^2$. If we
consider a kinematics for which $k_T \gg m_q$ we can omit $m_q$ in (24).
For this case $M_0^2=M_X^2$ in (30), (32) and $A_{LL}$ does not depend
on $M_0^2$ too. 
But these properties take place for the spectator graphs only (shown in
Fig.~1a, 1b). We see
also from (32) that the order of magnitude of $A_{LL}$ for the case
under consideration when $A_5=B_5=0$ is 
\begin{equation}
A_{LL}\sim |A_4|^2/|A_1|^2 \sim  
\frac{P_y^{12}}{P_s^{12}|\lambda_P|^6} 
\leq   10^{-12}.
\end{equation}
For our estimate of $A_1$ we have used the first term in (23) and for
$A_4$ we have used the last term. 
We can see from (19) that $|\lambda_P|$ increases with $s$. If we make 
use of the  value for $|\lambda_P|$ at $s=s_0$ and take
the values of $P_y$ and $P_s$ from Table~1 we get the numerical estimate
given by relation (33). The authors of \cite{BST}, \cite{BLSTM} stated
that the value
$P_y=0$ is compatible with experimental data. Considering the value of
$P_y$ presented in Table~1 as one standard deviation we have for
$\tilde{P}_y=3P_y$ (three standard deviations) instead of (33) an estimate
$A_{LL} \leq 10^{-6}$. As $\lambda _P \sim \ln s$ in accordance with (19)
at asymptotically high energy, then $A_{LL} \sim \ln ^{-6} s_{\gamma p}$
where $s_{\gamma p}$ is the square of the photon-nucleon center-of-mass
energy. 
We would like to make some remarks. It is easy
to see from formula (32) that the numerator does not vanish. Indeed, the
amplitude $A_4$ depends on z as $s_{12}=(p_1+p_2)^2\approx zs_{\gamma p}$
and 
$B_4$ depends on $(1-z)$ as $s_{13}=(p_1+p_3)^2 \approx (1-z)s_{\gamma
p}$.
Due to the positive signature of the pomeron the amplitude for
quark-nucleon
scattering is equal to the antiquark-nucleon amplitude at the same
collision energy. Hence the quantity $M_0^2 (2z-1)[|B_4|^2-|A_4|^2]$ does
not change  sign after the replacement $z \rightarrow (1-z)$ 
and the numerator in (32) is not equal to zero if $A_4$ and $B_4$ depend
on $z$. In all our further considerations of properties of some 
quantities under the transformation $z \rightarrow (1-z)$ 
we shall omit $M_0^2$ as it is invariant under this
transformation. When $A_4$ and $B_4$ are 
independent of the collision energy, then $A_{LL}=0$.
But we can see from (18), (20), (21), (22) that if $\alpha _P(0)=1$, then
the dependence of the amplitudes $A_j$ on $z$ is due to the dependence of
$\lambda_P$ on  $s$ which is logarithmic in accordance with (19).
In reality (see Table~1) $\alpha _P(0)$ is very close to  unity and the
s-dependence of the amplitudes of the vacuum pole exchange is rather
feeble. Therefore we have an additional suppression compared to our rough
estimate (33). Hence we conclude that in the high energy limit when
contributions of all known Regge trajectories except the vacuum one are
suppressed, the spin-spin
asymmetry in the diffractive dissociation of the virtual photon with high
$Q^2$ is much less than $10^{-6}$ for the spectator diagram contribution.
In our simple model with parameters
extracted from the experimental data on hadron-hadron scattering at
energies 10 - 100 GeV we have an even lower limit at $\Delta 
_T=0$ $A_{LL} \leq 10^{-12}$.

\section{Spectator graphs. Secondary Regge trajectory contributions} 

Let us consider the contributions of the $\rho,\;f,\;A_2,\;\omega$
reggeons. It is easy to see from Table~1 that for any reggeon $c \neq P$ 
$\alpha _P(t)-\alpha _c(t) \approx 0.5$, hence the amplitude of 
exchange with a reggeon $c$ is proportional to the small parameter
$\varepsilon =\sqrt{s_0/s_{\gamma p}}$ compared with the pomeron exchange
amplitude. We remind that  $s_{\gamma p}$ is the square of the
center-of-mass
energy of the $\gamma p$-system. We decompose
all amplitudes, cross sections and asymmetries into power a series in
$\varepsilon$ up to $\varepsilon ^2$ terms included. Now the amplitude of
quark-nucleon scattering looks like \cite{BLST}
\begin{equation}
\hat{A}=\sum_n \hat{A}_P^{(n)}+\frac{1}{1!}\sum_{n,c}
\hat{A}_{c,P}^{(n-1)}+\frac{1}{2!}\sum_{n,c,d}
\hat{A}_{c,d,P}^{(n-2)}+\cdots
\end{equation}
with $c,\;d\;\neq P$. In (34) $\hat{A}_{c,P}^{(n-1)}$ denotes the
amplitude of
exchanges with a reggeon $c$ and $n-1$ pomerons, and the amplitude
$\hat{A}_{c,d,P}^{(n-2)}$ describes contributions of reggeons $c$, $d$ and
$n-2$
pomerons. The formul\ae$\;$ for the amplitudes $\hat{A}_{c,P}^{(n-1)}$ and 
$\hat{A}_{c,d,P}^{(n-2)}$ read \cite{BLST}
\begin{eqnarray}
\nonumber
\hat{A}_{c,P}^{(n-1)}=
\frac{ i^{n-1}}{\pi^{n-1} (n-1)!} \int N_1^{n}(c) N_2^{n}(c)
G_c(\vec{\Delta}_1,s) G_P(\vec{\Delta}_2,s)...G_P(\vec{\Delta}_n,s)\\
\nonumber
\delta(\sum _{i=1}^n \vec{\Delta} _i-\vec{\Delta}) d^2\vec{\Delta}_1
d^2 \vec{\Delta}_2...d^2 \vec{\Delta}_n\;,\\
\nonumber
\hat{A}_{c,d,P}^{(n-2)}=
\frac{ i^{n-1}}{\pi^{n-1} (n-2)!} \int N_1^{n}(c,d) N_2^{n}(c,d)
G_c(\vec{\Delta}_1,s)
G_d(\vec{\Delta}_2,s)G_P(\vec{\Delta}_3,s)...G_P(\vec{\Delta}_n,s)\\
\delta(\sum _{i=1}^n \vec{\Delta} _i-\vec{\Delta}) d^2\vec{\Delta}_1
d^2 \vec{\Delta}_2...d^2 \vec{\Delta}_n\;,
\end{eqnarray}
with the $n$ reggeon vertices given by
\begin{eqnarray}
\nonumber
N_1^{n}(c)=C_{sh}^{(n)} \bigl \{C(\Delta _1)P(\Delta _2)...P(\Delta _n)
\bigr
\}\;,\\
\nonumber
N_2^{n}(c)=\bigl \{c(\Delta _1)p(\Delta _2)...p(\Delta _n) \bigr
\}\;,\\
\nonumber
N_1^{n}(c,d)=C_{sh}^{(n)} \bigl \{C(\Delta _1)D(\Delta
_2)P(\Delta _3)...P(\Delta _n) \bigr
\}\;,\\
N_2^{n}(c,d)=\bigl \{c(\Delta _1)d(\Delta _2)p(\Delta _3)...p(\Delta _n)
\bigr \}\;,
\end{eqnarray}
where $b(\Delta )$, $B(\Delta )$ ($b=c,\;d$; $B=C,\;D$) in (36) are the
vertex functions of a reggeon
$b$ analogous to $p(\Delta )$, $P(\Delta )$ defined in (3), (5),
respectively. Relation (7) is valid for any reggeon with isospin
$T=0$. For the $\rho$ and $A_2$ mesons having  isospin $T=1$ the
vertices contain the isospin Pauli matrices $\tau _j$. For example, 
the vertex for the
emission of the $\rho _j$ meson ($\rho^{\pm} =(\rho _1\pm  i \rho
_2)/\sqrt{2},\;\rho^0=\rho _3$) is proportional to $\tau _j$.
One can easily get the relation 
\begin{equation}
b_s=B_s\;,\;\;\;b_y=\frac{3}{5}B_y\;
\end{equation}
for the case $T=1$ with the aid of the wave functions (6).
Formul\ae$\;$ for the antiquark-nucleon scattering can be obtained from
(35), (36)
with the substitutions $\hat{A}_{c,P}^{(n-1)} \rightarrow
\hat{B}_{c,P}^{(n-1)}$, $\hat{A}_{c,d,P}^{(n-2)} \rightarrow
\hat{B}_{c,d,P}^{(n-2)}$ the vertex functions $b(\Delta _1)$
(with $b=c,\;d$) being  multiplied by the factor $\sigma _b (-1)^{T_b}$
where $\sigma _b$, $T_b$ denote  signature and  isospin of 
reggeon $b$.

Let us consider the first order contributions ($\sim \varepsilon$)
at $\vec{\Delta} _T=0$. There is only one nonvanishing amplitude $A_1$ of
the
one reggeon exchange. Hence the inclusion of the pole
contributions of secondary Regge trajectories does not lead to any
polarization phenomena. It follows from the general formul\ae$\;$ 
presented in
Appendix that the two reggeon exchange amplitudes $A_3^{(2)}$, $A_4^{(2)}$ 
at $\vec{\Delta} _T=0$ looks like
\begin{eqnarray}
A_3^{(2)}=A_4^{(2)}=-\frac{i\eta _c(0) \eta
_P(0)}{2(\lambda_c+\lambda_P)^2}
\Bigl (\frac{s}{s_0} \Bigr )^{\alpha _c(0)+\alpha
_P(0)-2}(c_yp_s-p_yc_s)(C_yP_s-P_yC_s) C_{sh}^{(2)}\;,
\end{eqnarray}
where $c_s,\;c_y,\;C_s \equiv C_s(0),\;C_y\equiv C_y(0)$ are the vertex
constants for a reggeon $c$ which are analogous to 
$p_s,\;p_y,\;P_s(0),\;P_y(0)$ for the pomeron defined
by (3), (5) (We denote vertices with the same letter as the letter
denoting a reggeon. For
example, the vertex constants for the $\omega$ reggeon will be
denoted  $\omega_s,\;\omega_y,\;\Omega_s,\;\Omega_y$ etc.).

As for the secondary trajectories $\alpha _c(0) \approx 0.5$ it follows
from (38) that even for $\alpha _P(0) = 1$ $A_4^{(cP)} \sim \sqrt{s_0/s}
\sim z^{-0.5}$ and for the antiquark-nucleon scattering amplitude
$B_4^{(cP)} \sim (1-z)^{-0.5}$. We denote by $A_4^{(cP)}$
($B_4^{(cP)}$) the part of the amplitude $A_4$ describing exchanges with
the reggeons $c$ and $P$. Hence the numerator 
in (32) does not vanish as the expression $(2z-1)(|B_4|^2-|A_4|^2)$ 
conserves its sign under the transformation $z \rightarrow (1-z)$. 
For reggeons with  negative signature $\sigma$ ($\rho$ and $\omega$ 
reggeons), their contribution of the first order in 
$\varepsilon$ to the numerator in (32) is equal to zero.
For example, considering  pomeron and $\omega$ exchanges we have
$A_4^{(2)}(z)=A_4^{(PP)}(z)+A_4^{(\omega P)}(z)$ and
$B_4^{(2)}(1-z)=A_4^{(PP)}(1-z)-A_4^{(\omega P)}(1-z)$. As the first order
contribution to $(2z-1)(|B_4|^2-|A_4|^2)$ equal to 
\begin{eqnarray}
\nonumber
2(2z-1) \{
\Re[A_4^{*(PP)}(z) A_4^{(\omega P)}(z)]+\Re[A_4^{*(PP)}(1-z) A_4^{(\omega
P)}(1-z)] \}
\end{eqnarray}
is an odd function under transformation $z \rightarrow (1-z)$, hence after
integration over $z$ we get a zero contribution to the numerator in (32).
The general formul\ae$\;$ for the two and three reggeon exchange
amplitudes
presented in Appendix show that the contribution to $A_5$ of the first
order in $\varepsilon$ is equal to zero. 

We start discussion of terms of the second order in $\varepsilon$ with
consideration of the contribution of the amplitudes $A_5^{(3)}$ and
$B_5^{(3)}$ in (32) as the two reggeon exchange amplitudes $A_5^{(2)}$
and $B_5^{(2)}$ vanish at $\Delta _T=0$. We
can see from the general formul\ae$\;$ presented in Appendix that the
amplitude $A_5^{(3)}$ can be nonzero if and only if two exchanged reggeons
(say $c$ and $h$) have isospins equal to 1 (the third reggeon is the
pomeron as we consider contributions of the second order in
$\varepsilon$). Indeed, the formula for $A_5^{(3)}$ at 
$\vec{\Delta} _T =0$ reads
\begin{equation}
A_5^{(3)}=\tilde{A}_5^{(3)}(\vec{\tau} _1 \cdot \vec{\tau} _j)\;,
\end{equation}
where $\vec{\tau} _1$ and $\vec{\tau} _j$ are the Pauli matrices acting on
the isospin variables of the nucleon and quark ($j=2$) or antiquark
($j=3$) and $\tilde{A}_5^{(3)}$ is 
\begin{eqnarray}
\nonumber
\tilde{A}_5^{(3)}=\frac{\eta _P(0)\eta _C(0)\eta _H(0)}{9
(\lambda_P\lambda_C  +\lambda_P\lambda_H+\lambda_C\lambda_H)^2} 
\Bigl (\frac{s}{s_0} \Bigr )^{\alpha
_P(0)+\alpha
_C(0)+\alpha _H(0)-3} p_yP_yc_yC_yh_yH_y \\ 
\Bigl \{\frac{C_s}{C_y}\Bigl(\frac{c_s}{c_y}-\frac{p_s}{p_y}\Bigr )
+\frac{H_s}{H_y}\Bigl(\frac{h_s}{h_y}-\frac{p_s}{p_y}\Bigr )  
+\frac{c_s}{c_y}\Bigl(\frac{H_s}{H_y}-\frac{P_s}{P_y}\Bigr )
+\frac{h_s}{h_y}\Bigl(\frac{C_s}{C_y}-\frac{P_s}{P_y}\Bigr )
+\frac{p_s P_s}{p_y P_y} \Bigr\}C_{sh}^{(3)}\;.
\end{eqnarray}
If a reggeon $c$ has the same signature as $h$, then the amplitude
$\tilde{A}_5^{(3)}$ is invariant under the charge
conjugation transformation $c_s \rightarrow \pm c_s, \;
c_y \rightarrow \pm c_y,\;h_s \rightarrow \pm h_s, \;
h_y \rightarrow \pm h_y$. We do not consider in the present paper 
charge exchange of the proton, hence we can write 
$(\tau _{1z} \tau _{jz})$ instead of 
$(\vec{\tau} _1 \cdot \vec{\tau} _j)$ in (39). Acting on the quark
($q=u,\;d$) and antiquark $\tau _{jz}$ gives values with opposite signs
therefore $B_5^{(3)}(s)=-A_5^{(3)}(s)$. Returning to (32) we see that
\begin{eqnarray}
(2z-1)\Re \{(A^*_1+B^*_1)(A_5-B_5) \}\\
\nonumber
=(2z-1)\Re
\{[A^*_1(z)+A^*_1(1-z)][\tilde{A}^{(3)}_5(z)+\tilde{A}^{(3)}_5(1-z)] \}
\end{eqnarray}
is an odd function with respect to the transformation $z \rightarrow 1-z$,
hence the integral over $z$ for term (41) in the numerator in (32)
vanishes. We have taken into account in (41) that $A_1$ contains 
pomeron exchanges only as $\tilde{A}^{(3)}_5,\;\tilde{B}^{(3)}_5 \sim
\varepsilon ^2$. It is easy to conclude that the amplitudes contribute
to the spin-spin asymmetry if they contain secondary reggeon exchanges
with opposite sign signatures. For the Regge trajectories considered in
the present paper such an amplitude is the amplitude of  
$P\rho A_2$ exchange.  

The second order contributions of the difference $|B_4|^2-|A_4|^2$ to the
numerator in (32) which do not vanish after integration over $z$ can be
divided into two groups. The first one 
\begin{eqnarray}
\nonumber
|A_4^{f}(1-z)+A_4^{A_2}(1-z)|^2-|A_4^{f}(z)+A_4^{A_2}(z)|^2\\ +
|A_4^{\rho}(1-z)+A_4^{\omega}(1-z)|^2-|A_4^{\rho}(z)+A_4^{\omega}(z)|^2
\end{eqnarray}
contains squares of first order amplitudes where $A_4^{f}$ denotes
the part of the
amplitude $A_4$ describing exchange with the  reggeon $f$ and some
number
of pomeron exchanges. The amplitudes
$A_4^{A_2},\;A_4^{\rho},\;A_4^{\omega}$ have an analogous meaning.
Relation
(42) shows that  the $f,\;\rho,\;\omega,\; A_2$ reggeons contribute to the
spin-spin asymmetry, the interference terms for reggeons with positive
($f,\; A_2$) and negative ($\rho,\;\omega$) signatures are absent. The
second
group of contributions to the numerator in (32) not vanishing after
integration over $z$ looks like
\begin{eqnarray}
\nonumber
\Re \Bigl \{A_4^{*P}(1-z) [A_4^{\rho \rho}(1-z)+A_4^{ff}(1-z)+
A_4^{\omega \omega}(1-z)\\
\nonumber
+A_4^{A_2A_2}(1-z)+A_4^{\rho \omega}(1-z)
+A_4^{fA_2}(1-z) \Bigr \}\\
-\Re \Bigl \{A_4^{*P}(z) [A_4^{\rho \rho}(z)+A_4^{ff}(z)+
A_4^{\omega \omega}(z)
+A_4^{A_2A_2}(z)+A_4^{\rho\omega}(z)+A_4^{fA_2}(z) \Bigr \}\;.
\end{eqnarray}
Expression (43)  represents the interference term of the amplitude $A_4$
of the second
order in $\varepsilon$ with $A_4^P$ which contains pomeron exchanges only
and starts  with the three pomeron exchange amplitude as 
 has been explained above. As for expression (42), the
contributions of exchanges of
two secondary Regge trajectories (and some number of pomeron exchanges)
with positive and negative signatures vanish after integration over $z$.
 
\section{Contributions of nonspectator graphs to $A_{LL}$}

The amplitude of the non-spectator graphs shown in Fig.~2a and 2b can be
written as a sum of two terms 
$F_{es}^q(\vec{\Delta})+F_{che}^q(\vec{\Delta})$ where
\begin{eqnarray}
\nonumber
F_{es}^q(\vec{\Delta})=e e_q\frac{i}{2 \pi} \int \bigl [\hat{A}_{m_q}
(\vec{\Delta} _1)
\hat{B}_{-m_q}(\vec{\Delta} _2)+\hat{B}_{-m_q}(\vec{\Delta}
_2)\hat{A}_{m_q}
(\vec{\Delta} _1)+\hat{D}(\vec{\Delta} _1,\vec{\Delta} _2) \bigr ]\\
\psi_{\gamma}^{(m)}(\vec{k}_T+\vec{\Delta} _2,z) \delta(\vec{\Delta}-
\vec{\Delta} _1-\vec{\Delta} _2) d^2\vec{\Delta} _1
d^2\vec{\Delta} _2\;,\\
\nonumber
F_{che}^q(\vec{\Delta})=e \tilde{e}_q\frac{i}{2 \pi} \int \bigl
[\hat{A}_{che}
(\vec{\Delta} _1)
\hat{B}_{che}(\vec{\Delta} _2)(2-4m_q)+\hat{B}_{che}(\vec{\Delta}
_2)\hat{A}_{che}  
(\vec{\Delta} _1)(2+4m_q)\\
+\hat{E}(\vec{\Delta} _1,\vec{\Delta} _2)\bigr]
\psi_{\gamma}^{(m)}(\vec{k}_T+\vec{\Delta} _2,z) \delta(\vec{\Delta}-
\vec{\Delta} _1-\vec{\Delta} _2) d^2\vec{\Delta} _1
d^2\vec{\Delta} _2\;,
\end{eqnarray}
In (44)  $F_{es}^q(\vec{\Delta})$ is the amplitude of elastic scattering
both
of the quark with flavour $q$ and its antiquark  on the proton and
$F_{che}^q(\vec{\Delta})$ in (45)
describes  quark charge exchange ($u\bar{u} \rightarrow d\bar{d}$ or
$d\bar{d} \rightarrow u\bar{u}$) with the same flavour $q$ in the final
state as for elastic
scattering. We remind that charge exchange of
the
proton is not considered here. The electric charge of the final
quark is related to the third component of its isospin $m_q$ by the
formula
\begin{equation}
e_q=\frac{1}{6}+m_q\;.
\end{equation}
The electric charge of the initial quark before charge exchange
process is equal to
\begin{equation}
\tilde{e}_q=\frac{1}{6}-m_q\;.
\end{equation}   
Let us compare the graphs shown in Fig.~1a and Fig.~2a. The transverse
component of the initial quark momentum in Fig.~1a is $\vec{k}_T$ and 
the final transverse momentum
  is equal to $\vec{k}_T+\vec{\Delta}$. The final momenta for
both graphs in Fig.~1a and Fig.~2a are equal to each other. Hence the
transverse component of the initial quark momentum in Fig.~2a is
$\vec{k}_T+\vec{\Delta}-\vec{\Delta}_1=\vec{k}_T+\vec{\Delta}_2$ which is
the first argument of the photon wave function $\psi_{\gamma}^{(m)}$ for
the
nonspectator diagrams shown in Fig.~2a and Fig.~2b. When we consider
exchanges of reggeons with  isospin $T=1$, then amplitude of elastic 
$qN$ scattering becomes an operator in  isospin space
\begin{equation}
\hat{A}=\hat{A}_{s}+\hat{A}_{che}(\vec{\tau}_1 \cdot \vec{\tau}_2)\;.
\end{equation}
For elastic quark-proton scattering the amplitude looks like
\begin{equation}
\hat{A}_{m_q}=\hat{A}_{s}+\hat{A}_{che}(2m_q)
\end{equation}   
and for antiquark elastic scattering on the proton it is described by
the formula
\begin{equation}
\hat{B}_{-m_q}=\hat{B}_{s}+\hat{B}_{che}(-2m_q)\;       
\end{equation}
as the third component of the antiquark isospin is equal to $-m_q$.
We remind that $\hat{B}_{s}$ and $\hat{B}_{che}$ can be obtained from
$\hat{A}_{s}$ and $\hat{A}_{che}$ by multiplying every quark-quark-reggeon
vertex by the factor $\sigma _b (-1)^{T_b}$ for a reggeon $b$. Relations
(46), (47), (48), (49), (50) explain the meaning of all quantities in the
main formul\ae$\;$ (44), (45) except $\hat{D}(\vec{\Delta} _1,\vec{\Delta}
_2)$
and $\hat{E}(\vec{\Delta} _1,\vec{\Delta} _2)$ which are presented in
Appendix. They are equal to zero for  two reggeon exchanges. Their
existence is related with the fact that the amplitudes of
three and more reggeon exchanges
cannot be presented even at fixed $\Delta _1$ and $\Delta _2$ 
as a sum of products of the $qN$ and $\bar{q}N$
amplitudes (the first two terms in the brackets in (44) and (45))
multiplied by the wave function of the $q\bar{q}$-pair. To
understand why this is so let us consider, for example, the graph (in the
eikonal approximation) shown in Fig.~4a. It is obvious, that the
amplitude of such a  subprocess
with fixed $\vec{\Delta} _1$ and $\vec{\Delta} _2$ can be presented as a
product of the  $qN$ and $\bar{q}N$ amplitudes (multiplied by the photon
wave function). After some permutation of vertices along the proton and
quark lines we can get the graph shown in Fig.~4b. The symmetry
property of  many reggeon emission vertices implies an existence of the
graph
shown in Fig.~4b. But the latter diagram is irreducible and cannot be
presented for fixed momenta $\vec{\Delta} _1$ and $\vec{\Delta} _2$ as
a product of the $qN$ and $\bar{q}N$ amplitudes and the photon wave
function.   

We start our discussion of properties of the nonspectator diagram
contributions to the spin-spin asymmetry with the case of two reggeon
exchanges. Formul\ae$\;$ (44) and 
\vspace{0.5cm}
\begin{center}
\begin{tabular}{cc}
\mbox{\epsfig{file=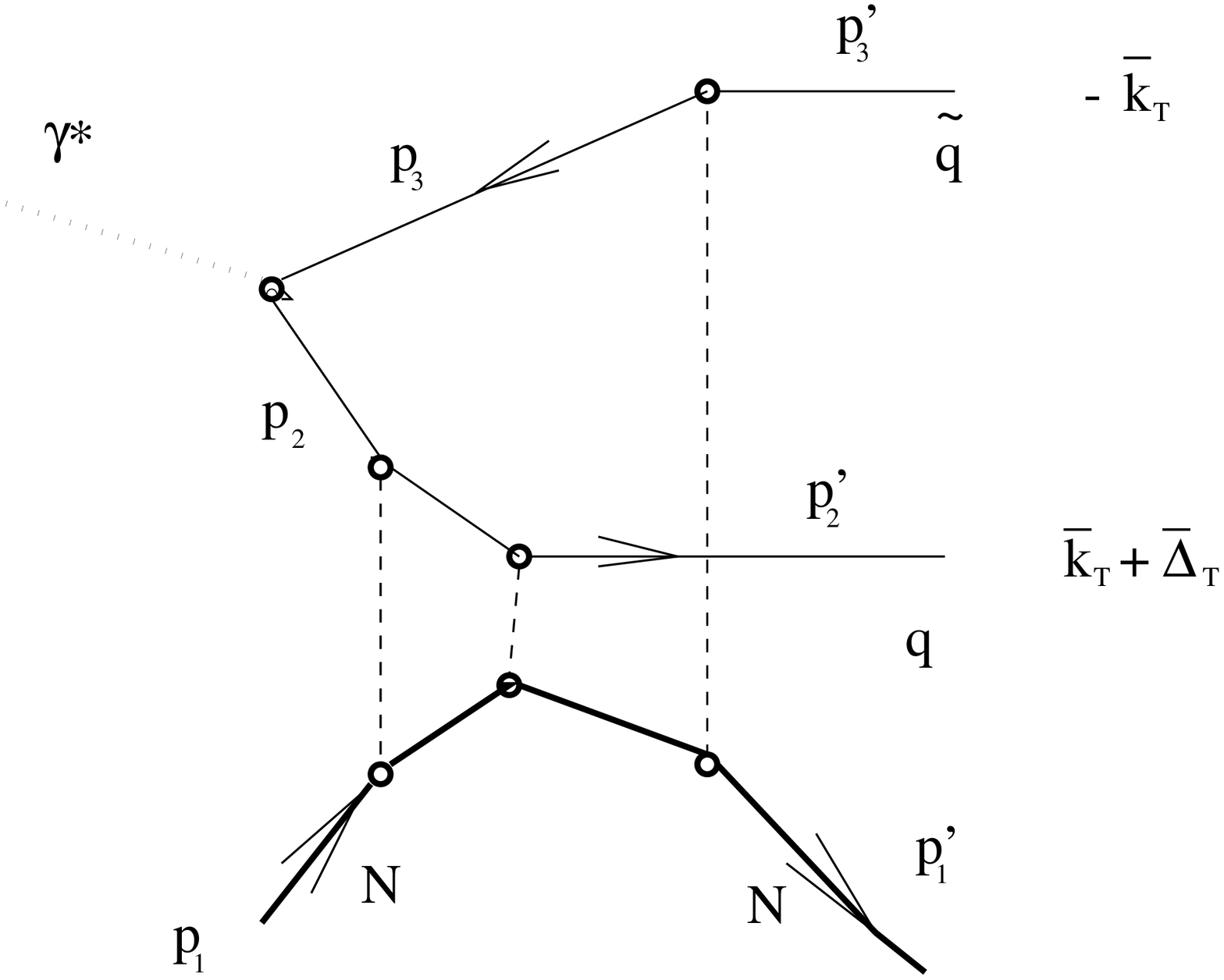,height=5cm,width=5cm,angle=270}}
\vspace{2mm}
\noindent
\small
&
\mbox{\epsfig{file=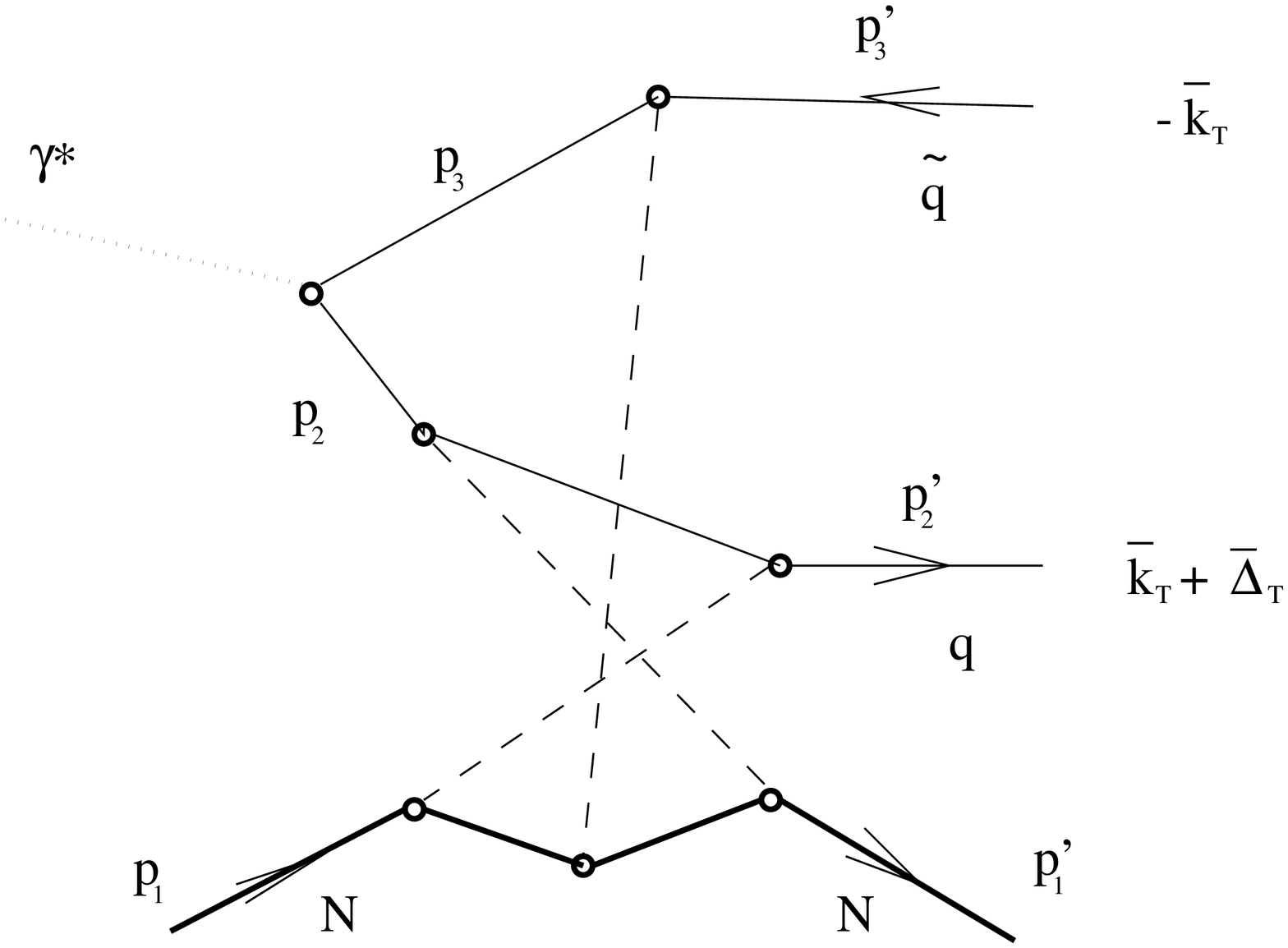,height=5cm,width=5cm,angle=270}}
\vspace{2mm}
\noindent
\small
\\
\begin{minipage}{7cm}
{\sf Fig.~4a:}
Non-spectator graph in eikonal approximation
with three reggeon exchanges.
Lines have  the same meaning as in Fig.~1a.
\end{minipage}
 \normalsize  
&
\begin{minipage}{7.5cm}
{\sf Fig.~4b:}
Non-spectator graph obtained from Fig.~4a after permutation of vertices.
Lines have  the same meaning as in Fig.~1a.
 \end{minipage}
\normalsize
\end{tabular}
\end{center}
\vspace{0.5cm}
(45) show that due to the integration over $\vec{\Delta} _1$ 
and $\vec{\Delta} _2$ the factorization into the $qN$, $\bar{q}N$
amplitudes and the wave function $\psi_{\gamma}^{(m)}$ which has been 
used for the case of the spectator graphs is lost. For $\vec{\Delta}_T=0$
 we have $\vec{\Delta} _1=-\vec{\Delta} _2$ in (44), (45) and 
integration runs over typical momentum transfers of soft scattering,
$\vec{\Delta}_1^2 \sim 1/|\lambda _P| \leq 1$ (GeV/c)$^2$. If we consider
large $Q^2$ and $k_T^2$ ($Q^2 \gg 1$ (GeV/c)$^2$, $k_T^2 \geq k_{min}^2
\gg 1$ (GeV/c)$^2 \gg  m_q^2$),
then for not too low $z(1-z)$ we can omit $\vec{\Delta} _2$ and $m_q$
($k_{min} \gg m_q$) in 
$\psi_{\gamma}^{(m)}(\vec{k}_T+\vec{\Delta} _2,z)$ as  is easy to see
from
(24). For this case we present $F_{es}^q$ by the approximate formula
\begin{equation}
F_{es}^q(0)=ie e_q \hat{f} \psi_{\gamma}(\vec{k}_T,z)\;,
\end{equation}
and for $\hat{f}$ it is easy to get the following relation from (44) and
(17)
\begin{eqnarray}
\nonumber
\hat{f}=f_1+f_2(\vec{\sigma }_{2T} \cdot \vec{\sigma }_{3T})+
f_3(\vec{\sigma }_{2T} \cdot \vec{\sigma }_{1T})+
f_4(\vec{\sigma }_{3T} \cdot \vec{\sigma }_{1T})\\
+f_5(\vec{\sigma }_2 \cdot \vec{l})(\vec{\sigma }_3 \cdot \vec{l})+
f_6(\vec{\sigma }_2 \cdot \vec{l})(\vec{\sigma }_1 \cdot \vec{l})+
f_7(\vec{\sigma }_3 \cdot \vec{l})(\vec{\sigma }_1 \cdot \vec{l})\;,
\end{eqnarray}
where $f_j$ are related to $A_i$ and $B_i$ as
\begin{eqnarray}
\nonumber
f_1=\int_0^{\infty} [A_1(\Delta _1)B_1(\Delta _1)-
A_6(\Delta _1)B_6(\Delta _1)] d \Delta _1^2\;,\\
\nonumber
f_2=\frac{1}{2} \int_0^{\infty} [A_3(\Delta _1)B_3(\Delta _1)
+A_4(\Delta _1)B_4(\Delta _1)
-A_2(\Delta _1)B_2(\Delta _1)] d \Delta _1^2\;,\\
\nonumber
f_3=\frac{1}{2} \int_0^{\infty} [A_3(\Delta _1)B_1(\Delta _1) 
+A_4(\Delta _1)B_1(\Delta _1)
-A_2(\Delta _1)B_6(\Delta _1)] d \Delta _1^2\;,\\
\nonumber
f_4=\frac{1}{2} \int_0^{\infty} [A_1(\Delta _1)B_3(\Delta _1)
+A_1(\Delta _1)B_4(\Delta _1)
-A_6(\Delta _1)B_2(\Delta _1)] d \Delta _1^2\;,\\
\nonumber
f_5=\int_0^{\infty}A_5(\Delta _1)B_5(\Delta _1) d \Delta _1^2\;,\\
\nonumber
f_6=\int_0^{\infty}A_5(\Delta _1)B_1(\Delta _1) d \Delta _1^2\;,\\
f_7=\int_0^{\infty}A_1(\Delta _1)B_5(\Delta _1) d \Delta _1^2\;.
\end{eqnarray}
We have omitted the flavour index $q$ in $f$,  $f_j$, $A_j$, and $B_j$ to
have simpler
notation.
Formula (51) shows that the factorization under discussion is restored
and we can use formul\ae$\;$ (28), (25), (26), (31) to get the spin-spin
asymmetry given by the nonspectator graphs only
\begin{eqnarray}
\nonumber
A_{LL}=2 \sum _{q=u,d,...}\int_0^1(2z-1)\Bigl
\{|f_4|^2-|f_3|^2+\Re[(f_1-f_5)(f_6-f_7)^*]
\Bigr \} \theta(M_1^2)dz\\
\nonumber
/\sum _{q=u,d,...}\int_0^1\Bigl
\{|f_1|^2+4|f_2|^2+2|f_3|^2+2|f_4|^2+|f_5|^2\\
\nonumber
+|f_6|^2+|f_7|^2-\Re(f_1f_5^*+f_6f_7^*)][z^2+(1-z)^2]\\
+4\Re(f_1f_2^*-f_5f_2^*+f_3f_4^*)z(1-z)  \bigr \}
\theta(M_1^2)dz\;
\end{eqnarray}
with $M_1^2=M_X^2-k_{min}^2/[z(1-z)]$.
We shall discuss contributions to the numerator in (54). Putting (18) and
(20) into (53) we obtain the relations 
\begin{equation}
f_5=f_6=f_7=0
\end{equation}
valid for the two reggeon exchange contributions. It follows from (55) and
(54) that only $f_3$ and $f_4$ can contribute to the numerator of the
expression for $A_{LL}$ in the two reggeon exchange approximation for the
nonspectator graph contributions. It is the contribution of $f_3$ and
$f_4$ which will be discussed below. We have for the pure 
pomeron contributions in addition to (55) the relations
\begin{equation}
f_3=f_4=0\;.
\end{equation}
Combining (55), (56) with (54) we conclude that the two pomeron exchanges
do not contribute to $A_{LL}$ as for the spectator graphs. 

Let us apply more detailed notations for $f_j$. We denote by
$f_j^{hc}(z)$
that part of the amplitude $f_j$ which is determined by exchanges of
some reggeons $c$ and $h$. For example, for $f_6(z)$ we have from (53)
\begin{equation}
f_6^{hc}(z)=\int _0^{\infty} [A_5^c(z) B_1^h(1-z)+A_5^h(z) B_1^c(1-z)]
d \Delta _1^2\;,
\end{equation}
where we have taken into account that the square of the center-of-mass
energy for  antiquark-proton scattering is $(1-z)s_{\gamma p}$ and hence
$B_j$ depends on $1-z$. We have omitted the $\Delta _1$-dependence of
$A_j$, $B_j$ in (57) as this is not essential for our consideration now.
Using
the relation $B_j^a(z)= \sigma _a A_j^a(z)$ for any reggeon $a$ we can
easily get the relation of interest 
\begin{equation}
f_3^{ch}(z)=\sigma _c \sigma _h f_4^{ch}(1-z)
\end{equation}
from (53). Let us consider the contributions of the first order 
amplitudes $f_3^{cP}$, $f_3^{hP}$ and $f_4^{cP}$, $f_4^{hP}$ to the
numerator in (54) which looks like 
\begin{equation}
\Re \bigl [ f_4^{*cP}(z)f_4^{hP}(z)-f_3^{*cP}(z)f_3^{hP}(z) \bigr ]\;.
\end{equation}
Applying (58) to the second term in (59) we transform it to the relation
\begin{equation}
\Re \bigl [ f_4^{*cP}(z)f_4^{hP}(z)-\sigma _c \sigma _h
f_4^{*cP}(1-z)f_4^{hP}(1-z) \bigr ]\;.
\end{equation}
It follows from (60) and (54) that the integral over $z$ vanishes  when 
$\sigma _c \sigma _h=-1$ and it is generally speaking nonzero if $\sigma
_c
\sigma _h=1$. As  has been pointed out the amplitudes of  pure pomeron
exchanges $f_3^{PP}$ and $f_4^{PP}$ are equal to zero hence there
are no contributions of the first order in $\varepsilon$ to $A_{LL}$ in
the two reggeon exchange approximation.
It follows from (60) and (54) that any secondary reggeon trajectory $c$
contributes
to the numerator of $A_{LL}$ (the term with $h=c$ in (60)) but there are
no interference terms for any
two reggeons $c$ and $h$ with opposite signatures $\sigma _c$ and $\sigma
_h$. As  can be seen from the general formul\ae$\;$ presented in the
Appendix the contributions $\sim \varepsilon ^0$ to $f_3$ 
and $f_4$ start from the three
pomeron exchange contribution. Denoting such amplitudes as $f_3^{PPP}$ 
and $f_4^{PPP}$ we can easily check that their contribution to $A_{LL}$
does not vanish. The first order contribution to $A_{LL}$ due to the
interference terms between the three pomeron exchange amplitudes
$f_3^{PPP}$,  $f_4^{PPP}$ and  $f_3^{hP}$, $f_4^{hP}$ looks like
\begin{equation}
2 \Re \bigl [ f_4^{*PPP}(z)f_4^{hP}(z)-
f_3^{*PPP}(z)f_3^{hP}(z) \bigr ]\;.
\end{equation}
Using (53) we can transform (61) to the relation
\begin{equation}
2 \Re \bigl [ f_4^{*PPP}(z)f_4^{hP}(z)-\sigma _h
f_4^{*PPP}(1-z)f_4^{hP}(1-z) \bigr ]\;
\end{equation}
as (59) was transformed into (60). Formula (62) shows 
that contributions to $A_{LL}$
of the first order in $\varepsilon$ vanish for reggeons with
negative signatures. It is easy to conclude that contributions of the
second order due to interference terms $f_3^{*PPP}f_3^{hc}$ and
$f_4^{*PPP}f_4^{hc}$ ($c,\;h \neq P$) vanish if reggeons $c$ and $h$ have
opposite signatures.

Comparison of (27) with (52) shows that the total amplitude $F_{tot}$ of
$q\bar{q}$-pair scattering on the proton for large $Q^2$ and $k_T^2$ at
$\Delta _T=0$ has the same form (52) as for $\hat{f}$ and we can formally
take into account the spectator graph amplitude if we make the
following replacements: $f_1 \rightarrow f_1-iA_1-iB_1,\;f_3 \rightarrow
f_3-iA_4,\;f_4 \rightarrow f_4-iB_4,\;f_6 \rightarrow f_6-iA_5,\;
f_7 \rightarrow f_7-iB_5$ and $f_2,\;f_5$ being unchanged. The main
conclusion that there are no contributions of the first order to the
numerator of expression (54) for the spin-spin asymmetry due to 
exchanges with a negative signature reggeon plus some number of pomerons 
remains valid. Contributions of the second order in $\varepsilon$
of any signature reggeons are not suppressed but interference terms for
opposite signature reggeon contributions to the numerator of $A_{LL}$ are
absent. In reality at $\Delta _T =0$ the spin-dependent amplitude of 
$q \bar{q}$-pair scattering off the proton due to  pomeron exchange is
numerically very small. Indeed, the ratio of it to the spin-independent
part of the pomeron exchange amplitude is proportional to a high 
power of the small quantity $P_y/(P_s \sqrt{|\lambda _P|})$ (compare for 
example the fourth term in (23) with the first one). Hence all
interference terms in the numerator in (54) 
of the first order ($\sim \varepsilon$) with the spin-dependent
amplitudes of pomeron exchanges  can be at energies 
achieved experimentally up to now at HERA numerically much less
than contributions of the second order which are not suppressed by some
selection rules. As this is not excluded experimentally that $P_y=p_y=0$,
then in this case the lowest order contributions to $A_{LL}$ are the
second order ones.

Last but important remarks. If we restrict our consideration by
demanding $k_T \geq k_{min}$, then
due to the $\delta$-function in (28) $z(1-z)\geq
(k_{min}^2+\mu_q^2)/M_X^2$
and we
integrate in (28) and (54) over $z$ from $z_{min}$ up to $1-z_{min}$ with
$z_{min}\approx (k_{min}^2+\mu_q^2)/M_X^2$. The square of the
center-of-mass energy of $qp$ or $\bar{q}p$ scattering is greater than
$s_{min} \approx s_{\gamma p}(k_{min}^2+\mu_q^2)/M_X^2$. 
If $s_{min} \gg s_0$ we
suppress the low energy parts of contributions to $A_{LL}$. But this is
the region of integration over $z$ where secondary reggeon trajectory
contributions are most important. Hence increasing the value of 
$k_{min}$ we can decrease the value of $A_{LL}$ appreciably. 
If $s_{min} \leq s_0$ we are out of the
applicability of the Regge phenomenology hence we should avoid
kinematics with very small $z$. Integrals (28) and (54)
are formally divergent at $z \rightarrow 0$ and $z \rightarrow 1$ as for
a contribution
 of some secondary reggeon $h$ with $\alpha _h(0)=0.5$ to quark-proton
scattering the energy
dependence
looks like $(zs_{\gamma p}/s_0)^{\alpha _h(0)-1} \sim z^{-1/2}$ and for 
antiquark-proton scattering the amplitude depends on $z$ as
$(1-z)^{\alpha _h(0)-1} \sim (1-z)^{-1/2}$. But the formula
$z_{min}\approx (k_{min}^2+\mu_q^2)/M_X^2$ shows that even for
$k_{min}=0$ 
$z_{min}> 0$ due to the mass of the constituent quark. Hence we cannot
neglect $\mu_q$ in the kinematics with low or zero $k_T$ to avoid the
infrared divergence of the integrals for the
cross section and for the longitudinal spin-spin asymmetry.

\section{Numerical results and discussion}

For the numerical calculation performed both for $\Delta _T^2=0$ and at
nonzero momentum transfers we have not used the approximations applied
in previous sections to discuss some qualitative properties of the
longitudinal
spin-spin asymmetry. For the amplitude corresponding to the spectator
graphs $F^q_{sp}(\vec{\Delta} _T)$ we have used the formula
\begin{equation}
F^q_{sp}(\vec{\Delta} _T)=ee_q[\hat{A}_{m_q}(\vec{\Delta} _T)
\Psi _{\gamma}^{(m)}(\vec{k} _T, z)+
\hat{B}_{-m_q}(\vec{\Delta} _T) \Psi _{\gamma}^{(m)}(\vec{k}
_T+\vec{\Delta}
_T, z)] 
\end{equation}
where $\hat{A}_{m_q}(\vec{\Delta})$ is the full amplitude of elastic
quark-proton
scattering containing six invariant amplitudes $A_j$ defined in (17). The
isospin structure of the amplitude is given by (48) and the meaning of the
notation $\hat{A}_{m_q}(\vec{\Delta})$ is explained by relation (49). The
amplitude of
antiquark-proton scattering $\hat{B}_{-m_q}(\vec{\Delta})$ has been
defined in
(50).
At $\Delta _T=0$ formula (63) reduces to
\begin{eqnarray}   
\nonumber
F^q_{sp}(0)=M^q_{sp}(0)\Psi _{\gamma}^{(m)}(\vec{k} _T, z)
\end{eqnarray}
with $M^q_{sp}(0)$ defined in (27).
For the contributions of the nonspectator graphs we have applied
formul\ae$\;$ (44) and (45) without using the approximate relation (51). 
We put the total
amplitude $F^q(\vec{\Delta} _T)$ for the diffractive production of the
quark-antiquark pair with a flavour $q$ in $\gamma p$-scattering  
\begin{equation}
F_q(\vec{\Delta} _T)=F^q_{sp}(\vec{\Delta} _T)+
F^q_{es}(\vec{\Delta} _T)+F^q_{che}(\vec{\Delta} _T)
\end{equation}
into the formula for the cross section of production of a hadronic system
with the total mass $M_X$
\begin{eqnarray}
\frac{d \sigma _m}{dt dM_x^2}=4 \pi n_c  \sum_{q=u,d,...}
\int F^+_q(\vec{\Delta} _T) F_q(\vec{\Delta} _T) \rho _1
\delta \Bigl (M^2_X-\frac{\mu ^2_q+
[\vec{k}_T+\vec{\Delta}_T (1-z)]^2}{z(1-z)} \Bigr ) dz d^2 k_T
\end{eqnarray}
where the proton spin density matrix $ \rho _1$ has been defined in (29)
and $F^q_{es}(\vec{\Delta} _T)$, $F^q_{che}(\vec{\Delta} _T)$
have been given by (44), (45).
We have omitted in (64) an index $m$ for all amplitudes but kept it
for the cross section where $m$ denotes the virtual photon helicity. The
longitudinal spin-spin asymmetry has been calculated with the aid of
the general formula (31).

Fig.~5 shows the results of the calculation of $A_{LL}$ when both the
spectator and non-spectator graphs have been taken into account. We
see that the
pure pomeron part of the spin-spin asymmetry shown by the dash-dotted
line is very small ($< 10^{-15}$ at $s_{\gamma p} \geq 100$ GeV$^2$).
We put in the calculation the mass of the constituent quark 
$\mu _q=0.35$ GeV/c$^2$, the minimal value of $k_T$ equal to 0.2 GeV/c,
the mass of the produced hadrons $M_X=10$ GeV/c$^2$ and the heavy photon
virtuality $Q^2=10$ (GeV/c)$^2$. The first order ($\sim
\epsilon=\sqrt{s_0/s_{\gamma p}}$) contribution to $A_{LL}$ is negative
and it represents an interference term of the pomeron exchange amplitude
with the amplitude of a sum of  $\rho(770),\;f_2(1270),\;A_2(1320)$ and
$\omega(782)$ exchanges (and some
number of  pomeron exchanges). Due to the smallness of the
spin-dependent part of the pomeron exchange amplitude the first order
contribution to $A_{LL}$ (the dashed curve) is much smaller
than the second order one shown with the dotted line in Fig.~5. 
We would like to remind that the phenomenological analysis of
hadron-hadron scattering \cite{BST} and \cite{BLSTM} is compatible with
the zero value of the spin-dependent vertex of pomeron exchange.
For this case the zeroth and first order contributions to $A_{LL}$
vanish. But even if they do not vanish they are not important 
at $s_{\gamma p} \leq 9 \cdot 10^4$ GeV$^2$ (the highest energy of $ep$
scattering at the collider HERA) since the
second order contribution dominates very much as one can see from
Fig.~5. The solid curve in Fig.~5 shows the behaviour of the
longitudinal spin-spin asymmetry calculated with the aid of formul\ae$\;$
(63), (64), (44), (45), (65) and (31) without decomposition of $A_{LL}$
into a power series in $\epsilon$. We shall call such $A_{LL}$  "the
total asymmetry". We remind that the calculated amplitudes contain terms
$\epsilon ^0$, $\epsilon $, $\epsilon ^2$ only and we have taken into
account one, two and three reggeon exchanges only.    

At first sight the zeroth order contribution to $A_{LL}$ has to be
a constant  and the first and the second orders should behave as
$\epsilon \sim 1/\sqrt{s_{\gamma p}}$ and 
$\epsilon ^2\sim 1/s_{\gamma p}$ as functions of the energy of hard
$\gamma p$-scattering. The reasons why it is not true in our
calculations are as follows. We start our consideration with the
amplitudes of  pure pomeron exchanges. We see from (18), (20), (21),
(22) that, with the exception of  the one pomeron exchange amplitudes,
 all the quark-proton amplitudes at $\Delta_T=0$
depend on $\lambda _P$ which is a linear function
of $\ln(s/s_0)$ in accordance with (19). More over as $\alpha _P(0)>1$
(see Table 1) the factors $(s/s_0)^{n \alpha _P(0)-n}$
in (18) are energy dependent and they are
different for different $n$, hence an interference between
amplitudes of  $m$ and $n$ pomeron exchanges ($m \neq n$) depends on
$s$. As the signature factor $\eta _P (0) \approx i$ for  pomeron
exchange, then the product $i^{n-1}[\eta _P (0)]^n \approx i (-1)^{n+1}$
(see (12) and (13)) changes its sign which leads to  destructive
interference between
exchanges with odd and even numbers of  pomerons. As a result
of such an interference  $A_{LL}$  can change a sign as a function of the
collision energy which can lead to irregular behaviour of  
$\ln|A_{LL}|$ versus $\ln s_{\gamma p}$ (see, for example, Fig.~10). Such
an
interference becomes especially
important when we add nonspectator graph contributions.
We show in all the figures presented in this section a behaviour 
of the absolute value of $A_{LL}$ smoothed near 
points in which $A_{LL}=0$ and hence $\ln|A_{LL}|$ tends to $- \infty$.
 If we put $d \alpha _a(t)/dt =0$ at $t=0$ in (19) for all reggeons under
consideration and $\alpha _P(0)=1$, then the zeroth order contribution to
$A_{LL}$
vanishes since $B_4$ and $A_4$ become independent of the collision energy
(and hence of $z$) and therefore the integral in the numerator in (32)
vanishes. The dependence of the first and 
\linebreak
\begin{center}
\mbox{\epsfig{file=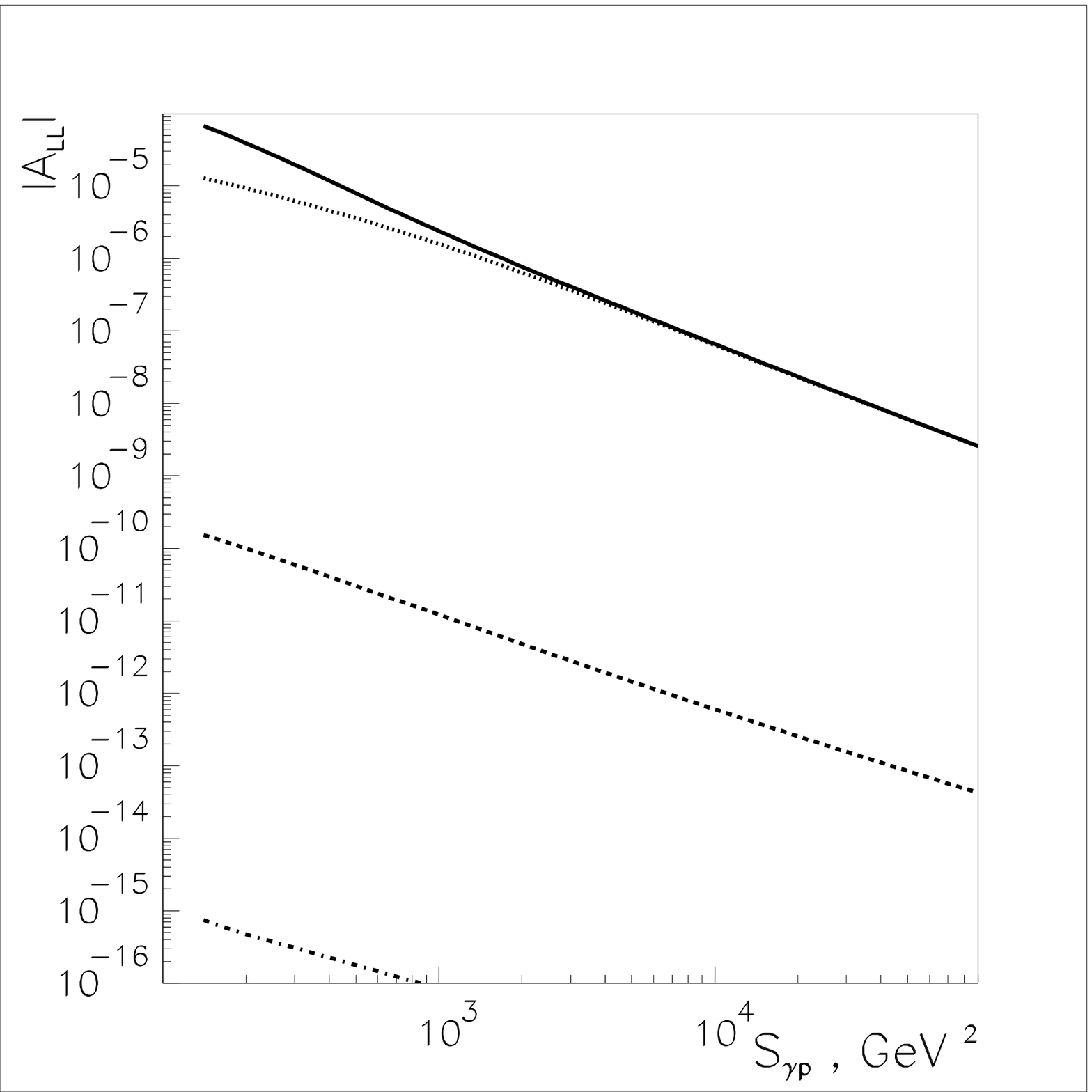,height=7.5cm,width=9.cm}}
\noindent
\small
\begin{minipage}{14.5cm}
\vspace{7mm}
{\sf Fig.~5:}
Dependence of longitudinal spin-spin asymmetry on $s_{\gamma p}$.
Dash-dotted and dashed curves show pure pomeron 
(zeroth order in $\epsilon =\sqrt{s_0/s_{\gamma p}}$) and
interference of secondary reggeons with pomeron ( 
$\sim \epsilon ^1$) contributions to the absolute value of
$A_{LL}$, respectively.
Dotted curve shows
contribution  $\sim \epsilon ^2$ to $|A_{LL}|$ and solid line represents
"total
asymmetry" (no decomposition into a power series in $\epsilon$). For all
curves $M_X=10$ GeV/c$^2$, $Q^2=$ 10 (GeV/c)$^2$, 
\linebreak
$\Delta _T =0$ , $k_T\geq k_{min}=0.2$ GeV/c.
\end{minipage}
\end{center}
\vspace{0.3cm}
\normalsize
\begin{center}
\mbox{\epsfig{file=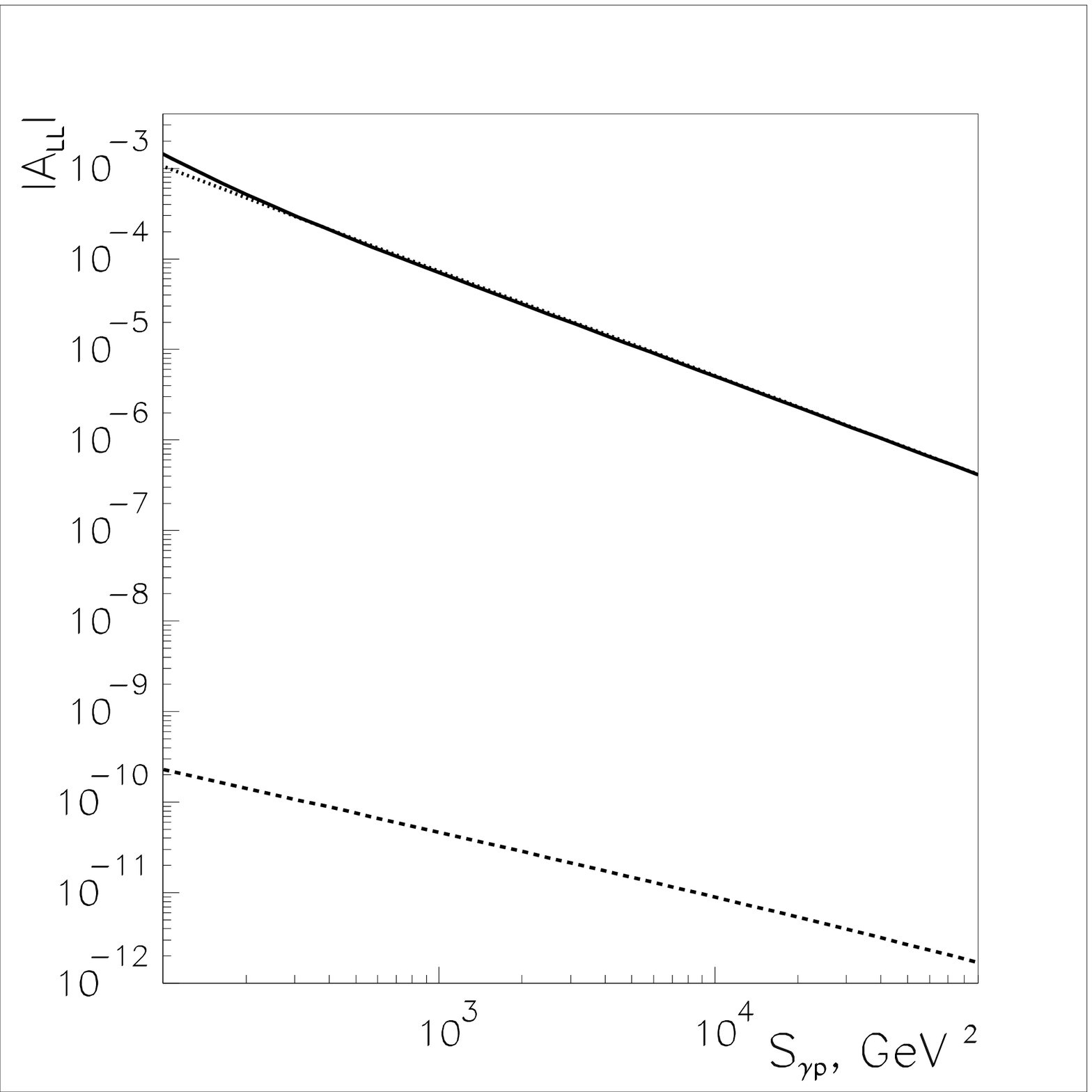,height=7.5cm,width=9.cm}}
\noindent
\small
\begin{minipage}{14.5cm}
\vspace{7mm}
{\sf Fig.~6:}
Spectator graph contributions to longitudinal spin-spin asymmetry.
Curves are the same as in Fig.~5 but $\alpha _P(0)=1,\; d \alpha
_a(t)/dt=0$ at $t=0$ for all reggeons $a=P,\;\rho,\;f,\;A_2,\;\omega$.
Pure pomeron contribution to $A_{LL}$ is equal to zero.
For all
curves $M_X=5$ GeV/c$^2$, $Q^2=$ 10 (GeV/c)$^2$,
$\Delta _T =0$, $k_T \geq k_{min}=0.2$ GeV/c.
\end{minipage}
\end{center}  
\vspace{0.3cm}
\normalsize   
second order contributions to
the longitudinal spin-spin asymmetry becomes linear with a high accuracy
if we plot $\ln|A_{LL}|$ versus $\ln(s_{\gamma p})$ and corresponds
approximately to the behaviour $\sim s_{\gamma p}^{-1/2}$
and $\sim s_{\gamma p} ^{-1}$, respectively. 
This statement is illustrated with the curves presented in
Fig.~6. If we take into account the dependence of all $ \lambda _a$ on  
$s_{\gamma p}$ and the interference of contributions of different numbers 
of reggeon exchanges the behaviour of $A_{LL}$ becomes more involved. We
can see 
this comparing the solid, dashed and dotted curves in Fig.~5 with the
curves
of the same kind in Fig.~6.

Contributions of exchanges with different reggeons to the total amplitude
of diffractive photoproduction can interfere, so the total result
could in principle be much
smaller than the asymmetry for the calculation of which exchanges
with one reggeon $c$ ($c=\rho,\;f,\;\omega,\;A_2$) and the pomeron have
been taken into account. In
this case a small value of $A_{LL}$ is very sensitive to the values of the
phenomenological parameters presented in Table~1. Any change
of the parameters would destroy the cancellation of the different
reggeon contributions and change the value of $A_{LL}$ crucially. The     
dotted curve in Fig.~7 shows the second order contributions
($\sim \epsilon ^2$) to $A_{LL}$ due to exchanges with the pomeron and
$A_2$-reggeon. For the calculation of the dashed curve we have taken into
account pomeron and $\rho$-reggeon exchanges. We see also from Fig.~7 that
contributions of the $\omega$-meson trajectory (dash-dotted curve) and the
$f$-trajectory (bold line) with some number of pomeron exchanges are less
than the $A_2$- and $\rho$-reggeon exchange contributions. We are to
compare these curves with the solid line representing contributions of all
reggeons discussed above ($P,\;\rho,\;A_2,\;\omega,\;f$). We see that
$A_{LL}$ corresponding to the solid line is  not much smaller than the
longitudinal spin-spin asymmetries shown with other curves. Hence there is
no destructive interference
discussed above. This conclusion remains true 
\vspace{0.3cm}
\begin{center}
\mbox{\epsfig{file=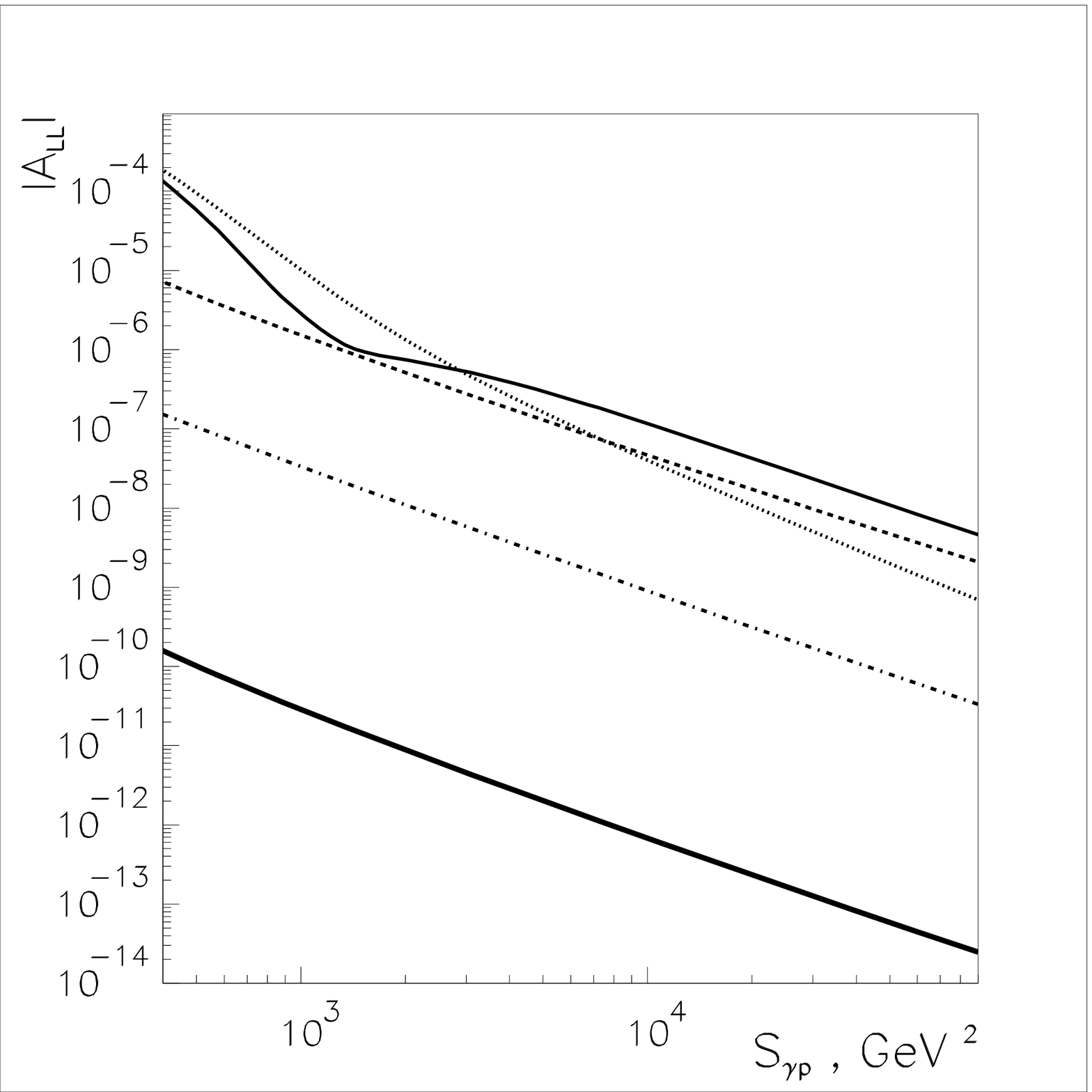,height=7.7cm,width=9.cm}}
\noindent
\small
\begin{minipage}{14.2cm}
\vspace{7mm}
{\sf Fig.~7:}
Second order contributions of different reggeons to $A_{LL}$.
Dashed ($P+\rho$), dotted ($P+A_2$), dash-dotted ($P+\omega$), 
bold ($P+f$) and solid 
\linebreak
($P+\rho+A_2+\omega+f$) curves are calculated
with $M_X=20$ GeV/c$^2$, $Q^2=$ 10 (GeV/c)$^2$, $\Delta _T =0$,
$k_T\geq k_{min}=0.2$ GeV/c for spectator graphs only.
\end{minipage}
\end{center}
\vspace{0.3cm}
\normalsize
if we add nonspectator graph
contributions. So we see from a comparison of the curves
presented in Fig.~7 that our estimate of $A_{LL}$ does not depend
crucially on values of the parameters found in \cite{BST},  \cite{BLSTM}.

As discussed above the spectator graph contributions to $A_{LL}$  
do not depend on $Q^2$. Figure~8 shows that $A_{LL}$ is not very sensitive
to the value of $Q^2$ for $s_{\gamma p}>400$ GeV$^2$ even if we add the
nonspectator graph contributions to the spectator ones. We see that the
difference is less than 4$\%$ for 4 (GeV/c)$^2 \leq Q^2 \leq 100$ 
(GeV/c)$^2$. In contrast to 
the practical independence on $Q^2$ at $s_{\gamma p} > 400$ GeV$^2$ 
the longitudinal spin-spin asymmetry is very sensitive to the value of the 
total mass $M_X$ of hadrons produced in the hard 
\vspace{0.5cm}
\begin{center}
\mbox{\epsfig{file=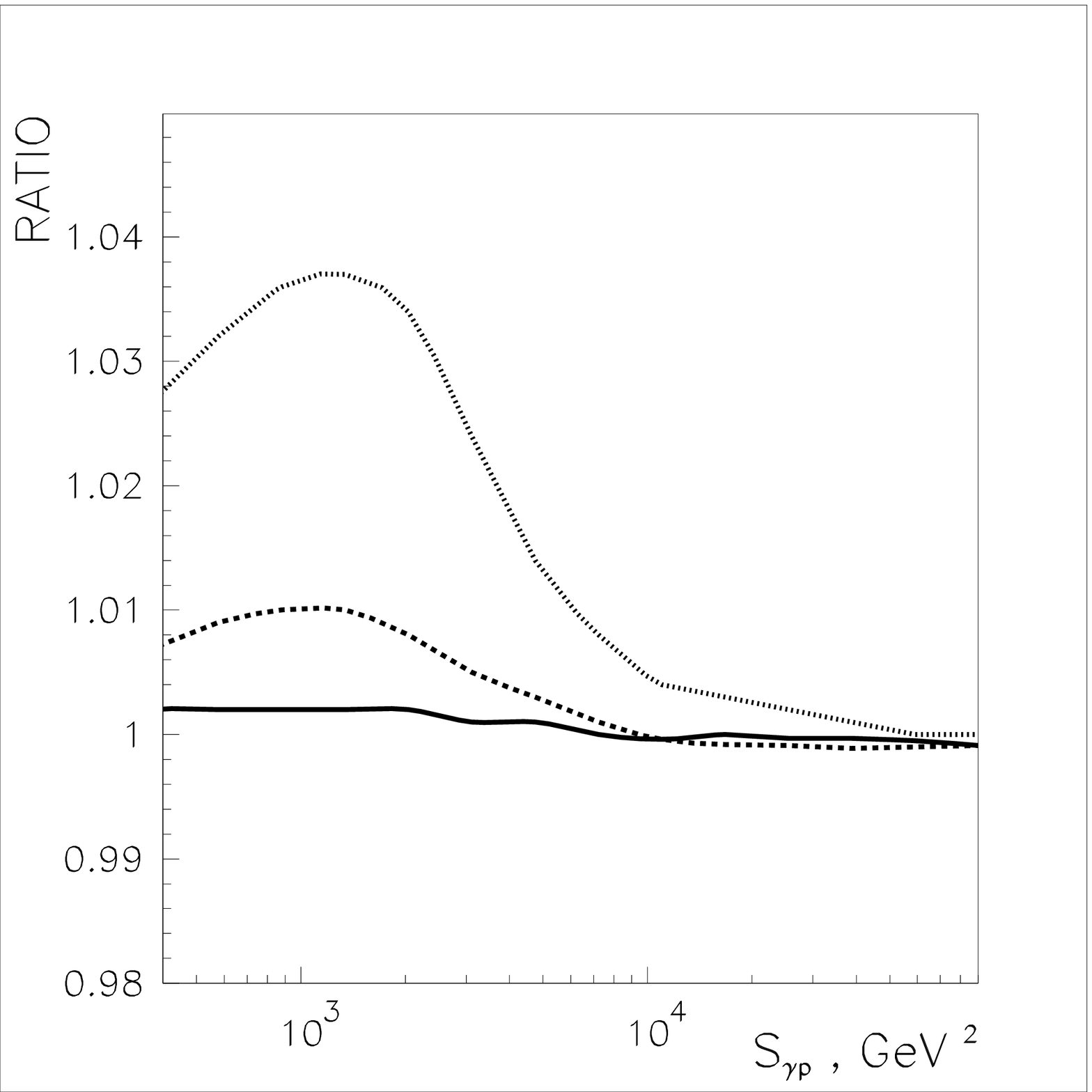,height=7.cm,width=9.cm}}
\noindent
\small
\begin{minipage}{14.2cm}
\vspace{7mm}
{\sf Fig.~8:}
Dependence of longitudinal spin-spin asymmetry on $Q^2$.
Solid, dashed, dotted curves show ratios of spin-spin asymmetries
calculated for $Q^2=10$,
20, 100 (GeV/c)$^2$ to $A_{LL}$ obtained at $Q^2=4$ (GeV/c)$^2$.
All curves are calculated with
\linebreak
$M_X=20$ GeV/c$^2$, $\Delta _T =0$,
$k_T \geq k_{min}=0.2$ GeV/c.
\end{minipage}
\end{center}  
\vspace{0.5cm}
\normalsize
$\gamma p$-collision as we can see from Fig.~9a. Indeed, when $M_X$
changes from 3 GeV/c$^2$ to 40 GeV/c$^2$, then $A_{LL}$ increases by more
than an order of magnitude. The
explanation is the following. The dominant contribution of secondary
reggeon trajectories to the longitudinal spin-spin asymmetry originates
from low energy collisions of the quark and antiquark off the proton. The
square of the center-of-mass energy of quark-proton scattering is
$zs_{\gamma p}$
where at $\Delta _T=0$ $z$ is a root of the equation
\begin{equation}
\nonumber
z(1-z)=\frac{k_T^2+\mu_q^2}{M_X^2}
\end{equation}
which makes the argument of the $\delta$-function in (65) equal to zero.
If $k_T^2 \ll M_X^2$ and $\mu_q^2 \ll M_X^2$ we have the approximate
solution
\begin{equation}
z \approx \frac{k_T^2+\mu_q^2}{M_X^2}
\end{equation}

\vspace{0.5cm}
\begin{center}
\begin{tabular}{cc}
\mbox{\epsfig{file=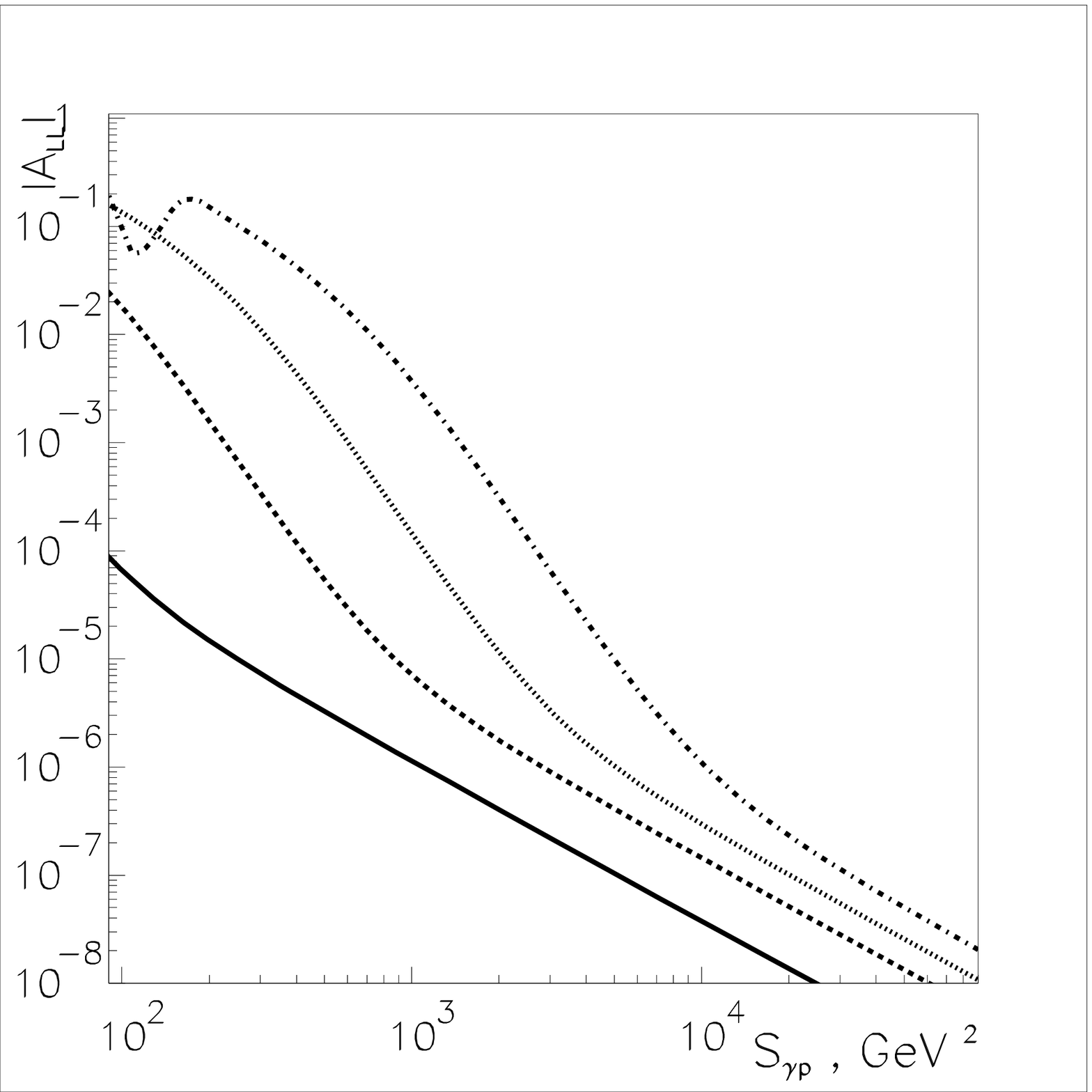,height=7.5cm,width=7.5cm}}   
\vspace{2mm}
\noindent
\small
&
\mbox{\epsfig{file=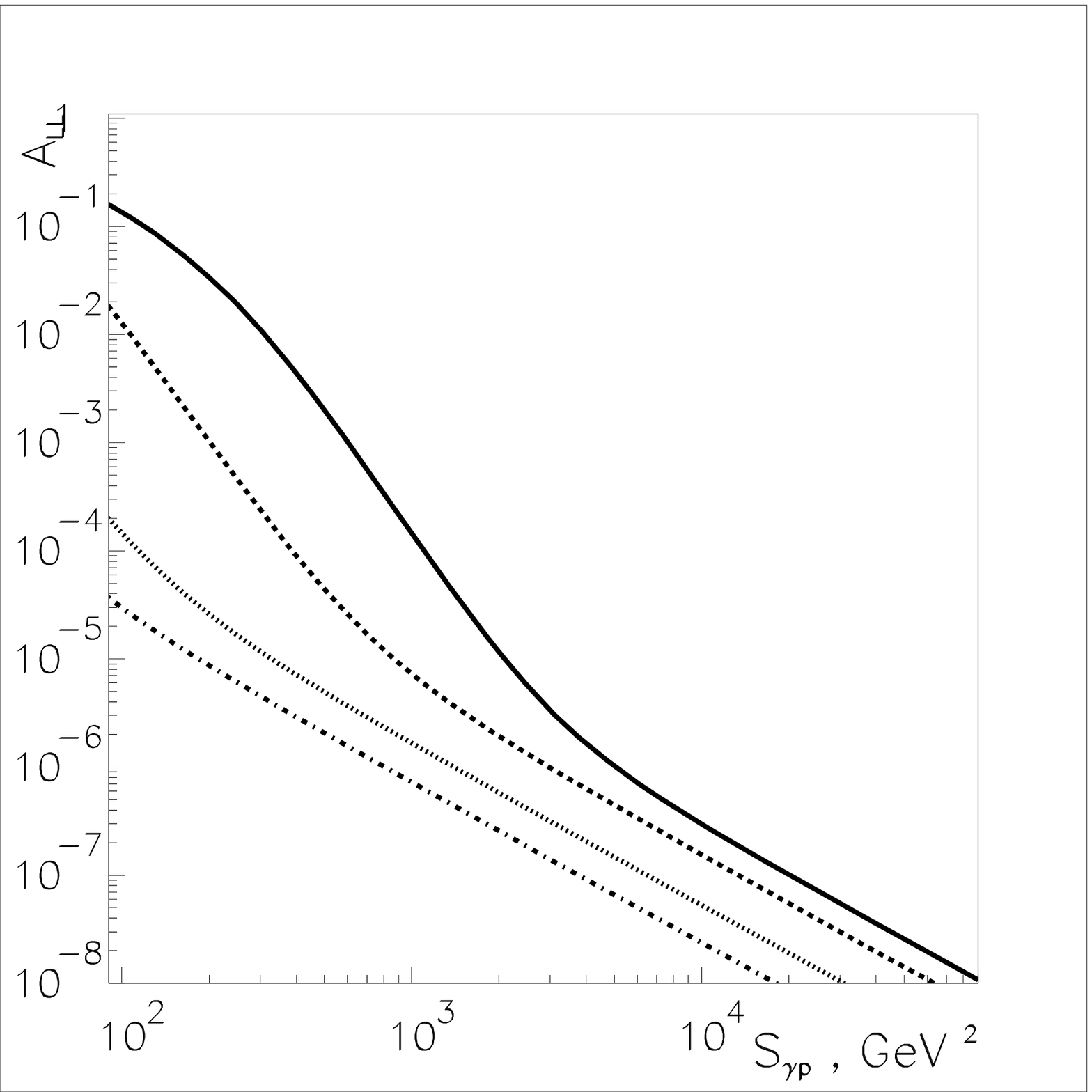,height=7.5cm,width=7.5cm}}
\vspace{2mm}
\noindent
\small
\\
\begin{minipage}{7cm}
{\sf Fig.~9a:}
Dependence of longitudinal spin-spin asymmetry on $M_X$.
Solid, dashed, dotted, and dash-dotted curves are calculated for $M_X=3$,
10,
20, 40 GeV/c$^2$
all curves being computed
with $Q^2=10$ (GeV/c)$^2$, $\Delta _T =0$,
$k_{min} = 0.2$ GeV/c.
\end{minipage}
 \normalsize
&
\begin{minipage}{7.5cm}
{\sf Fig.~9b:}
Dependence of longitudinal spin-spin asymmetry on the minimal value of
$k_T$.
Solid, dashed, dotted, and dash-dotted curves calculated for
$k_{min}=0.2$,
1, 3, 5 GeV/c.
All curves are computed
with $Q^2=10$ (GeV/c)$^2$, $\Delta _T =0$,
$M_X=20$ GeV/c$^2$.
 \end{minipage}
\normalsize
\end{tabular}
\end{center}   
\vspace{0.5cm}
which shows that the greater is $M_X^2$ the smaller is $z$ other things
being equal. But contributions of secondary reggeon exchanges to $A_{LL}$
increase with a decrease of the collision energy. When $1-z$ is small we
have instead of (67)   
\begin{equation}
\nonumber
1-z \approx \frac{k_T^2+\mu_q^2}{M_X^2}\;.
\end{equation}
But $(1-z)s_{\gamma p}$ is just the square of the center-of-mass
energy of antiquark-proton
scattering which decreases with an increase of $M_X^2$. We conclude that
the contribution of secondary Regge trajectories  to $A_{LL}$ increases
with an increase of $M_X^2$ for the
antiquark-proton collision too. The minimal value
of $z$ corresponds to $k_T=0$. Putting the mass of the constituent quark
$\mu _q=350$ MeV/c$^2$ and $M_X=10$ GeV/c$^2$ in (66) we have 
$z_{min}=\mu _q^2/ M_X^2 \approx 0.001$. This means that for 
$s_{\gamma p} \approx 10^3$ GeV$^2$ we have for the square of the
quark-proton collision energy the relation 
$z_{min}s_{\gamma p} \approx 1$  GeV$^2$. Hence for $s_{\gamma p} \ll
10^3$ GeV$^2$ we are out of the applicability of the Regge phenomenology
used in the present paper. 
Strictly speaking this is true if the dominant
contribution to the integral in (65) comes from those values of $z$ for
which $z(1-z)$ is close to $z_{min}$. 
Let us consider integral (65) with the lower and upper limits of
$z$-integration $z_0$ and $1-z_0$, respectively. 
Using this auxiliary integral we can study 
which region of $z$ gives the dominant contribution to integral (65).
The calculations with parameters $Q^2=10$ (GeV/c)$^2$, $M_X=10$
GeV/c$^2$, $s_{\gamma p}=2000$ GeV$^2$ show that $A_{LL}$ lost more
than $50\%$ of its value if $z_0=z_{1/2}=9\cdot z_{min}$ and more than
$70\%$ for
$z_0=20\cdot z_{min}$ where $z_{min} \approx 0.001$. This is a typical
example of the behaviour of integral (65) as a function of $z_0$. 
We see from the consideration of the auxiliary integral that
 most  of the contribution to $A_{LL}$ comes from $z$ close to
$z_{min}$ or $1-z$ close to $z_{min}$.
We conclude that a boundary value of $s_{\gamma p}$ for an
applicability of our approach is $s_{\gamma p} \sim s_0/z_{1/2} \approx
100$ GeV$^2$ since $s_0 \sim 1$ GeV$^2$. 

The main aim of our discussion is to find kinematical conditions where
the contributions to $A_{LL}$ of the soft processes are suppressed. For 
this  case the hard process contributions dominate and we can reliably
predict the
longitudinal spin-spin asymmetry within the framework of perturbative QCD.
As we can conclude from the considerations of Fig.~9a we are to decrease
the mass of the hadronic system  to decrease the soft process contribution
to $A_{LL}$. Another possibility to suppress this contribution follows
immediately
from formula (66). When we select events with $k_T \geq k_{min}$, then 
\begin{equation}
\nonumber
z_{min} \approx \frac{k_{min}^2+\mu_q^2}{M_X^2}\;.
\end{equation}
Hence for relatively large $k_{min}$ the minimal value of the quark-proton
collision energy in the center-of-mass system
($\sqrt{z_{min}s_{\gamma p}}$) can be much greater than
$\sqrt{s_0}$ (we assume that $s_{\gamma p} \gg s_0$). 
As a result the contributions of  secondary Regge
trajectories
to $A_{LL}$  are suppressed. As the maximal value of z is equal to
$1-z_{min}$ the antiquark-proton scattering energy is greater than
$\sqrt{(1-z_{max})s_{\gamma p}}=\sqrt{z_{min}s_{\gamma p}}$ and the
soft
process contribution to $A_{LL}$ in antiquark-proton scattering is  
suppressed too. Figure~9b illustrates this possibility to decrease the
soft process contribution to the asymmetry. We see that increasing
$k_{min}$ from 0.2 GeV/c to 1 GeV/c we decrease 
$A_{LL}$ by more than an order of  magnitude at 
$s_{\gamma p} \sim 10^2$ GeV$^2$ where the longitudinal spin-spin
asymmetry has the largest values. At higher energies the suppression
of the reggeon exchange contributions to $A_{LL}$ is significant too when
we apply the cut $k_T \geq k_{min} \geq 1$ GeV/c. The last
kinematical condition means that we consider experimental events with two
jets having the difference between their transverse momenta greater than
$2 k_{min}$. If we can reliably measure longitudinal hadron momenta we can
exclude events with a very low value of $z(1-z)$ (when $z s_{\gamma p}\sim
s_0$ or $(1-z) s_{\gamma p} \sim s_0$ ). This
way suppresses essentially the contributions to $A_{LL}$ of secondary
reggeon trajectories too.  We conclude from the discussion of Figs.~9a and
9b that decreasing $M_X$ or increasing $k_{min}$ we increase $z_{min}$
(and decrease $z_{max}=1-z_{min}$ too). As a result we suppress the
contributions of $\rho,\;\omega,\;f,\;A_2$ to $A_{LL}$.

The results of the calculations of $A_{LL}$ at  $\Delta _T>0$ are
presented in Fig.~10. We see
from the comparison of the curves
 in Fig.~10a that the pure pomeron contribution to
$A_{LL}$ increases significantly with
an increase of $\Delta _T$ and at $\Delta _T=1.5$
GeV/c takes on a value $\sim 10^{-6}$ which is much greater than at
$\Delta _T=0$.
It is easy to understand the increase of the pomeron exchange
contributions to $A_{LL}$ for the spectator graphs. We see from
formul\ae$\;$
(18), (20), (21), (22) that $A_5=0$ even at $\Delta _T \neq 0$ ($A_5$ is
defined in (17)). Hence the longitudinal spin-spin correlations
$(\vec{\sigma} _1 \cdot\vec{l}) (\vec{\sigma} _2 \cdot\vec{l})$ and
$(\vec{\sigma} _1 \cdot\vec{l}) (\vec{\sigma} _3 \cdot\vec{l})$ in the
product
$F^+_q(\vec{\Delta} _T) F_q(\vec{\Delta} _T)$ in (65) which give $A_{LL}$
can arise in the product of the amplitudes $A_3$ and $A_4$ as
\begin{center}
\mbox{\epsfig{file=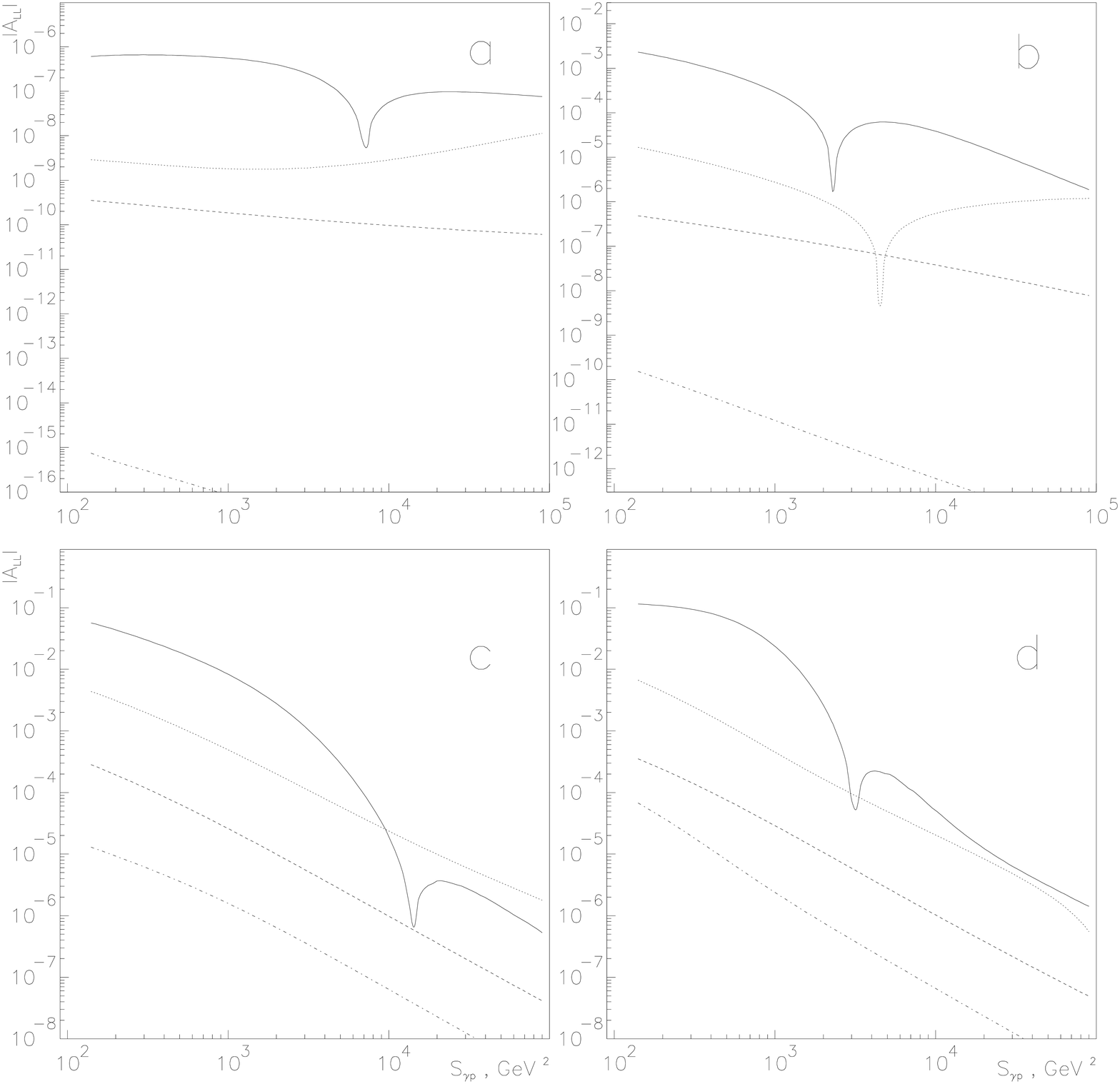,height=19.cm,width=17.cm}}
\noindent
\small
\begin{minipage}{15.cm}
\vspace{7mm}
{\sf Fig.~10:}
Dependence of longitudinal spin-spin asymmetry on $\Delta _T $.
Figs.~10a and 10b show pure pomeron (zeroth order in 
$\epsilon=\sqrt{s_0/s_{\gamma p}}$)
contribution to $|A_{LL}|$ and interference of contributions
of secondary reggeons with pomeron 
($\sim \epsilon ^1$), respectively.
Figs.~10c and 10d show
second order contribution to $|A_{LL}|$ and "total
asymmetry" (no decomposition into a power series in $\epsilon$).   
Dash-dotted, dashed, dotted  and solid curves are calculated for
$\Delta _T =0,\;0.5,\;1.0,\;1.5$ GeV/c, respectively.
All curves are smoothed near points where $A_{LL}=0$. 
For all curves
$M_X=10$ GeV/c$^2$, $Q^2=$ 10 (GeV/c)$^2$,
$k_{min}= 0.2$ GeV/c.
\end{minipage}
\end{center}
\begin{equation}
\nonumber
(\vec{\sigma} _1 \cdot\vec{m}) (\vec{\sigma} _j \cdot\vec{m})
(\vec{\sigma} _1 \cdot\vec{n}) (\vec{\sigma} _j \cdot\vec{n})=-
(\vec{\sigma} _1 \cdot\vec{l}) (\vec{\sigma} _j \cdot\vec{l})
\end{equation}
where $j=2$ in (70) for the quark-proton collision and $j=3$ for 
antiquark-proton scattering. But $A_3$ and $A_4$ at $\Delta _T=0$ are
equal to
zero for the one and two pomeron exchange amplitudes and the three
pomeron exchange amplitudes $A_3$ and $A_4$ are very small compared with
$A_1$ (they contain the small factor $p_y^6/(243 \lambda _P^5)$ as
one can see from (22)). 
The amplitudes $A_3^{(n)}$ of the one ($n=1$) and two
($n=2$) pomeron exchanges are proportional to $\Delta _T^2$  and increase 
when $\Delta _T$ increases from zero to some value. It is the increase of
$A_3$ which causes the increase of the pure pomeron contribution to
$A_{LL}$ shown in  Fig.~10a. 
We see also from a comparison of the
curves presented in Fig.~10b that the first order contribution to    
$A_{LL}$ at $\Delta _T>0.5$ GeV/c is much greater than at $\Delta _T=0$.
It is proportional to the sum of products of the amplitudes of  pure
pomeron exchanges ($\sim \epsilon ^0$) by the amplitudes 
($\sim \epsilon ^1$) of exchanges with
one secondary reggeon and one or two pomerons. The
first order contribution to $A_{LL}$ increases with $\Delta _T$ when the
absolute value of the pure pomeron exchange amplitudes increase. 
A comparison of the  curves calculated at different $\Delta _T$ in
Figs.~10c and 10d  
shows an increase of the second order contributions to $A_{LL}$ and 
"the total asymmetry" with $\Delta _T$.

Up to now we have considered Regge trajectories containing resonances with
 natural parity $\pi= \sigma =(-1)^J$ where $\pi$ and $J$ denote  parity
and 
total spin of a resonance. It follows from the parity conservation and the
$T$-invariance of the strong interaction that the dependence on the spin
variables of the $qqB$- and $NNB$-vertices for a reggeon $B$ with 
natural parity has the form given by (3) and (5) (see, for instance, 
\cite{IW}, \cite{KAIDALOV}, \cite{GK}). For the Regge
trajectories under discussion the one reggeon exchange amplitudes $A_4$
and $A_5$ of $qN$- and $\bar{q}N$-scattering in (17) are equal to zero 
hence the longitudinal spin-spin correlation in the cross section
$(\vec{\sigma _1}\cdot\vec{l})(\vec{\sigma _j}\cdot\vec{l})$ is absent
($j=2$ (or $j=3$) for scattering of a quark (or an antiquark)  
on the proton). The nonzero
amplitudes $A_4$ and $A_5$ can appear for  many reggeon exchanges only
and their contribution to $A_{LL}$ becomes nonzero.  For unnatural parity
trajectories (with $\sigma \pi=-1$) there are two alternatives. The $NNB$-
and $qqB$-vertices look like either 
\begin{eqnarray} B=\Delta _T
B_x(\Delta _T^2)(\vec{\sigma} _1\cdot\vec{m})\;,\;\;\;\;  
b=\Delta _T b_x(\vec{\sigma} _j\cdot\vec{m})  
\end{eqnarray} 
or 
\begin{eqnarray}
B=B_z(\Delta _T^2)(\vec{\sigma} _1\cdot\vec{l})\;,\;\;\;\;
b=b_z(\vec{\sigma} _j\cdot\vec{l})\;
\end{eqnarray}
which is also a consequence of the parity conservation and the time
reverse invariance \cite{IW}, \cite{KAIDALOV}.
It follows from the $G$-parity conservation (see relations for the
helicity amplitudes in \cite{KAIDALOV}, \cite{GK})
that formula (71) is applicable for  $\pi$-reggeon exchange
($T^G=1^-,\;J^{\pi}=0^-$) and for exchange with the $A_1(1260)$-meson
($T^G=1^-,\;J^{\pi}=1^+$) the vertices are described by relation (72). For
the pion trajectory we have $\alpha _{\pi}(0)\approx 0$. The intercept of
the $A_1$-trajectory is not well established but the region for it is the
following:  $-0.25 \leq \alpha _{A_1}(0) \leq 0$ \cite{IW},
\cite{AAGABK}. 
For
the $\pi$ and $A_1$ Regge trajectories the one reggeon exchange amplitudes
are at least quantities $\sim \epsilon ^2 \sim s_0/s$ 
where $s$ is the square of the center-of-mass energy
of colliding particles. Since the $\pi$- and $A_1$-reggeon exchange
amplitudes decrease with the collision energy more rapidly 
($\sim s^{-n}$, $n\geq 1$) than the amplitudes of 
$f$, $\rho$, $\omega$, $A_2$ ($\sim 1/\sqrt{s}$) and  pomeron exchanges,
hence their contributions to $A_{LL}$ are suppressed compared with the
natural parity reggeon contributions at $s \rightarrow \infty$. But they
contribute to the longitudinal spin-spin 
asymmetry even  in the one reggeon exchange approximation. 
Indeed, the pomeron-pion interference term in the cross section gives the
longitudinal spin-spin correlation $(\vec{\sigma}_1\cdot\vec{l})
(\vec{\sigma} _j\cdot\vec{l})$ at nonzero $\Delta _T$ 
due to (70). Recall that the one pomeron exchange amplitude 
contains the term $A_3 (\vec{\sigma}_1\cdot\vec{n})
(\vec{\sigma}_j\cdot\vec{n})$ (see (20)) and the one $\pi$-reggeon 
exchange amplitude  has the
term $A_4(\vec{\sigma} _1\cdot\vec{m})(\vec{\sigma} _j\cdot\vec{m})$
according to (71).  The correlation term $(\vec{\sigma} _1\cdot\vec{l})
(\vec{\sigma}_j\cdot\vec{l})$ is equal to zero if the spin-dependent
vertex of the pomeron
$p_y$ or $P_y$ in (3), (5) is zero. 
On the other hand for exchange with $A_1(1260)$
the amplitude $A_5$ is nonzero as this follows from (72). Hence the
longitudinal
spin-spin term $(\vec{\sigma} _1\cdot\vec{l})
(\vec{\sigma} _j\cdot\vec{l})$ exists and produces nonzero $A_{LL}$
even if the spin-dependent part of the pomeron vertex is equal to zero.

For the numerical calculations we have used the expression \cite{BST} 
\begin{eqnarray}
\nonumber
A_4^{(\pi)}= \frac{G^2_{\pi NN}}{16 \pi m_N E_0}
\frac{0.6t}{(t-m_{\pi}^2)}
\frac{s}{s-m_N^2-m_q^2}
\Big (\frac{s}{s_0}\Big )^{\alpha _{\pi}(m_{\pi}^2)-1}\\
\exp \Bigl\{-\Big[\frac{r^2_{\pi}}{2}+\alpha ^{'}_{\pi}(m_{\pi}^2)
\Big (\ln \Big(\frac{s}{s_0}\Big)-i\frac{\pi}{2}\Big)\Big]
(m_{\pi}^2-t) \Bigr\}
(\vec{\tau} _1 \cdot\vec{\tau} _j)
\end{eqnarray}
for the one $\pi$-reggeon exchange amplitude where 
$G^2_{\pi NN}/(4 \pi)=14.6$, $G_{\pi NN}$ is the pion-nucleon
constant, $t \approx -\Delta _T^2$, $r^2_{\pi}=3$ (GeV/c)$^{-2}$,
$\alpha ^{'}_{\pi}(m^2_{\pi})=1$ (GeV/c)$^{-2}$ 
\cite{BST},  $m_{\pi}$ and $m_N$ denote the 
pion and nucleon masses, respectively. The additional factor 0.6 in (73)
compared with the $NN$-scattering amplitude is due to the relation
between the $qqB$- and $NNB$-vertices for $B=\pi$ which looks like 
$b_x=0.6B_x$. This relation is in agreement with the
second relation (37) since $\pi$ has  isospin $T=1$.
We have written $r^2_{\pi}/2$ in (73) instead of $r^2_{\pi}$ as we
consider
the $qq \pi$-vertex as the vertex of
reggeon emission by a point-like quark.
As usual $j=2$ in (73) corresponds to  the quark-proton collision and
$j=3$
does to  antiquark-proton scattering. 
We parameterize the $A_1(1260)$-reggeon exchange amplitude as in \cite{IW}
\begin{eqnarray}
\nonumber
A_5^{(A_1)}=-\frac{1}{16 \pi s}A_{1z}(\vec{\sigma} _1\cdot\vec{l})a_{1z}
(\vec{\sigma} _j\cdot\vec{l})[ 1-\exp \{-i\pi \alpha _{A_1}(t) \}] \\
\Gamma(1-\alpha _{A_1}(t))(\alpha ^{'}_{A_1}(0) s)^{\alpha _{A_1}(t)}
(\vec{\tau} _1\cdot\vec{\tau} _j)\;.
\end{eqnarray}
In (74) $\Gamma(x)$ denotes the Euler gamma
function. 
The vertex parameter $A_{1z}$ in (74) is put equal to $6.2/\sqrt{2}$ in
accordance with \cite{IW}, $a_{1z}=0.6A_{1z}$ in agreement with (37). 
The results of our estimate  of the $A_1(1260)$-exchange 
contribution to the longitudinal spin-spin asymmetry  are shown in
Fig.~11a for the spectator graphs only. The curves show a behaviour of
the ratios of $A_{LL}$ calculated with and without contributions of 
one $A_1$-reggeon exchange,  exchanges with $P,\;\rho,\;f,\;A_2$
and $\omega$ reggeons being taken into  
account for all the
curves in Fig.~11a. We have calculated ratios under
discussion for $\alpha _{A_1}(0)=0$ and $\alpha _{A_1}(0)=-0.2$ finding
the slope of the linear $A_1$-trajectory $\alpha ^{'}_{A_1}(0)$ from
the relation
\begin{eqnarray}
\nonumber
1=\alpha _{A_1}(0)+\alpha ^{'}_{A_1}(0) m^2_{A_1}
\end{eqnarray}
with the mass of the $A_1$-meson $m_{A_1}$ equal to $1.23$ GeV/c$^2$. We
see from Fig.~11a that the relative contribution of $A_1$ at $100
<s_{\gamma p} <9\cdot10^4$ GeV$^2$ is less than 8$\%$.
Though the 
\vspace{0.5cm}
\begin{center}
\begin{tabular}{cc}
\mbox{\epsfig{file=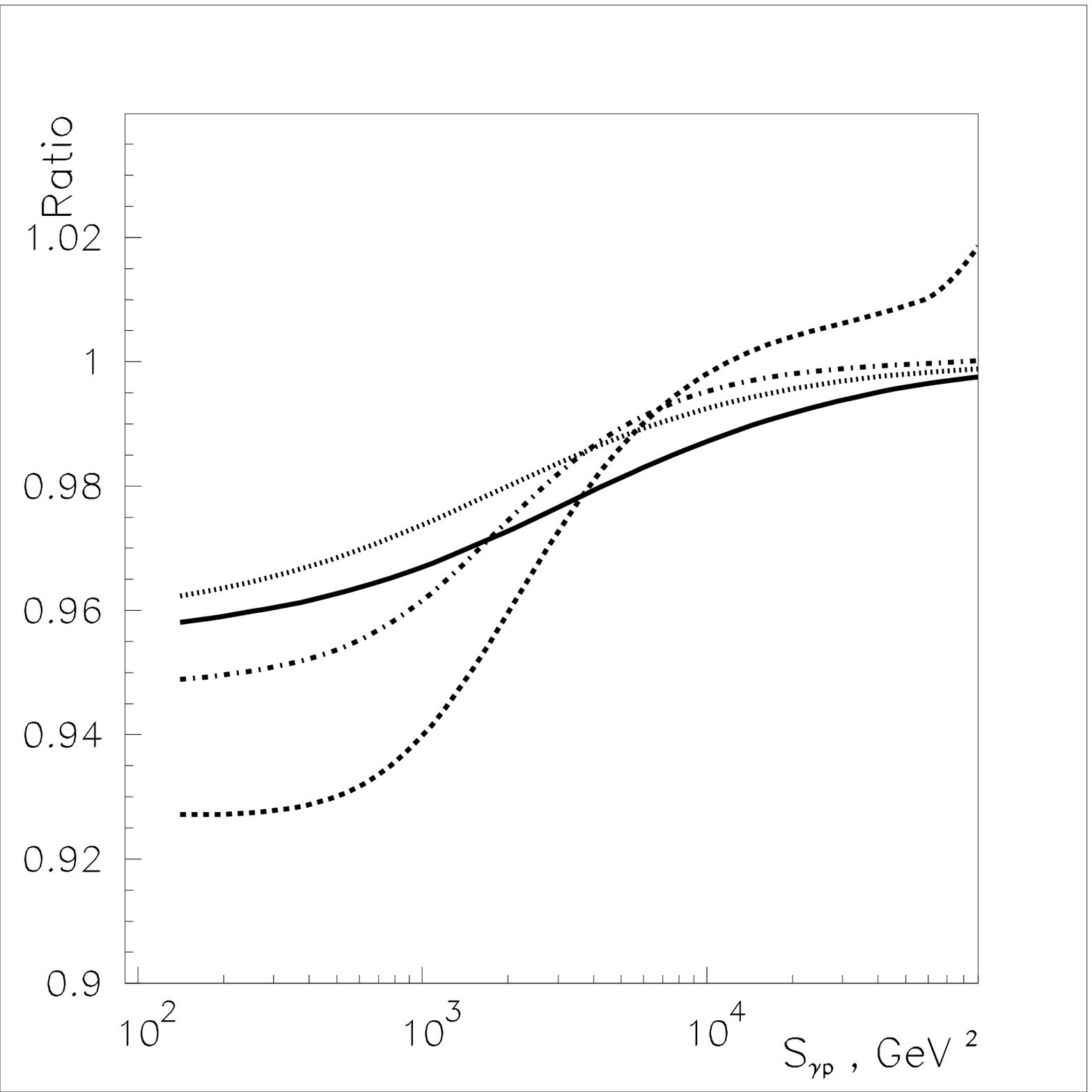,height=7.5cm,width=7.5cm}}
\vspace{2mm}
\noindent 
\small
&
\mbox{\epsfig{file=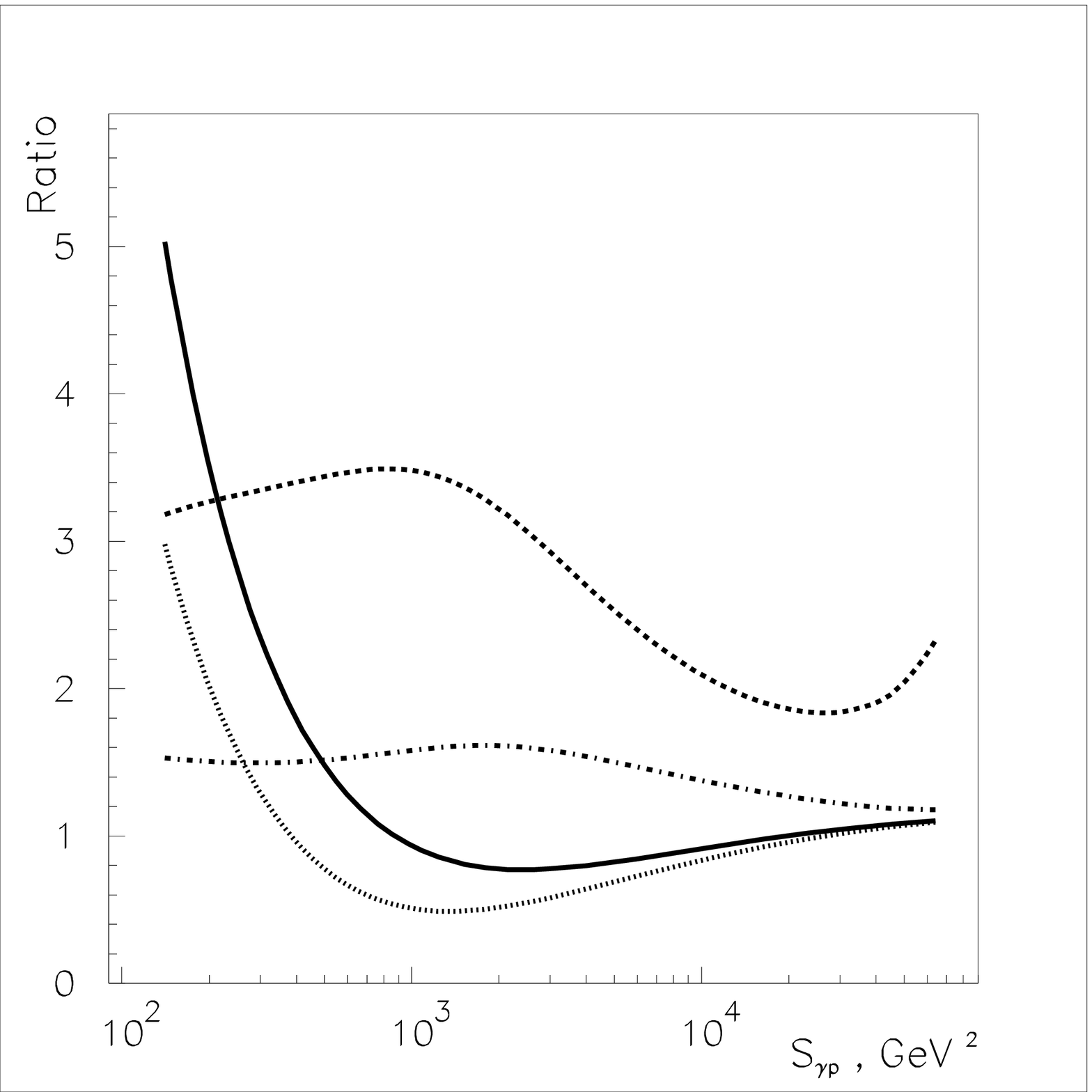,height=7.5cm,width=7.5cm}}
\vspace{2mm}
\noindent
\small
\\
\begin{minipage}{7cm}
{\sf Fig.~11a:}
Relative contribution of
\linebreak
$A_1(1260)$-reggeon to $A_{LL}$.
All curves show ratios of $A_{LL}$ calculated with and without
one
$A_1$-reggeon exchange contribution.
Solid and dashed
curves are
calculated for $\alpha _{A_1}(0)=0$ at $\Delta _T=0.5$ GeV/c and 1 GeV/c,
respectively. Dotted and dash-dotted curves are
obtained for $\alpha _{A_1}(0)=-0.2$ at $\Delta _T=0.5$ GeV/c and
1 GeV/c, respectively. Spectator graphs have been considered  
 only. For all curves contributions of
$P,\;f,\;\rho,\;\omega,\;A_2$ are taken into account for
$Q^2=10$ (GeV/c)$^2$, $M_X=10$ GeV/c$^2$, $k_{min}=0.2$ GeV/c.
\end{minipage}
 \normalsize
&     
\begin{minipage}{7.5cm}
{\sf Fig.~11b:} 
Relative contribution of $\pi$-reggeon to longitudinal spin-spin
asymmetry.
Curves show ratios of $A_{LL}$ calculated with and without
$\pi$-reggeon exchange contribution, spectator and nonspectator graphs
being taken into account. Solid, dotted curves are
calculated at $\Delta _T=0.5$  GeV/c and dashed, dash-dotted curves
are obtained for $\Delta _T=1.0$ GeV/c. Computing solid and dashed curves
we take into account $\pi$ and $\pi+P$ exchange amplitudes. In
obtaining  dotted and dash-dotted curves exchanges with $\pi$, $\pi+P$
and $\pi+P+P$ are included into calculation.
Parameters $Q^2,\;M_X,\;k_{min}$ are the same as in Fig.~11a and
contributions of $P,\;f,\;\rho,\;\omega,\;A_2$ are taken into account.
 \end{minipage}
\normalsize
\end{tabular}
\end{center}   
\vspace{0.5cm}
amplitude of  $A_1$-exchange has the spin-spin term
$(\vec{\sigma} _1\cdot\vec{l})(\vec{\sigma} _j\cdot\vec{l})$
most suitable to produce $A_{LL}$ and
interferes with the spin-independent part of
the pomeron exchange amplitude (which is large) the contribution of the
$A_1$-trajectory to $A_{LL}$ is small due to two
factors: $-0.2 \leq \alpha _{A_1}(0) \leq
0$ and the negative signature $\sigma$ of $A_1$. Due to the unequality 
$\alpha _{A_1}(0) \leq 0$ the $A_1$ exchange amplitude contains the
factor $(s/s_0)^{\alpha _{A_1}(t)-1}\leq s_0/s$ which is small at large
energies. The factor 
$1+\sigma \exp \{-i\pi \alpha _{A_1}(t) \}$ in (74) suppresses the
contribution of the $A_1(1260)$-reggeon since $\sigma =-1$ and  
$\alpha _{A_1}(t)$ is small (this factor is especially small for
$\alpha _{A_1}(0)=0$). 
Pion exchange gives more appreciable contribution 
to $A_{LL}$ than $A_1$ as we see from Fig.~11b. 
For the pion having $\sigma =1$ the factor
$1+\sigma \exp \{-i\pi \alpha _{A_1}(t) \}$ is of the order of unity
besides the constant $G_{\pi NN}$ in (73) is large. This two facts explain
why the contribution of $\pi$ to $A_{LL}$ is greater than the
$A_1$-reggeon exchange contribution. 
Since  $\pi$-reggeon exchange gives the contribution to $A_{LL}$ 
of the same order of magnitude as
the $\rho$-, $f$-, $A_2$-, $\omega$-contributions for the
energies achieved at HERA we have to take into account in our numerical
calculations not only the pole term but the branch cats too. The intercept
$\alpha _{\pi}(0) \approx 0$ hence 
$(s/s_0)^{\alpha _{\pi}(0)-1} \sim s^{-1} \sim \epsilon ^2$
therefore we are
to include into consideration $\pi$, $\pi P$ and $\pi P P$ exchanges only.
This has been done in obtaining curves presented in Fig.~11b. 
They are calculated in two approximations. In the former approximation
we take into account the amplitudes of pion and pion $+$ pomeron
exchanges. In the latter approximation we add to these amplitudes the
amplitude of $\pi+P+P$ exchanges.
The applied formul\ae$\;$ are presented in the Appendix. Since
the contribution of the $A_1$ Regge trajectory is small we restrict our
consideration with the pole amplitude given by relations (17) and (74).

Let us analyse the results presented in Figs.~5, 8, 9, 10, 11 from the    
point of view of the main idea of the present paper. We remind that we try
to find kinematical conditions to suppress the reggeon contributions to
the longitudinal spin-spin asymmetry making them much smaller than
perturbative QCD contributions to $A_{LL}$. The latter contributions can
be more reliably predicted  theoretically and compared with experimental
data than the contributions of the soft processes considered in the
present paper.  We see from Fig.~10d that $|A_{LL}| < 10^{-5}$ 
at $s_{\gamma p} \geq 10^4$ GeV$^2$ and $\Delta _T < 1$ GeV/c.
It follows from Figs.~11a and 11b that even if we include in the
calculation the contributions of the $\pi$- and $A_1(1260)$-reggeons 
$|A_{LL}|$ will be smaller than $10^{-4}$. We see from Figs.~9a and 9b
that
we reduce a value of $A_{LL}$ decreasing $M_X$ or increasing the
minimal value of $k_T$ (the demand $k_T>k_{min}$ means that one selects
events with a difference between transverse momenta of quark jets greater
than $2 k_{min}$). In these cases we increase a minimal fraction of the
virtual photon
momentum $z$ carried by a quark/antiquark. This follows immediately from  
formula (69). The quantity of $z_{min}$ is the most important variable
which can be changed experimentally to suppress the reggeon
contributions to $A_{LL}$. Values of
$|A_{LL}| < 10^{-4}$ cannot  presumably be measured by modern
experimental technique since one needs much more than $10^8$ events to
have
statistical errors of the $\gamma p$ cross sections much less than  
$10^{-4}$. If $|A_{LL}|$ predicted within the framework
of perturbative QCD  is greater than $10^{-4}$ we do not
need any additional kinematical cuts at $s_{\gamma p}\geq 10^4$ GeV$^2$ 
and  $\Delta _T < 1$ GeV/c  to suppress
the reggeon contributions to the 
longitudinal spin-spin asymmetry. We have argued above
that our approach is applicable at $s_{\gamma p} \geq 10^3$ GeV$^2$. We   
see from the presented numerical results that $A_{LL}$ at
$s_{\gamma p} = 10^3$ GeV$^2$ can be $\sim 10^{-2}$.
For the region $10^3 \leq s_{\gamma p} \leq 10^4$ GeV$^2$ we have to take
into account the contributions to $A_{LL}$ not only of
$P,\;f,\;\rho,\;\omega,\;A_2$ Regge trajectories but $A_1$- and
$\pi$-reggeons as well. As we have told the most important parameter
influencing a
value of the longitudinal spin-spin asymmetry is $z_{min}$. Figure~12
shows the dependence of $A_{LL}$ on the lower limit ($z_{min}$) in the
integral over $z$ in (65)  for $s_{\gamma p}=10^3$ GeV$^2$ 
the upper limit in the integral
being put equal to $z_{max}=1-z_{min}$. 
The contributions both of $P,\;f,\;\rho,\;\omega,\;A_2$ and $\pi,\;A_1$
reggeons are taken into account in the calculations. We see that   
$|A_{LL}|$ is less than $10^{-4}$ for $\Delta _T =0.5$ GeV/c. The
absolute value of the longitudinal spin-spin asymmetry at
$\Delta _T = 1$ GeV/c becomes less than $10^{-4}$  for $z_{min}>0.025$ .
Remembering formula (69) we get the region  $k_T^2 \geq 0.025 M_X^2$ 
at $s_{\gamma p}\geq 10^3$ GeV$^2$  and  
$\Delta _T \leq 1$ GeV/c in which $|A_{LL}| \leq 10^{-4}$. 
In this kinematical regions we can reliably
compare the perturbative QCD predictions for $A_{LL}$ with experimental   
data to 
\begin{center}
\mbox{\epsfig{file=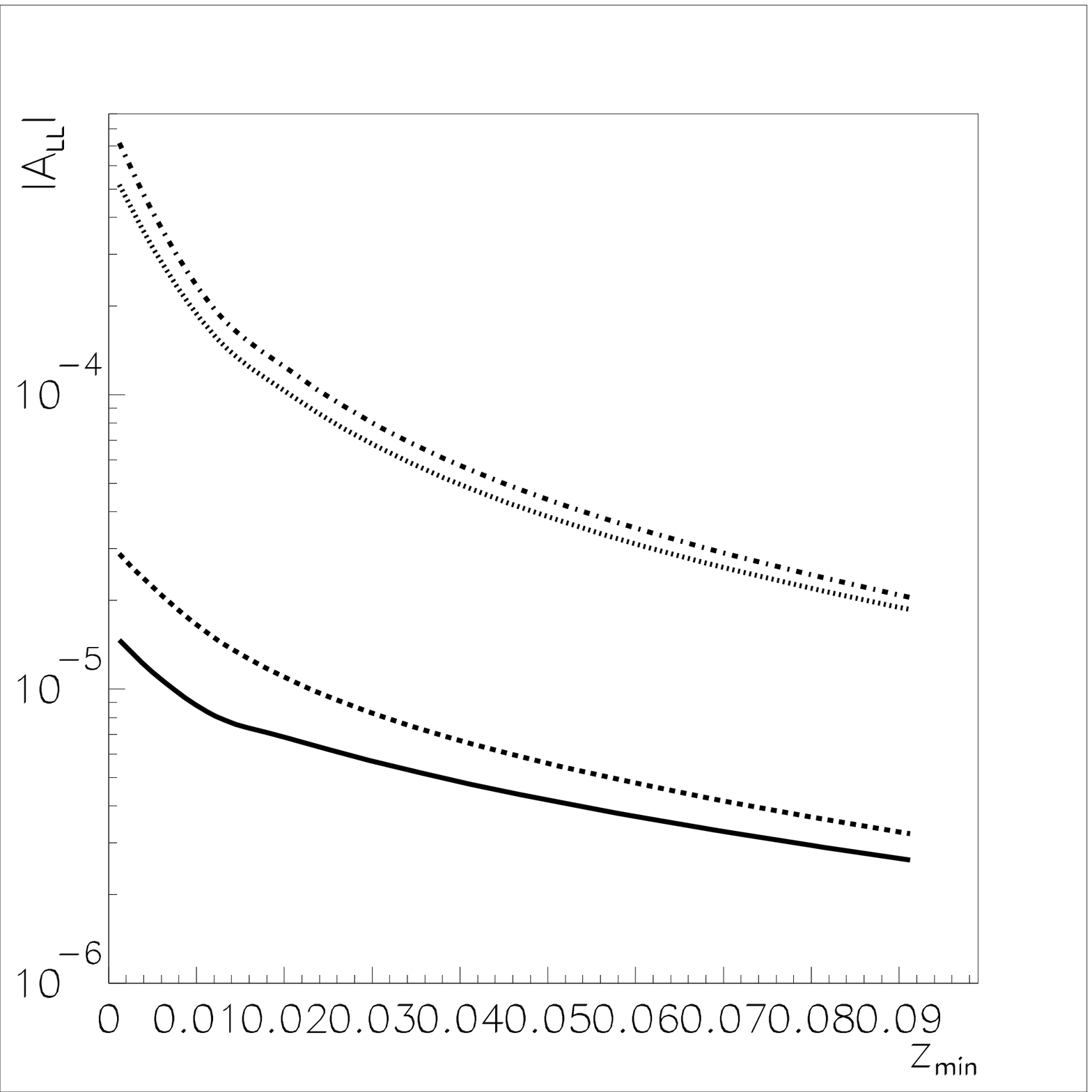,height=8.cm,width=9.cm}}
\noindent
\small
\begin{minipage}{14.2cm}
\vspace{7mm}
{\sf Fig.~12:}
Dependence of longitudinal spin-spin asymmetry on $z_{min}$.
Dashed and solid curves show contributions $\sim \epsilon ^2$ to
$A_{LL}$ and "total asymmetry"  
at $\Delta _T=0.5$ GeV/c.
Dotted and dash-dotted lines are the same as dashed and solid curves
but  for $\Delta _T=1.0$ GeV/c.
For all curves $s_{\gamma p}=10^3$ GeV$^2$, $M_X=10$ GeV/c$^2$, 
$Q^2=$ 10 (GeV/c)$^2$,  $k_{min}= 0.2$ GeV/c.
\end{minipage}
\end{center}
\vspace{0.5cm}
\normalsize
be obtained in the nearest future at the HERA collider in
scattering of the polarized electrons/positrons off the polarized protons.

\section{Conclusions}  

We have calculated the contributions of 
$P,\;\rho(770),\;f_2(1270),\;A_2(1320),\;\omega(782),\;\pi$ and
$A_1(1260)$  reggeon exchanges 
to the longitudinal spin-spin
asymmetry, $A_{LL}$ in the diffractive hadron production in hard 
$\gamma p$-scattering at energies accessible at HERA ($10^2 \leq 
s_{\gamma p} \leq 10^5$ GeV$^2$). Our numerical predictions are obtained
with
the aid of the phenomenological parameters found 
in \cite{IW}, \cite{BST}, \cite{BLSTM}  from the study of
hadron-hadron scattering within the framework of the Regge theory.
We restrict our consideration with
the amplitudes of exchanges with one, two, and three reggeons only and
decompose $A_{LL}$ into a power series in $\epsilon$ ($\epsilon =
\sqrt{s_0/s_{\gamma p}}$). We study terms $\sim \epsilon ^0,\; \epsilon
^1,\;\epsilon ^2$ and the total longitudinal spin-spin asymmetry (no
decomposition into power series in $\epsilon$) . It is shown that the dominant
 contributions to
$A_{LL}$ at $10^2 \leq s_{\gamma p}\leq10^5$ GeV$^2$ are the second order
contributions ($\sim \epsilon ^2$) in spite of a validity of the
inequality $\epsilon ^0 \gg \epsilon ^1 \gg \epsilon ^2$. The pure pomeron
exchange contributions ($\sim \epsilon ^0$) to $A_{LL}$ are very small
($|A_{LL}|< 10^{-6}$ at $\Delta _T \leq 1.5$ GeV/c, $M_X \leq 10$
GeV/c$^2$)
at energies achieved at the collider HERA. One can neglect them with
respect to the other contributions. Exchanges with one secondary reggeon
and some number of the pomerons ($\sim \epsilon ^1$) are comparable with
the second order contributions to  $A_{LL}$ for momenta transferred to the
proton $\Delta _T \sim 1$ GeV/c and are negligible at $\Delta _T =0$.
The dominant contribution to the numerator in the formula for $A_{LL}$
comes from those $z$ (the heavy photon momentum fraction carried by a
quark) for which $z$ (or $1-z$) is close to its lower limit $z_{min}$.
In this $z$-region, the center-of-mass energy of quark-proton
(antiquark-proton)     
scattering can be rather low ($\sim 1$ GeV) and 
$\pi$-reggeon exchange becomes very important though the intercept of the
pion Regge trajectory ($\alpha _{\pi}(0) \approx 0$) is smaller than the
intersepts of $P,\;\rho,\;f_2,\;A_2,\;\omega$ trajectories. The
$A_1$-reggeon has the suitable spin structure of 
the $qqA_1$-vertex to produce the longitudinal spin-spin 
asymmetry even in the one reggeon exchange approximation when the 
spin-dependent pomeron vertex is equal to zero. Nevertheless the relative 
contribution of the $A_1(1260)$-trajectory to $A_{LL}$ is less than 10$\%$
since $A_1$ has the negative signature and the small coupling constant.

The main purpose of the present paper is to find kinematical conditions
where $|A_{LL}|< 10^{-4}$ (the lowest limit for $A_{LL}$ which can 
presumably be
measured by modern experimental technique). It is shown that 
$|A_{LL}|< 10^{-4}$ at $s_{\gamma p}\geq 10^3$ GeV$^2$,  $\Delta _T \leq
0.5$ GeV/c, and the mass of produced hadrons, $M_X \leq 10$ GeV/c$^2$. 
The longitudinal spin-spin asymmetry at $Q^2 \geq4$
(GeV/c)$^2$ is practically insensitive to $Q^2$ hence we do not need any
additional cuts for $Q^2$. For $\Delta _T \leq 1$ GeV/c  (other conditions
are as before) we need some cuts to reduce $A_{LL}$. For this aim
we have to make $z_{min}$ higher (to increase 
the quark-proton and antiquark-proton collision energy)  than 0.03.
Selecting experimental events with $k_T^2 \geq 0.03 M_X^2$ we make
$|A_{LL}|$ smaller than $10^{-4}$. If the perturbative QCD contribution to
$A_{LL}$ is greater than $10^{-4}$, then making the cuts to suppress the
soft Regge process contributions one can reliably compare the perturbative
QCD predictions for $A_{LL}$ with the experimental data which can be
obtained in the
future at HERA in hard scattering of the polarized electrons/positrons 
off the polarized protons.

\section*{Acknowledgement}
I am grateful to Prof. J.~Bartels, Prof. N.~N.~Nikolaev, Dr. T.~Gehrmann
for useful discussions and especially to
Prof. M.~G.~Ryskin for his constant interest in the present work. 
I am much obliged to Prof. V.~B.~von~Schlippe whose help  improved
the present paper significantly.
I would like to gratefully acknowledge the hospitality of
the DESY Theory Group and the financial support of the Volkswagen
Stiftung.

\section*{Appendix}
\subsection*{Spectator graphs}
In the Appendix we present the general formul\ae$\;$ for the contributions
of one,
two and three
reggeon exchanges both for the spectator and non-spectator graphs. 
For the spectator graphs the amplitude of the $q \bar{q}$-pair
production and its rescattering on the proton is given by the general 
relation (63).
If  isospin of a reggeon $b$ is equal to zero, then invariant
amplitudes $A_j^{(1)}$ corresponding to  exchange with the reggeon $b$ 
can be described by relations (18), (19), (20)
where we are to substitute $b_s,\;b_y,\;B_s,\;B_y$ instead of
$p_s,\;p_y,\;P_s,\;P_y$. When isospin of the reggeon $b$ is equal to
1 we have the relations
 
\begin{eqnarray}
\nonumber
a_1^{(1)}=b_sB_s(\vec{\tau} _1 \cdot\vec{\tau} _j)\;,\\
\nonumber
a_2^{(1)}=i\Delta _T b_y B_s (\vec{\tau} _1 \cdot\vec{\tau} _j)\;,\\
\nonumber
a_6^{(1)}=i\Delta _T B_y b_s (\vec{\tau} _1 \cdot\vec{\tau} _j)\;,\\
a_3^{(1)}=-\Delta ^2 _T b_y B_y (\vec{\tau} _1 \cdot\vec{\tau} _j)\;,
\end{eqnarray}
instead of (20) where $j=2$ in (75) for $qN$-scattering and $j=3$ for the
$\bar{q}N$-collision. The total one reggeon exchange amplitude is a sum
over all reggeons $b$ which can contribute to the process under
discussion.

For  exchange with two reggeons, say, $b$ and $h$, both having 
isospin $T=0$ the invariant amplitudes are given by the formul\ae$\;$
\begin{eqnarray}
\nonumber
A_l^{(2)}= \sum_{b,h}\frac{f_I(b,h)}{2!}C_{sh}^{(2)} \eta _b(0) 
\eta _h(0) (s/s_0)^{\alpha _b(0)-\alpha _h(0)-2} 
\exp \{ - \lambda \Delta _T ^2 \} a_l^{(2)}(b,h) \;,\\
\nonumber
a_1^{(2)}(b,h)=\frac{i}{\lambda _1} [ b_s h_s B_s H_s-(b_yh_yB_sH_s+
B_yH_yb_sh_s)(\lambda \Delta _T^2-1)/ \lambda _1  \\
\nonumber
+b_yh_yB_yH_y(\lambda ^2 \Delta _T ^4 + \lambda _1 \Delta _T ^2/2 -4
\lambda \Delta _T^2 +2)/\lambda _1^2 ]\;, \\
\nonumber
a_2^{(2)}(b,h)=- \frac{\Delta _T}{\lambda _1^2} \{\lambda _h b_y h_s
+\lambda _b h_y b_s)[B_s H_s-B_y H_y(\lambda \Delta _T^2 -1)/ \lambda _1
]\\
\nonumber
-\frac{\lambda _2}{2 \lambda _1} B_y H_y (b_y h_s -h_y b_s) \}\;,\\
\nonumber
a_6^{(2)}(b,h)=- \frac{\Delta _T}{\lambda _1^2} \{\lambda _h B_y H_s
+\lambda _b H_y B_s)[b_s h_s-b_y h_y(\lambda \Delta _T^2 -1)/ \lambda _1
]\\
\nonumber
-\frac{\lambda _2}{2 \lambda _1} b_y h_y (B_y H_s -H_y B_s) \}\;,\\
\nonumber
a_3^{(2)}(b,h)=- \frac{i}{\lambda _1^2} \bigl
[\frac{1}{2}(b_yh_s-h_yb_s)(B_yH_s-H_yB_s)+\frac{\Delta _T^2}{\lambda _1}
(\lambda _h b_yh_s+ \lambda _b h_y b_s)\\
\nonumber
(\lambda _h B_y H_s+\lambda _b H_y B_s) \bigr ]\;, \\
a_4^{(2)}(b,h)=- \frac{i}{2\lambda _1^2}(b_yh_s-h_yb_s)(B_yH_s-H_yB_s)\;.
\end{eqnarray}
In (76) we have presented the nonzero amplitudes $a_j^{(2)}(b,h)$ only
and applied the short notations
\begin{eqnarray}
\nonumber
\lambda =\lambda _b \lambda _h /(\lambda _b +\lambda _h)\;,\\
\nonumber
\lambda _1 =\lambda _b +\lambda _h\;,\\
\lambda _2 =\lambda _b -\lambda _h\;,
\end{eqnarray}
where $\lambda _a$ for any reggeon $a$ has been defined by (19). The
$qqa$-vertex is given by the relation
\begin{equation}
a(\Delta )=a_s+ia_y(\vec{\sigma}_2\cdot\vec{l}\times\vec{\Delta}_T)\;,
\end{equation}
and the $NNa$-vertex is 
\begin{equation}
A(\Delta)=A_s(\Delta _T)+
iA_y (\Delta  _T) (\vec{\sigma}_1\cdot\vec{l}\times\vec{\Delta}_T)\;.
\end{equation}

The isospin factor $f_I(b,h)=1$ in (76) when the  reggeons $b$ and
$h$ have isospins $T=0$. Formul\ae$\;$ (76) can be applied for
nonzero
isospins. When isospin of one reggeon only, say $b$, is equal
to 1, then the isospin factor is
\begin{equation}
f_I(b,h)=(\vec{\tau} _1 \cdot\vec{\tau} _j)
\end{equation}
with $j=2$ for $qN$-scattering and $j=3$ for $\bar{q}N$-scattering. As we
do not consider  nucleon charge exchange, then the isospin factor is
equal to
\begin{equation}
f_I(b,h)=2 m_j  
\end{equation}
where $m_j= \pm \frac{1}{2}$ denotes the third component of 
quark ($j=2$) isospin (or antiquark ($j=3$) isospin if we consider 
antiquark-proton scattering). When isospins of the  reggeons
$b,\;h$ are equal to 1, then the factor $f_I(b,h)$ in (76) is
\begin{equation}
f_I(b,h)=3\;.     
\end{equation} 
For this case the amplitude $a_5^{(2)}(b,h)$ becomes nonzero
\begin{equation}
a_5^{(2)}(b,h)=\frac{i \Delta _T^2}{3 \lambda _1^2} b_y h_y B_y H_y
(\vec{\tau} _1 \cdot \vec{\tau} _j)\;.
\end{equation}
As charge exchange of the proton is not considered here we can replace
$(\vec{\tau} _1 \cdot\vec{\tau} _j)$ by $2 m_j$ and hence the final
formula for 
$a_5^{(2)}(b,h)$ becomes as follows:
\begin{equation}
a_5^{(2)}(b,h)=\frac{2i m_j \Delta _T^2}{3 \lambda _1 ^2} b_yh_yB_yH_y\;.
\end{equation}

It is interesting to point out that the amplitude $A_4^{(2)}$ which
contributes to the spin-spin asymmetry (even if we consider the spectator
graphs only) is equal to zero identically when $b=h$. This means that the 
two pomeron  exchange contribution to $A_{LL}$ vanishes. We see also
from (76) that  $a_4^{(2)}(b,h)$ is nonzero 
for exchanges with two different reggeons and is
equal to $a_3^{(2)}(b,h)$ at $\Delta _T=0$. Hence for this case the sum of 
the amplitudes $A_3$ and $A_4$ is proportional to 
$(\vec{\sigma} _{1T} \cdot \vec{\sigma} _{2T})$ 
in accordance with (17). It follows from formula (32)
for $A_{LL}$ that the amplitude $A_5$ can in principle contribute to the
numerator but for exchanges with two reggeon $A_5=0$ at $\Delta
_T =0$ what
one can see from (76) and (84). 

Let us consider three reggeon exchanges. In the present paper we restrict
ourselves with the case when not more than two of exchanged reggeons are
not the pomeron as we consider terms $\sim \epsilon ^0,\; \epsilon
,\;\epsilon ^2$ only. Hence one reggeon among three reggeons is the
pomeron.
For a
compact representation of the final formul\ae$\;$ let us introduce short
notations
\begin{eqnarray}
\nonumber
S=b_sh_sc_sB_sH_sC_s,\;\;\;Y=b_yh_yc_yB_yH_yC_y,\\
\nonumber
S_y=b_sh_sc_sB_yH_yC_y,\;\;\;Y_s=b_yh_yc_yB_sH_sC_s,\\
\nonumber
a=b_y/b_s+h_y/h_s+c_y/c_s,\;\;\;A=B_y/B_s+H_y/H_s+C_y/C_s\;,\\
\nonumber
c=b_s/b_y+h_s/h_y+c_s/c_y,\;\;\;C=B_s/B_y+H_s/H_y+C_s/C_y\;,\\
\nonumber
w=b_y/(b_s \lambda _b)+h_y/(h_s\lambda _h)
+c_y/(c_s\lambda _c)\;,\\
\nonumber
W=B_y/(B_s\lambda _b)+H_y/(H_s\lambda _h)
+C_y/(C_s \lambda _c)\;,\\
\nonumber
e=\lambda _b b_s/b_y+\lambda _h h_s/h_y
+\lambda _c c_s/c_y,\;\;\;E=\lambda _b B_s/B_y+\lambda
_h H_s/H_y+\lambda _c C_s/C_y\;, \\
\nonumber
U=b_yB_y/(b_sB_s \lambda _b)+h_yH_y/(h_sH_s \lambda _h)+
c_yC_y/(c_sC_s \lambda _c)\;, \\
\nonumber
V=\lambda _b b_yB_s/(b_sB_y)+\lambda _h h_yH_s/(h_sH_y)
+\lambda _c c_yC_s/(c_sC_y)\;, \\
\nonumber
v=\lambda _b B_yb_s/(B_sb_y)+\lambda _h H_y h_s/(H_s h_y)
+\lambda _c C_yc_s/(C_sc_y)\;, \\
\nonumber
T=\lambda _b b_s B_s/(b_y B_y)+\lambda _h h_s H_s/(h_y H_y)+
\lambda _c c_s C_s/(c_y C_y)\;, \\
\nonumber
R=b_s B_s/(b_y B_y)+h_s H_s/(h_y H_y)+c_s C_s/(c_y C_y)\;, \\ 
\nonumber
d=b_s/(b_y \lambda _b)+h_s/(h_y \lambda _h)+c_s/(c_y \lambda _c)\;, \\
\nonumber
D=B_s/(B_y \lambda _b)+H_s/(H_y \lambda _h)+C_s/(C_y \lambda _c)\;, \\
1/ \Lambda =1/ \lambda _b+1/ \lambda _h+1/ \lambda _c\;,
X=\lambda _b \lambda _h \lambda _c\;. 
\end{eqnarray}
The formula for the amplitudes of three reggeon exchanges reads 
\begin{eqnarray}
\nonumber
A_j^{(3)}=\frac{C_{sh}^{(3)}}{3!} \sum_{b,c,h} f_I(b,c,h) 
\eta _b(0) \eta _h(0) \eta _c(0)
(s/s_0)^{\alpha _b(0)+\alpha _h(0)+\alpha _c(0)-3} \\
\exp \{- \Lambda \Delta
_T^2 \}
a_j^{(3)}(b,h,c)\;.
\end{eqnarray}
When all three reggeons have isospin $T=0$, then the isospin 
factor $f_I(b,h,c)=1$ and the nonzero amplitudes $a_j^{(3)}(b,h,c)$ are
\begin{eqnarray}
\nonumber
a_1^{(3)}(b,h,c)=-\frac{\Lambda}{X} \bigl \{S+\frac{\Lambda}{X}(1- \Lambda
\Delta _T^2)(Y_se+S_yE) \\
\nonumber
+\frac{Y \Lambda}{2 X}[T \Delta _T^2+
(1- \Lambda \Delta _T^2)(2R-cC)
+\frac{2 \Lambda}{X}(2-4 \Lambda \Delta
_T^2+\Lambda ^2\Delta _T^4) e E] \bigr \}\;, \\
\nonumber
a_2^{(3)}(b,h,c)=-i \Delta _T\frac{\Lambda ^2}{X}
\bigl \{Sw+\frac{\Lambda}{X}(2-\Lambda \Delta _T^2)(Y_s+S_yEw)+
\frac{S_y}{2 X}(V-aE) \\
\nonumber
+\frac{Y}{3X}[-(5/4-\Lambda \Delta _T^2)C+\Lambda
(2-\Lambda \Delta _T^2)D+\frac{2\Lambda ^2}{X}
(6-6\Lambda \Delta _T^2+\Lambda ^2\Delta _T^4)E ] \bigl \}\;,\\
\nonumber
a_6^{(3)}(b,h,c)=-i \Delta _T \frac{\Lambda ^2}{X}
\bigl \{SW+\frac{\Lambda}{X}(2-\Lambda \Delta _T^2)(S_y+Y_s eW)+
\frac{Y_s}{2 X}(v-Ae) \\
\nonumber
+\frac{Y}{3X}[-(5/4-\Lambda \Delta _T^2)c+\Lambda
(2-\Lambda \Delta _T^2)d+\frac{2 \Lambda ^2}{X}(6-6 \Lambda \Delta _T^2+
\Lambda ^2 \Delta _T^4) e] \bigl \}\;,\\
\nonumber
a_3^{(3)}(b,h,c)=-\frac{\Lambda }{2X} \bigl [-SU+S \Lambda w W (1-2
\Lambda \Delta _T^2)+\frac{\Lambda }{3X}(-2+3\Lambda \Delta
_T^2)(Y_sA+S_ya) \\
\nonumber
+\frac{\Lambda ^2}{X}(2-7\Lambda \Delta _T^2+2
\Lambda  ^2\Delta _T^4)(Y_sW+S_yw) \bigr ]-\frac{Y \Lambda }{36X^2}
\bigl [-2-3 \Lambda \Delta _T^2-2 \Lambda ^2 \Delta _T^4 \\
\nonumber
\frac{2\Lambda ^2}{X}(\lambda _b+\lambda _h+\lambda _c)(-8+25 
\Lambda \Delta _T^2-7\Lambda ^2 \Delta _T^4)+
\Lambda ^2
(\lambda _b^{-2}+\lambda _h^{-2}+\lambda _c^{-2})\\
\nonumber
(2-7\Lambda \Delta _T^2+2\Lambda ^2\Delta _T^4)
+18 \frac{\Lambda ^3}{X}(6-30 \Lambda \Delta _T^2
+17 \Lambda ^2 \Delta _T^4-2 \Lambda ^3 \Delta _T^6) \bigr ]\;, \\
\nonumber
a_4^{(3)}(b,h,c)=-\frac{\Lambda }{2X} \bigl \{-SU+S \Lambda w W 
+\frac{\Lambda }{3X}(2-\Lambda \Delta _T^2) \\
\nonumber
[3\Lambda (Y_sW+S_y w)-Y_s
A-S_y a ] \\
\nonumber
+\frac{Y}{18X}[-2-\Lambda \Delta _T^2-\frac{2 \Lambda ^2}{X}
(\lambda _b+\lambda _h+\lambda _c)(8-7\Lambda \Delta _T^2+
\Lambda ^2\Delta _T^4) \\+
\Lambda ^2(\lambda _b^{-2}+\lambda _h^{-2}+\lambda _c^{-2})(2-\Lambda
\Delta _T^2)+18\frac{\Lambda ^3}{X}
(6-6\Lambda \Delta _T^2+\Lambda ^2\Delta _T^4) ] \bigr \}\;.
\end{eqnarray}

When only one reggeon has isospin $T=1$, then the isospin factor
$f_I(b,c,h)$ in (87) is to be put equal to expression (80) or (81). When
two reggeons (say $b$ and $h$) have $T=1$, then $f_I(b,c,h)$ is given by
(82) and the new
nonzero amplitude $a_5^{(3)}(b,h,c)$ appears. The formula for it reads
\begin{eqnarray}
a_5^{(3)}(b,h,c)=-\frac{Y \Lambda ^2}{27 X ^2} \bigl \{
T \Delta _T^2+\bigl [2 \bigl
(R+\frac{b_sH_s}{b_yH_y}+\frac{B_sh_s}{B_yh_y} \bigr)-cC \bigr ] 
(1-\Lambda \Delta _T^2)(\vec{\tau} _1 \cdot\vec{\tau} _j)\;.
\end{eqnarray}
It has been already pointed out that we can substitute $2 m_j$ instead of 
$(\vec{\tau} _1 \cdot\vec{\tau} _j)$ in (88).

\subsection*{Non-spectator graphs}

The contributions of the non-spectator graphs are given by (44) and (45)
the latter formula describing charge exchange in which
the $u\bar{u}$-pair 
is transformed into the $d\bar{d}$-pair and vice versa. We start
with discussion of the contributions given by (44). In the first two terms
in the brackets in (44) we are to rest  two and three reggeon exchanges
only not to go beyond  our approximation. As we do not consider in the
present paper  nucleon charge exchange we can make use of the
substitutions $(\vec{\tau} _1 \cdot\vec{\tau} _2) \rightarrow 2 m_q$ in
the amplitude of quark-nucleon elastic scattering $A_{m_q}$ and 
$(\vec{\tau}_1 \cdot\vec{\tau} _3) \rightarrow -2 m_q$ in the
antiquark-nucleon  scattering amplitude 
$B_{-m_q}$ since $m_{\bar{q}}=-m_q$. The third term in
the brackets in (44) is nonzero for three reggeon exchanges only. It can
be represented by the sum of two terms   
\begin{equation}
\hat{D}(\vec{\Delta} _1, \vec{\Delta} _2)=\hat{D}_{2+1}(\vec{\Delta} _1,
\vec{\Delta} _2)+
\hat{D}_{1+2}(\vec{\Delta} _1, \vec{\Delta} _2)\;.
\end{equation}
In (89) $\hat{D}_{2+1}(\vec{\Delta} _1, \vec{\Delta} _2)$ corresponds to
the graphs
with exchanges of two reggeons $b$ and $h$ between the quark and the
proton (the total momentum transferred to the nucleon is equal to $\Delta
_1$), the antiquark and the proton interacting through  exchange with
a reggeon $c$ having the momentum $\Delta _2$. The term 
$\hat{D}_{1+2}(\vec{\Delta} _1, \vec{\Delta} _2)$ 
describes exchanges with one reggeon $c$ having the momentum
$\Delta _1$ between the quark and the proton and with two reggeons $b,\;h$
between the antiquark and proton (with the total momentum of the $b+h$
system equal to $\Delta _2$). The formula for 
$\hat{D}_{2+1}(\vec{\Delta}_1, \vec{\Delta} _2)$ reads
\begin{eqnarray}
\nonumber
\hat{D}_{2+1}(\vec{\Delta}_1, \vec{\Delta} _2)=
D_1(\vec{\sigma} _1 \cdot\vec{l}\times \vec{\Delta} _2)+
D_2(\vec{\sigma} _1 \cdot\vec{l}\times \vec{\Delta} _1)+
D_3(\vec{\sigma} _1 \cdot\vec{l}\times \vec{\Delta} _2)
(\vec{\sigma} _2 \cdot\vec{l}\times\vec{\Delta} _1) \\+
\nonumber
D_4(\vec{\sigma} _1 \cdot\vec{l}\times \vec{\Delta} _1)  
(\vec{\sigma} _2 \cdot\vec{l}\times\vec{\Delta} _2)+
D_5(\vec{\sigma} _1 \cdot\vec{l}\times \vec{\Delta} _1)  
(\vec{\sigma} _2 \cdot\vec{l}\times\vec{\Delta} _1)+
D_6(\vec{\sigma} _{1T}\cdot \vec{\sigma} _{2T}) \\+
\nonumber
D_7(\vec{\sigma} _1 \cdot\vec{l})(\vec{\sigma} _2\cdot\vec{l})+
D_8(\vec{\sigma} _1 \cdot\vec{l}\times\vec{\Delta} _2)
(\vec{\sigma} _3 \cdot\vec{l}\times \vec{\Delta} _2)+
D_9(\vec{\sigma} _1 \cdot\vec{l}\times\vec{\Delta} _1)
(\vec{\sigma} _3 \cdot\vec{l}\times\vec{\Delta} _2) \\+
\nonumber
D_{10}(\vec{\sigma} _1 \cdot\vec{l}\times\vec{\Delta} _2)
(\vec{\sigma} _2 \cdot\vec{l}\times \vec{\Delta} _1)
(\vec{\sigma} _3 \cdot\vec{l}\times\vec{\Delta} _2)\\
\nonumber
+ D_{11}(\vec{\sigma} _1 \cdot\vec{l}\times\vec{\Delta} _1)
(\vec{\sigma} _2 \cdot\vec{l}\times\vec{\Delta} _2)   
(\vec{\sigma} _3 \cdot\vec{l}\times\vec{\Delta} _2) \\ +
\nonumber
D_{12}(\vec{\sigma} _1 \cdot\vec{l}\times\vec{\Delta} _1)
(\vec{\sigma} _2 \cdot\vec{l}\times\vec{\Delta} _1)
(\vec{\sigma} _3 \cdot\vec{l}\times\vec{\Delta} _2)+
D_{13}(\vec{\sigma} _{1T}\cdot \vec{\sigma} _{2T})
(\vec{\sigma} _3 \cdot\vec{l}\times\vec{\Delta} _2) \\ +
D_{14}(\vec{\sigma} _1\cdot\vec{l})(\vec{\sigma} _2\cdot\vec{l})
(\vec{\sigma} _3 \cdot\vec{l}\times\vec{\Delta} _2)
\end{eqnarray}
where $(\vec{\sigma} _{1T}\cdot\vec{\sigma} _{2T})$ has been defined after
relation (23). Every amplitude $D_j$ in (90) is a sum over
contributions
of the reggeons $b,\;c,\;h$
\begin{equation}
D_j=\sum_{b,h,c} f_I(b,h,c) D_j^{(b,h,c)}
(\vec{\Delta} _1,\vec{\Delta} _2)\;,
\end{equation}
the reggeon $c$ being exchanged between the proton and the antiquark,
the reggeons $b,\;h$ being emitted by the quark. When all exchanged
reggeons
have isospins equal to zero, then $D_j$ for $j \leq 7$ are given by
\begin{eqnarray}
\nonumber
D_1^{(b,h,c)}(\vec{\Delta} _1,\vec{\Delta} _2)
=id \bigl [ s_2(2\lambda \Delta _1^2-1)-y \Delta _1^2- 
\frac{y \lambda}{x}(2-7 \lambda \Delta _1^2+2 \lambda ^2 \Delta _1^4) 
\bigr ]\;, \\
\nonumber
D_2^{(b,h,c)}(\vec{\Delta} _1,\vec{\Delta} _2)=
i d (\vec{\Delta} _1\cdot\vec{\Delta} _2) \bigl \{-2s_2 \lambda +
y [1-\frac{2 \lambda ^2}{x}(3-\lambda \Delta _1^2) \bigr \}\;, \\
\nonumber 
D_3^{(b,h,c)}(\vec{\Delta} _1,\vec{\Delta} _2)=
-ds_2[O+O_y(2 \lambda \Delta _1^2-3)]\;, \\
\nonumber
 D_4^{(b,h,c)}(\vec{\Delta} _1,\vec{\Delta} _2)=-ds_2[O_y-O/2]\;, \\
\nonumber
D_5^{(b,h,c)}(\vec{\Delta} _1,\vec{\Delta} _2)=
2d s_2 \lambda (\vec{\Delta} _1\cdot\vec{\Delta} _2) O_y\;, \\
\nonumber
D_6^{(b,h,c)}(\vec{\Delta} _1,\vec{\Delta} _2)=
-d s_2(\vec{\Delta} _1\cdot\vec{\Delta} _2)[ O_y-O/2]\;, \\
D_7^{(b,h,c)}(\vec{\Delta} _1,\vec{\Delta} _2)=0\;
\end{eqnarray}
with the isospin factor in (91) $f_I(b,h,c)=1$ for this case. 
We have  applied the short notations in (92) 
\begin{eqnarray}
\nonumber
d=C_{sh}^{(3)}\frac{i \eta_b(0) \eta_h(0)\eta_c(0)}
{3(\lambda _b+\lambda _h)^2}
(zs/s_0)^{\alpha _b(0) +\alpha _h(0)-2}[(1-z)s/s_0]^{\alpha _c(0)-1} \\
\nonumber
B_y H_y C_y \tilde{c}_s \exp \{-\lambda \Delta _1^2
-\lambda _c \Delta _2^2 \}\;, \\
\nonumber
\lambda=\lambda _b \lambda _h /(\lambda _b + \lambda _h)\;,\\
\nonumber
s_2=b_sh_s\;, \\
\nonumber
y=b_yh_y\;, \\
\nonumber
x= \lambda _b  \lambda _h\;, \\
\nonumber
O=b_y/b_s+h_y/h_s\;, \\
O_y=\lambda [b_y/(b_s \lambda _b) +h_y/(h_s \lambda _h)]\;. 
\end{eqnarray}
For $8 \leq l \leq 14$ $D_l$ are given by the relation
\begin{equation}
D_{j+7}^{(b,h,c)}(\vec{\Delta} _1, \vec{\Delta} _2) =
iD_{j}^{(b,h,c)}(\vec{\Delta} _1, \vec{\Delta} _2)
\tilde{c}_y /\tilde{c}_s
\end{equation}
with $\bar{q}\bar{q}a$ vertices $\tilde{a}_s$, $\tilde{a}_y$ being equal
to ($a=b, \;h,\;c$)
\begin{equation}
\tilde{a}_s=(-1)^{T_a} \sigma _a a_s\;,\;\;\;\;\tilde{a}_y=(-1)^{T_a}
\sigma _a a_y\;.
\end{equation}
We would like to stress that the longitudinal components $D_7=D_{14}=0$ in
(90) when $T_b=T_h=T_c=0$. Formul\ae$\;$ for 
$\hat{D}_{1+2}(\vec{\Delta} _1, \vec{\Delta} _2)$
can be obtained from (90), (91), (92), (93), (94) through
transformations 
\begin{equation}
b_{s,y} \rightarrow \tilde{b} _{s,y}\;,
h_{s,y} \rightarrow \tilde{h} _{s,y}\;,
\tilde{c}_{s,y} \rightarrow c _{s,y}\;,
\Delta _1 \leftrightarrow \Delta _2\;,
\vec{\sigma} _2 \leftrightarrow \vec{\sigma} _3\;,
\vec{\tau} _2 \leftrightarrow \vec{\tau} _3
\end{equation}
(see definition of $\tilde{a} _s,\;\tilde{a} _y$ in (95)).

When only one reggeon has  nonzero isospin, then we have for the case of 
$\hat{D}_{2+1}(\vec{\Delta} _1, \vec{\Delta} _2)$ $f_I(b,h,c)=2 m_q$ in
(91) if $T_c=0$. For $T_c=1$
$f_I(b,h,c)=2 m_{\bar{q}}=-2 m_q$ hence for these two cases we can write 
$f_I(b,h,c)=2 m_q(-1)^{T_c}$. Vice versa   
$\hat{D}_{1+2}(\vec{\Delta} _1, \vec{\Delta} _2)$ contains the factor 
$(-2 m_q)$ when $T_c=0$ and $2m_q$ for $T_c=1$ hence 
$f_I(b,h,c)=-2 m_q(-1)^{T_c}$. Formul\ae$\;$ (92) for $D_j$ for all the
cases
discussed above remain valid. 
When the reggeon $c$ and one reggeon among
$b$ and $h$ have isospins equal to 1 we have $f_I(b,h,c)=-1$ both for 
$\hat{D}_{2+1}(\vec{\Delta} _1, \vec{\Delta} _2)$ and
$\hat{D}_{1+2}(\vec{\Delta} _1, \vec{\Delta} _2)$. When $T_b=T_h=1$ a
reggeon $c$ is the pomeron. In this case $f_I(b,h,c)=3$ but $D_7$ and
$D_{14}$ become nonzero. The formula for $D_7$ looks like
\begin{equation}
D_7=\frac{1}{3} y d (\vec{\tau} _1\cdot\vec{\tau} _j)
(\vec{\Delta}_1\cdot\vec{\Delta} _2)
(B_s/B_y+H_s/H_y)\;,
\end{equation}
where $j=2$ in (97) for 
$\hat{D}_{2+1}(\vec{\Delta} _1, \vec{\Delta} _2)$ and $j=3$ for
$\hat{D}_{1+2}(\vec{\Delta} _1, \vec{\Delta} _2)$. 
The quantity of $D_{14}$ for 
$\hat{D}_{2+1}(\vec{\Delta} _1, \vec{\Delta} _2)$
is given by (94). The expression for $\hat{D}_{1+2}(\vec{\Delta} _1,
\vec{\Delta} _2)$ can be obtained from the formula for 
$\hat{D}_{2+1}(\vec{\Delta} _1, \vec{\Delta} _2)$ through
transformations (96) as for the case when $T_b=T_h=T_c=0$.

The contribution of charge exchange is described by formula (45). In
the first and second terms in the brackets in (45) we are to rest the two
and three reggeon exchange contributions only. In the former case both 
reggeons emitted with the quark and the antiquark have isospins 
equal to 1. Their electric charges are of opposite signs to
conserve the electric charge of the proton. For  three reggeon exchanges 
we have additional pomeron emission either with the quark or with 
the antiquark. The third term in the brackets in (45) is absent for two
reggeon exchanges. For the three reggeon exchange contribution it can
be represented as the sum of two terms
\begin{equation}
\hat{E}(\vec{\Delta} _1, \vec{\Delta} _2)=
\hat{E}_{2+1}(\vec{\Delta} _1, \vec{\Delta} _2)+
\hat{E}_{1+2}(\vec{\Delta} _1, \vec{\Delta} _2)\;.
\end{equation}
The meaning of the terms $\hat{E}_{2+1}(\vec{\Delta} _1, \vec{\Delta} _2)$
and $\hat{E}_{1+2}(\vec{\Delta} _1, \vec{\Delta} _2)$ in (98) is
analogous
to the meaning of $\hat{D}_{2+1}(\vec{\Delta} _1, \vec{\Delta} _2)$ and
$\hat{D}_{1+2}(\vec{\Delta} _1, \vec{\Delta} _2)$ in (89). The amplitude 
$\hat{E}_{1+2}(\vec{\Delta} _1, \vec{\Delta} _2)$ can be obtained from
$\hat{E}_{2+1}(\vec{\Delta} _1, \vec{\Delta} _2)$ via transformations
(96). The isospin structure of 
$\hat{E}_{2+1}(\vec{\Delta} _1, \vec{\Delta} _2)$ is given by the relation
\begin{equation}
\hat{E}_{2+1}(\vec{\Delta} _1, \vec{\Delta} _2)=
\hat{\Theta}_{2+1}(\vec{\Delta} _1, \vec{\Delta} _2)
(\vec{\tau} _2\cdot \vec{\tau} _3)+
i \hat{\Phi}_{2+1}(\vec{\Delta} _1, \vec{\Delta} _2)
(\vec{\tau} _1\cdot\vec{\tau} _2\times\vec{\tau} _3)\;.
\end{equation}
The spin structure of $\hat{\Theta}_{2+1}(\vec{\Delta} _1,\vec{\Delta}_2)$
is the same as  $\hat{D}_{2+1}(\vec{\Delta} _1, \vec{\Delta} _2)$ 
which is given by (90). More over formul\ae$\;$ (90), (91), (92) after
replacements $\hat{\Theta}_{2+1}(\vec{\Delta} _1,\vec{\Delta}_2) 
\rightarrow \hat{D}_{2+1}(\vec{\Delta} _1,\vec{\Delta}_2)$,
$D_j \rightarrow  \Phi _j$,
$D_j^{(b,h,c)}(\vec{\Delta} _1, \vec{\Delta} _2)
\rightarrow \Phi _j ^{(b,h,c)}(\vec{\Delta} _1, \vec{\Delta} _2)$
remain true in which we are to put $f_I(b,h,c)=1$.  But we should rest 
now in the sum in (91) only such sets of $b$, $h$, $c$ when one reggeon 
among $b,\;h$ is the pomeron and other two reggeons in a set have 
isospins equal to 1. The spin structure of the second term in (99) 
$\hat{\Phi} _{2+1}(\vec{\Delta} _1, \vec{\Delta} _2)$ looks like
\begin{eqnarray}
\nonumber
\hat{\Phi} _{2+1}(\vec{\Delta} _1, \vec{\Delta} _2)=
\Phi _1 (\vec{\sigma} _1 \cdot\vec{l})+
\Phi _2 (\vec{\sigma} _1 \cdot\vec{l})
(\vec{\sigma} _2 \cdot\vec{l}\times\vec{\Delta} _1)+
\Phi _3 (\vec{\sigma} _1 \cdot\vec{l})
(\vec{\sigma} _2 \cdot\vec{\Delta} _2) \\+
\nonumber
\Phi _4 (\vec{\sigma} _1 \cdot \vec{l})
(\vec{\sigma} _2 \cdot\vec{\Delta} _1)+ 
\Phi _5 (\vec{\sigma} _1 \cdot \vec{l})
(\vec{\sigma} _3 \cdot\vec{l}\times\vec{\Delta} _2)+
\Phi _6 (\vec{\sigma} _1 \cdot\vec{l})
(\vec{\sigma} _2 \cdot\vec{l}\times\vec{\Delta} _1)
(\vec{\sigma} _3 \cdot\vec{l}\times\vec{\Delta} _2) \\
\nonumber
+\Phi _7 (\vec{\sigma} _1 \cdot\vec{l})
(\vec{\sigma} _2 \cdot\vec{\Delta} _2) 
(\vec{\sigma} _3 \cdot\vec{l}\times\vec{\Delta} _2)+
\Phi _8 (\vec{\sigma} _1 \cdot\vec{l})
(\vec{\sigma} _2 \cdot\vec{\Delta} _1) 
(\vec{\sigma} _3 \cdot\vec{l}\times\vec{\Delta} _2)\;,
\end{eqnarray}
where $\Phi_j$ for $j\leq 4$ are
\begin{eqnarray}
\nonumber
\Phi_j=\sum_{b,h,c} 
\Phi_j^{(b,h,c)} (\vec{\Delta} _1,\vec{\Delta} _2) \;,\\
\nonumber
\Phi _1^{(b,h,c)}(\vec{\Delta} _1,\vec{\Delta} _2)=
2i \kappa \frac{B_s}{B_y}
(\vec{l}\cdot\vec{\Delta} _1\times\vec{\Delta} _2) 
\{2 s_2 \lambda _b -y[1-2\frac{\lambda}{\lambda _p}
(2-\lambda \Delta _1^2)]\}\;, \\
\nonumber
\Phi _2^{(b,h,c)}(\vec{\Delta} _1,\vec{\Delta} _2)=
-4 \kappa \frac{y \lambda}{\lambda _p} \frac{B_s}{B_y}
(\vec{l}\cdot\vec{\Delta} _1\times\vec{\Delta} _2)
(\lambda _p p_s/p_y+\lambda _b b_s/b_y)\;, \\
\nonumber
\Phi _3^{(b,h,c)}(\vec{\Delta} _1,\vec{\Delta} _2)=
2 \kappa y \frac{B_s}{B_y} (p_s/p_y+b_s/b_y)\;, \\
\Phi _4^{(b,h,c)}(\vec{\Delta} _1,\vec{\Delta} _2)= 
\kappa y \frac{C_s}{C_y} (p_s/p_y+b_s/b_y)\;.
\end{eqnarray}
Formul\ae$\;$ (100) have been written for the case when the reggeon $h$ is
the pomeron and  reggeons $b$ and $c$ have isospins $T=1$. When $b=P$
and $T_h=T_c=1$ we are to replace in (100) $B_s,\;B_y,\;b_s,\;b_y$
with  $H_s,\;H_y,\;h_s,\;h_y$, respectively. For 
$\Phi _k ^{(b,h,c)} (\vec{\Delta} _1,\vec{\Delta} _2)$ with $k\geq 5$ we
are to use the relations
\begin{equation}
\Phi _{j+4}^{(b,h,c)} (\vec{\Delta} _1,\vec{\Delta} _2)=
i\Phi _j ^{(b,h,c)} (\vec{\Delta} _1,\vec{\Delta} _2) 
\tilde{c}_y /\tilde{c} _s
\end{equation}
where $j \leq 4$. The quantity $\kappa$ is given by the formula
\begin{eqnarray}
\nonumber
\kappa=i C_{sh}^{(3)} \frac{\eta _b(0) \eta _p(0)\eta _c(0)}
{6 x^2} 
(zs/s_0)^{\alpha _b(0) +\alpha _h(0)-2}[(1-z)s/s_0]^{\alpha _c(0)-1}
\lambda ^2 \tilde{c} _s \\
\nonumber\nonumber
B_y P_y C_y \exp \{-\lambda \Delta _1^2 - \lambda _c \Delta _2 ^2 \}
\end{eqnarray}
and the short notations $s_2,\;y,\;\lambda ,\; x$ have been defined in
(93).

\subsection*{Contributions of $\pi$-reggeon}

The amplitude of quark/antiquark scattering on the proton due to  one
$\pi$-reggeon exchange is given by (17) and (73). To get the amplitudes of
$\pi P$- and $\pi P P$-exchanges we have applied the representation 
\begin{eqnarray}
A_4^{(\pi)}(\vec{\sigma} _1\cdot\vec{m})(\vec{\sigma} _j\cdot\vec{m})=
D^{(\pi)} (\vec{\sigma} _1 \cdot \vec{\Delta} _T)
(\vec{\sigma} _j \cdot \vec{\Delta}_T) 
e^{-\lambda _{\pi} \Delta ^2_T }
\int _0 ^{\infty} 
\exp \{-\alpha (m^2_{\pi}+\vec{\Delta} ^2_T) \} d \alpha
\end{eqnarray}
for the pole amplitude instead of (73). 
In (102) $j=2$ ($j=3$) for $qN$ ($\bar{q}N$) scattering and 
\begin{eqnarray}
\lambda _{\pi}=\frac{r^2_{\pi}}{2}+\alpha '_{\pi}(m^2_{\pi})
\Big [\ln \Big (\frac {s}{s_0} \Big )-i\frac{\pi}{2} \Big ]\;, \\
D^{(\pi)}=\frac{3}{5} \frac{G^2_{\pi N N }}{16 \pi m_N E_0} 
\frac{s_0 e^{- \lambda _{\pi} m^2_{\pi}}}{s-m^2_N-m^2_q} 
(\vec{\tau} _1\cdot\vec{\tau} _j) \;.
\end{eqnarray}
Exponential representation (102) is convenient for calculations of
integrals over transverse momenta $\vec{\Delta} _1$, $\vec{\Delta} _2$,
..., $\vec{\Delta} _n$ in formul\ae$\;$ (35). Putting (102) into (35) one
can
easily get the expression for the total amplitude of $qN$/$\bar{q}N$
scattering due to $\pi P$-exchanges
\begin{eqnarray}
\nonumber
A^{(\pi P)}(\vec{\Delta} _T)=\frac{1}{2}C_{sh}^{(2)}D^{(\pi)} \eta _P(0)
\Big (\frac{s}{s_0} \Big )^{\alpha _P(0)-1} \\
\nonumber
\{ ip_s P_s [ J_2(\vec{\sigma} _{1T}\cdot\vec{\sigma} _{jT})+2 
\lambda ^2_P J_3
(\vec{\sigma} _{1}\cdot\vec{\Delta}_T)
(\vec{\sigma} _{j}\cdot\vec{\Delta}_T)] \\
-J_2[p_yP_s(\vec{\sigma} _{1}\cdot\vec{l}\times\vec{\Delta}_T)
+p_sP_y(\vec{\sigma} _{j}\cdot\vec{l}\times\vec{\Delta}_T)]
-i\Delta _T^2 p_y P_y J_2 \}\;,
\end{eqnarray} 
where we denote by $J_n$ ($n=2,\;3$) the integrals
\begin{eqnarray}
J_n=(m_{\pi})^{n-1} \int _0 ^1
\exp \{-\nu  \Delta ^2_T \} \frac{d \xi}{\varrho^n}  
\end{eqnarray}
with $\varrho$ and $\nu $ given by
\begin{eqnarray}
\nonumber
\varrho=(\lambda _P+\lambda _{\pi}) m^2_{\pi}-\ln \xi \;, \\
\nu =\lambda _P (\lambda _{\pi} m^2_{\pi} -\ln \xi)/ \varrho\;.
\end{eqnarray}
Comparing (105) with (17) one can easily get from 
$A^{(\pi P)}(\vec{\Delta} _T)$ the invariant amplitudes 
$A_j^{(\pi P)}(\vec{\Delta} _T)$ ($j=1,\;2,...,\;6$).

For the $\pi P P$-exchanges we can get with the aid of (102) and (35) the
invariant amplitudes $A_j^{(\pi P P)}(\vec{\Delta} _T)$
\begin{eqnarray}
\nonumber
A_1^{(\pi P P)}(\vec{\Delta} _T)=
-\frac{\chi^{(\pi)}p_yp_sP_yP_s \Delta ^2_T}
{\lambda _P} \int _0^1 e^{-\Upsilon \Delta ^2_T}
\frac{d \xi}{\beta ^2}\;,\\
\nonumber
A_2^{(\pi P P)}(\vec{\Delta} _T)=i \Delta _T \chi^{(\pi)}
\Big \{ \frac{p_s^2P_sP_y}{\lambda _P}
\int _0^1 e^{- \Upsilon \Delta ^2_T}
\frac{d \xi}{\beta ^2}+
\frac{p_y^2P_sP_y}{3 \lambda _P ^2}
\int _0^1 e^{- \Upsilon \Delta ^2_T}
(\lambda _{\pi} +\alpha ) \\
\nonumber
\Big [2- \lambda _{P} \Delta _T^2 (1+\Upsilon/ \lambda _P) \Big] 
\frac{d \xi}{\beta ^3} \Big \}\;,\\
\nonumber
A_3^{(\pi P P)}(\vec{\Delta} _T)=-\frac{\Delta _T^2 \chi^{(\pi)}}
{18 \lambda _{P}^2}
\int _0^1 e^{- \Upsilon \Delta ^2_T}
(\lambda _{\pi} +\alpha ) 
\Big \{ 6 \lambda _P (p_y^2P_s^2+p_s^2P_y^2) \\
\nonumber
+4 p_y^2P_y^2 \Big [1-\Upsilon \Delta ^2_T- \frac{\lambda _P}
{4(\lambda _{\pi} +\alpha )}-\frac{\Upsilon}{2 \beta}
(12-8\Upsilon \Delta ^2_T+\Upsilon ^2 \Delta ^4_T) \Big ] \Big \}
\frac{d \xi}{\beta ^3} +\delta\;, \\
\nonumber
A_4^{(\pi P P)}(\vec{\Delta} _T)=\Delta _T^2 \chi^{(\pi)}
\Big \{ \Big (\lambda _{P} p_s^2P_s^2+\frac{p_y^2P_y^2}{6 \lambda _P}
\Big ) \int _0^1 e^{- \Upsilon \Delta ^2_T}
\frac{d \xi}{\beta ^3} \\
\nonumber
+\frac{1}{3}(p_y^2P_s^2+p_s^2P_y^2)
\int _0^1 e^{- \Upsilon \Delta ^2_T}
(\lambda _{\pi} +\alpha ) (3-\Upsilon \Delta ^2_T)
\frac{d \xi}{\beta ^4} \Big \} +\delta \;, \\
\nonumber
A_5^{(\pi P P)}(\vec{\Delta} _T)=0\;,\\
\nonumber
A_6^{(\pi P P)}(\vec{\Delta} _T)=i \Delta _T \chi^{(\pi)}
\Big \{ \frac{p_s p_y P_s^2}{\lambda _P}
\int _0^1 e^{- \Upsilon \Delta ^2_T}   
\frac{d \xi}{\beta ^2}+
\frac{p_sp_y P_y^2}{3 \lambda _P ^2}
\int _0^1 e^{- \Upsilon \Delta ^2_T}
(\lambda _{\pi} +\alpha ) \\
\Big [2- \lambda _{P} \Delta _T^2 (1+\Upsilon/ \lambda _P) \Big]
\frac{d \xi}{\beta ^3} \Big \}\;,
\end{eqnarray}
where $\alpha$, $\beta$, $\Upsilon$ and $\chi^{(\pi)}$ in (108) denote
\begin{eqnarray}
\nonumber
\alpha= -\frac{1}{m_{\pi}^2} \ln \xi\;,\\
\beta=\lambda _P+2(\lambda _{\pi} +\alpha ) \;,\\
\nonumber
\Upsilon=\lambda _P (\lambda _{\pi} +\alpha )/\beta\;,\\
\nonumber
\chi^{(\pi)}=-\frac{C^{(3)}_{sh}D^{(\pi)} \eta ^2_P(0)}
{2! m_{\pi}^2}
\Big (\frac{s}{s_0} \Big )^{2 \alpha _P(0)-2}\;. 
\end{eqnarray}
The amplitude $\delta$ in (108) looks like
\begin{eqnarray}
\nonumber
\delta=\chi^{(\pi)} \Big \{\frac{p_s^2P_s^2}{\lambda _P}
\int _0^1 e^{- \Upsilon \Delta ^2_T} \frac{d \xi}{\beta ^2}+
\frac{p_y^2P_s^2+p_s^2P_y^2}{3 \lambda _P^2}
\int _0^1 e^{- \Upsilon \Delta ^2_T}
(\lambda _{\pi} +\alpha ) (2-\Upsilon \Delta^2_T)
\frac{d \xi}{\beta ^3} \\
\nonumber
+\frac{p_y^2P_y^2}{18 \lambda _P^3}
\int _0^1 e^{- \Upsilon \Delta ^2_T}
(\lambda _{\pi} +\alpha ) \\
\nonumber
\Big [2-4 \Upsilon \Delta^2_T+
\Upsilon ^2\Delta^4_T+4\frac{\lambda _P}{\Upsilon}+
\frac{\lambda _P^2 \Delta ^2_T}{\lambda _{\pi} +\alpha }
-\frac{\Upsilon }{\lambda _{\pi} +\alpha }
(6-6 \Upsilon \Delta^2_T +\Upsilon ^2\Delta^4_T) \Big ]
\frac{d \xi}{\beta ^3} \Big \}\;.
\end{eqnarray}

\end{document}